\documentclass{elsarticle}
\usepackage{lipsum}
\makeatletter
\def\ps@pprintTitle{%
 \let\@oddhead\@empty
 \let\@evenhead\@empty
 \def\@oddfoot{}%
 \let\@evenfoot\@oddfoot}
\makeatother
\usepackage[utf8]{inputenc}
\usepackage[english]{babel}
\usepackage{amsthm}

\usepackage{xcolor}
\usepackage{float}
\usepackage{subcaption}
\usepackage{amsmath}
\usepackage[toc,title,page]{appendix}
\usepackage{enumerate}
\usepackage{amssymb}
\usepackage[margin=1in]{geometry}
\usepackage{setspace}
\usepackage{lineno}
\usepackage{bbm}
\usepackage{tocbibind}
\usepackage[normalem]{ulem}
\usepackage{epsfig}
\usepackage{amsmath}
\usepackage{graphicx}
\usepackage{float}
\usepackage{vmargin}
\usepackage{mathrsfs}
\usepackage{mathbbol}
\usepackage{url}
\usepackage[parfill]{parskip}
\setmargrb{20mm}{20mm}{20mm}{20mm}

\usepackage{booktabs}
\usepackage{tabularx}
\newcommand*{\rom}[1]{\expandafter\@slowromancap\romannumeral #1@}
	\newcommand{\RNum}[1]{\uppercase\expandafter{\romannumeral #1\relax}}
\newcommand{\be}{\begin{equation}}
\newcommand{\ee}{\end{equation}}
\newcommand{\ba}{\begin{equation} \begin{aligned}}
\newcommand{\ea}{\end{aligned} \end{equation}}
\newcommand{\ddt}[1]{\frac{\mathrm{d}#1}{\mathrm{d}t}}
\newcommand{\indic}[1]{\mathbf{1}_{\{#1\}}}

\usepackage{titlesec}
\titleformat*{\section}{\Large\bfseries}
\titleformat*{\subsection}{\bfseries}
\titleformat*{\subsubsection}{\bfseries}

\usepackage{epstopdf}
\date{}
\graphicspath{{C:/Users/hscovert/Documents/PostDoc_manchester/Matlab/}{../groupfigures/}{submissionfigures/}}
\title{Using statistics and mathematical modelling to understand infectious disease outbreaks: COVID-19 as an example}
\author[1,11]{Christopher E. Overton\corref{cor1} \fnref{fn1}}
\author[1]{Helena B. Stage\corref{cor1}\fnref{fn1}}
\author[6,13]{Shazaad Ahmad}
\author[1]{Jacob Curran-Sebastian}
\author[5,14]{Paul Dark}
\author[1]{Rajenki Das}
\author[3]{Elizabeth Fearon}
\author[5,9]{Timothy Felton}
\author[1,12]{Martyn Fyles}
\author[4]{Nick Gent}
\author[1,4]{Ian Hall}
\author[1,2]{Thomas House}
\author[1,10]{Hugo Lewkowicz}
\author[1]{Xiaoxi Pang}
\author[1]{Lorenzo Pellis}
\author[2]{Robert Sawko}
\author[7,8]{Andrew Ustianowski}
\author[1]{Bindu Vekaria}
\author[1]{Luke Webb}
\address[1]{Department of Mathematics, University of Manchester}
\address[2]{IBM Research, Hartree Centre, SciTech Daresbury}
\address[3]{Department of Global Health and Development, London School of Hygiene and Tropical Medicine}
\address[4]{Emergency Response Department, Public Health England}
\address[5]{Division of Infection, Immunity and Respiratory Medicine, NIHR Biomedical Research Centre, University of Manchester}
\address[6]{Department of Virology, Manchester Medical Microbiology Partnership, Manchester Foundation Trust}
\address[7]{Regional Infectious Diseases Unit, North Manchester General Hospital}
\address[8]{School of Medical Sciences, University of Manchester}
\address[9]{Intensive Care Unit, Wythenshawe Hospital, Manchester University NHS Foundation Trust}
\address[10]{Department of Health Sciences, University of Manchester}
\address[11]{Department of Mathematical Sciences, University of Liverpool}
\address[12]{The Alan Turing Institute}
\address[13]{Manchester Academic Health Sciences Centre}
\address[14]{Critical Care Unit, Salford Royal Hospital, Northern Care Alliance NHS Group}

\cortext[cor1]{Corresponding authors}
\fntext[fn1]{These authors contributed equally to this paper.}
\begin{document}
\begin{abstract}
\noindent During an infectious disease outbreak, biases in the data and complexities of the underlying dynamics pose significant challenges in mathematically modelling the outbreak and designing policy. 
Motivated by the ongoing response to COVID-19, we provide a toolkit of statistical and mathematical models beyond the simple SIR-type differential equation models for analysing the early stages of an outbreak and assessing interventions. In particular, we focus on parameter estimation in the presence of known biases in the data, and the effect of non-pharmaceutical interventions in enclosed subpopulations, such as households and care homes. We illustrate these methods by applying them to the COVID-19 pandemic.

\end{abstract}

\begin{keyword}
COVID19 \sep epidemic modelling \sep parameter estimation \sep outbreak \sep bias \sep intervention
\end{keyword}

\maketitle

%%%% INTRODUCTION___________________________________________________
\section{Introduction}

Mathematical epidemiology is a well-developed field. Since the pioneering work of Ross in malaria modelling~\cite{Ross1910} and Kermack and McKendrick’s general epidemic models~\cite{Kermack1927}, there has been gathering interest in using mathematical tools to investigate infectious diseases. The allure is clear, since mathematical models can provide powerful insight into how these complex systems behave, which in turn can enable these problems to be better controlled/prevented.

Not only is the power of the mathematical tools increasing, but the availability of data on infectious diseases, whether this be a rapid release of data during an outbreak or detailed collection of data for endemic pathogens, is increasing. Rapid interpretation of epidemiological data is critical for the development of effective containment, suppression and mitigation interventions, but there are many difficulties to interpreting case data in real-time. These include interpreting symptom progression and fatality ratios with delay distributions and right-censoring, exacerbated by exponential growth in cases leading to the majority of case data being on recently infected individuals; lack of clarity and consistency in denominators; inconsistency of case definitions over time and the eventual impact of interventions and changes to behaviour on transmission dynamics. Mathematical and statistical techniques can help overcome some of these challenges to interpretation, aiding in the development of intervention strategies and management of care. Examining key epidemiological quantities alongside each other in a transmission model can provide quantitative insights into the outbreak, testing the potential impact of intervention strategies and predicting the risk posed to the human (or animal) host population and healthcare preparedness. 

Mathematical modelling has been used as part of the planning process during outbreak response by governments worldwide for many recent outbreaks. For example the UK Department of Health has a long established committee Scientific Pandemic Influenza group on Modelling, or SPI-M to advise on new and emerging respiratory infections~\cite{SPIM}. One of the largest instances of such an outbreak in recent history was the 2009 H1N1 pandemic. The World Health Organisation developed a network of modelling groups and public health experts to work on exploring various characteristic of the outbreak~\cite{Biggerstaff2020,Kerkhove2012}. These ranged from characterising the dynamics of the outbreak to investigating the effectiveness of different intervention strategies. This integration of mathematics into policy design indicates the important insights that modelling and statistics can provide. 

This paper is a collection of work-streams addressing various technical questions faced by the group as part of the ongoing response to COVID-19, and as such is written to be reflective of the experience we have gone and are currently going through. Therefore, to aid the reader each section includes results and a short discussion. Many of the questions and techniques presented here can be further developed as the availability of data and research interests evolves, but are compiled into this manuscript as an overview of methodology and scientific approaches beyond the standard SIR textbook model that benefit the ongoing efforts in tackling this and other outbreaks.

\subsection{COVID-19 pandemic background}
First documented in December 2019, an outbreak of community-acquired pneumonia began in Wuhan, Hubei Province, China. In January, this outbreak was attributed to a novel coronavirus, SARS-CoV2. The initial spread of the pathogen in Wuhan was fast, and after a period of case-finding and contact tracing, China moved to implement a ‘shutdown’ of Wuhan on January $23$, and other cities in China the following days, to try to suppress the growth of the epidemic. These measures may have succeeded at slowing down the rate at which cases have been seeded elsewhere, but in many countries initial importation of cases and transmission has not been contained. Countries around the world are now seeing outbreaks that are overwhelming, or have the potential to overwhelm, healthcare systems and cause a high number of deaths even in high-income countries~\cite{Remuzzi2020}.

While the majority of documented symptomatic cases are mild, characterised in many reports by persistent cough and fever, a significant proportion of these individuals go on to develop pneumonia, with some then developing acute respiratory failure and a small proportion of overall cases becoming fatal. Severity of symptoms has been observed to increase with age and with the presence of underlying health conditions such as diabetes~\cite{Fang2020} and cardiac conditions, with some evidence that severity of symptoms might depend on gender and ethnicity~\cite{Guan2020,Rimmer2020,WuJT2020,WuZ2020,Zhou2020}.

SARS-CoV2 has a fast doubling time (the time it takes for the number of cases in the region to double, estimated at approximately 3 days~\cite{pellis2020challenges}) and, potentially, a very large $R_0$ (the average number of infections caused by each infected individual, with estimates ranging from 1.4 to 6.47~\cite{Liu2020a,Mahasem308,Majumder2020b, whositrep}). It is possible that there is a significant degree of asymptomatic and/or pre-symptomatic transmission~\cite{LiR2020,Mizumoto2020,Nishiura2020}, though without robust serosurveys, this is difficult to quantify with certainty. These characteristics result in the pathogen being able to spread widely, rapidly and undetected, presenting a significant risk to public health. 

Typically, the aim of an intervention strategy
would be to push and keep the reproduction number $R_t$, defined as the average number of cases generated by a typical infective at time $t$, below 1. At this point each infected individual subsequently infects, on average, less
than one individual, such that the number of cases should decline. The basic reproduction number, $R_0$, represents the initial value of $R_t$, before any intervention is put in place. 

High $R_0$, fast growth, and possible pre- or asymptomatic infection make the design of potential interventions, and the modelling that would inform them, particularly challenging. Large values of $R_0$ mean a substantial amount of transmission needs to be halted; fast growth causes the number of cases in the absence of interventions to rise rapidly, so that the time scale of interventions to reduce $R_0$ must also be fast in order to effect substantive early changes on a population level; finally, the resulting interventions must encompass possible pre- and asymptomatic cases, a challenging prospect when in many instances these individuals are indistinguishable from healthy individuals. Consequently, we must consider the possibility of interventions that are massively disruptive to society and may have to be sustained for a long period of time in order to cause the number of infections to decline towards zero~\cite{Ferguson2020}. If infections remain, and the susceptible proportion of the population remains above the herd immunity threshold, these interventions must be upheld to prevent a second wave of the epidemic. There is not yet conclusive evidence as to the degree and duration of immunity conferred by infection with SARS-CoV2 nor the feasibility of a vaccine, the timeline for which is unlikely to be any time shorter than 18 months away at the time of writing~\cite{bbcvaccine}. Therefore, short term extreme interventions are not as effective as they might be in other circumstances, since after their removal there remains a long period of time in which cases can rise again. The longer these significantly suppressive and disruptive interventions are in effect, the more severe the effect on the economy, and broader societal health and well-being. Furthermore, adherence to interventions will likely vary with their duration and severity.

We are further challenged by the lack of transferable intuition. Early work looked at intuition gained from SARS and MERS outbreaks, also caused by coronaviruses. Some parameters do appear to be similar to these pathogens, such as the average length of the incubation period~\cite{Lauer2020,Varia2003,Virlogeux2016}. However, there are also clear differences, with both SARS and MERS being more fatal, but seemingly less efficient at spreading since they did not seed major global pandemics. Another complication is the spread of the infection during the Chinese Spring Festival, a time period during which movement, social, and contact patterns vary significantly. This presents significant challenges as experience and intuition from other studies regarding population mixing and spatial patterns must either be modified or are invalid. Furthermore, the pandemic has received a proportionately larger level of public attention than e.g.\ the 2009 H1N1 pandemic~\cite{Chew2010,Rubin2009}, largely boosted by social media. This greater level of public awareness, and the successive, staggered interventions placed to prevent disease spread are responsible for significant variations in behaviour~\cite{Butler2014,Funk2009} and adherence to public guidance both in China and abroad. 

The structure of this paper follows two main themes. In Section~\ref{sec:stats}, we discuss various biases that are present in outbreak data and techniques for estimating epidemiological parameters. Accounting for biases and producing robust parameter estimates is important throughout the duration of an epidemic, both for increasing our understanding of the underlying dynamics, and for feeding into models. Firstly, we discuss a bias-corrected method for estimating the incubation period, which can also be applied to serial intervals, onset-to-death time, and other delay distributions. We then present a method for estimating the true growth rate of the epidemic, accounting for the bias encountered since infected individuals may be exported from the region. Our next method is a tool for estimating the expected size of the next generation of infectives based on the rate of observed cases. This tool provides insight into the size of small outbreaks, which can inform decision making when trying to prevent a major outbreak taking off. 
We end this section with a selection of bias corrected estimates of: incubation period, doubling time, and onset-to-hospitalisation.

In Section~\ref{sec:models}, we propose a variety of mathematical models looking at disease impact and intervention strategies, with particular focus on non-pharmaceutical interventions due to the current lack of widely deployable, targeted pharmaceutical treatments. These models focus on enclosed populations, since this is the level at which most interventions are implemented. Since the disease is particularly fatal in the elderly and other at risk groups, we develop a care home model to investigate how the pathogen may spread through care homes. We also develop household models to investigate the impact of different intervention/control strategies. These models can inform policy design for mitigating or controlling epidemic spread. Finally, in the context of relaxing strong social distancing policies, we investigate the extinction probability of the pathogen. We first consider the extinction probability after lifting restrictions. We then develop a household-based contact tracing model, with which we investigate the extinction probability under weaker isolation policies paired with contact tracing, thus shedding light on possible combination of interventions that allow to feasibly manage the infection while minimising the social impact of control policies.

%%%%_________________________________________________________________________________________

%%%% BIASES___________________________________________________________________________________
\section{Biases and estimation during outbreaks}\label{sec:stats}
\subsection{Potential biases in the outbreak data}
\label{sec:data}
Techniques are constantly developing that enable higher volumes of more accurate data to be collected real-time during an epidemic. These data present a large opportunity for analysis to gain insight into the pathogen and the dynamics of the outbreak. However, although the quality of the data is constantly increasing, there are still many biases present. Some of these are due to the data collection methods, and in an ideal world we would be able to eliminate them, and some are simply due to the nature of the outbreak, and will be present regardless of data collection methods.

During an outbreak, many parameters depend on delay distributions (the length of time between two events), such as the time from infection to symptoms onset (the incubation period). If an individual can be followed indefinitely, it is easy to determine the length of these events. However, in reality only events that occur before a given date are observed. Therefore, the data is subject to censoring and truncation issues. In the incubation period, for example, censoring comes into play since, if we have observed an infection but the individual has not yet developed symptoms, we only have a lower bound on how long it will take them to develop symptoms. To account for this, we can instead condition on observing symptom onset before the cut-off date. However, this leads to a truncation issue, since individuals who were infected close to the cut-off date will only be observed if they have a short incubation period, which leads to an overexpression of short delays.

The number of cases tends to grow exponentially during the early stages of an outbreak, causing the force of infection and the number of reported cases to increase with time. This further complicates the truncation issue since not only are recent cases truncated but they also account for the majority of cases. 
The growing force of infection also needs to be accounted for, since if the potential time of infection is interval-censored rather than observed directly, the probability that the case was infected in each day of that interval is not constant.

In theory, both of these biases are relatively straightforward to account for. In practice however, there are other biases in the data. One of the major biases is the reporting rate. Although the total number of cases may be reasonably described as growing exponentially with a constant rate in the early stages of an outbreak, high-resolution data may exhibit more complex behaviour. This can be due to a variety of reasons, such as the workload becoming overwhelming, the availability of individual-level data decreasing, the laboratories or offices slowing down activity over the weekend, the case definition changing, the testing capabilities increasing, and so on.

Another uncertainty arises since generally only the date of each event is recorded rather than the time. This presents a large window of uncertainty in the length of the delay, since the time of each event can vary up to 24 hours, and for a delay distribution, which depend on two events, it could vary by up to 48 hours.

Travel rate is another bias present in the data. For example, this changes the density of observed cases in a region, which can change the apparent growth rate. Intervention strategies present a further bias because this can change the growth rate of the epidemic and the reporting rate. Additionally, estimates of certain parameters may vary depending on the interventions that are implemented, so these need to be considered carefully. 
%%%%_________________________________________________________________________________________

%%%% INCUBATION_______________________________________________________________________________________
\subsection{Incubation period}
\label{sec:incubation}
\begin{figure}
        \centering
\begin{subfigure}[t]{0.45\textwidth}
        \includegraphics[width=1\linewidth]{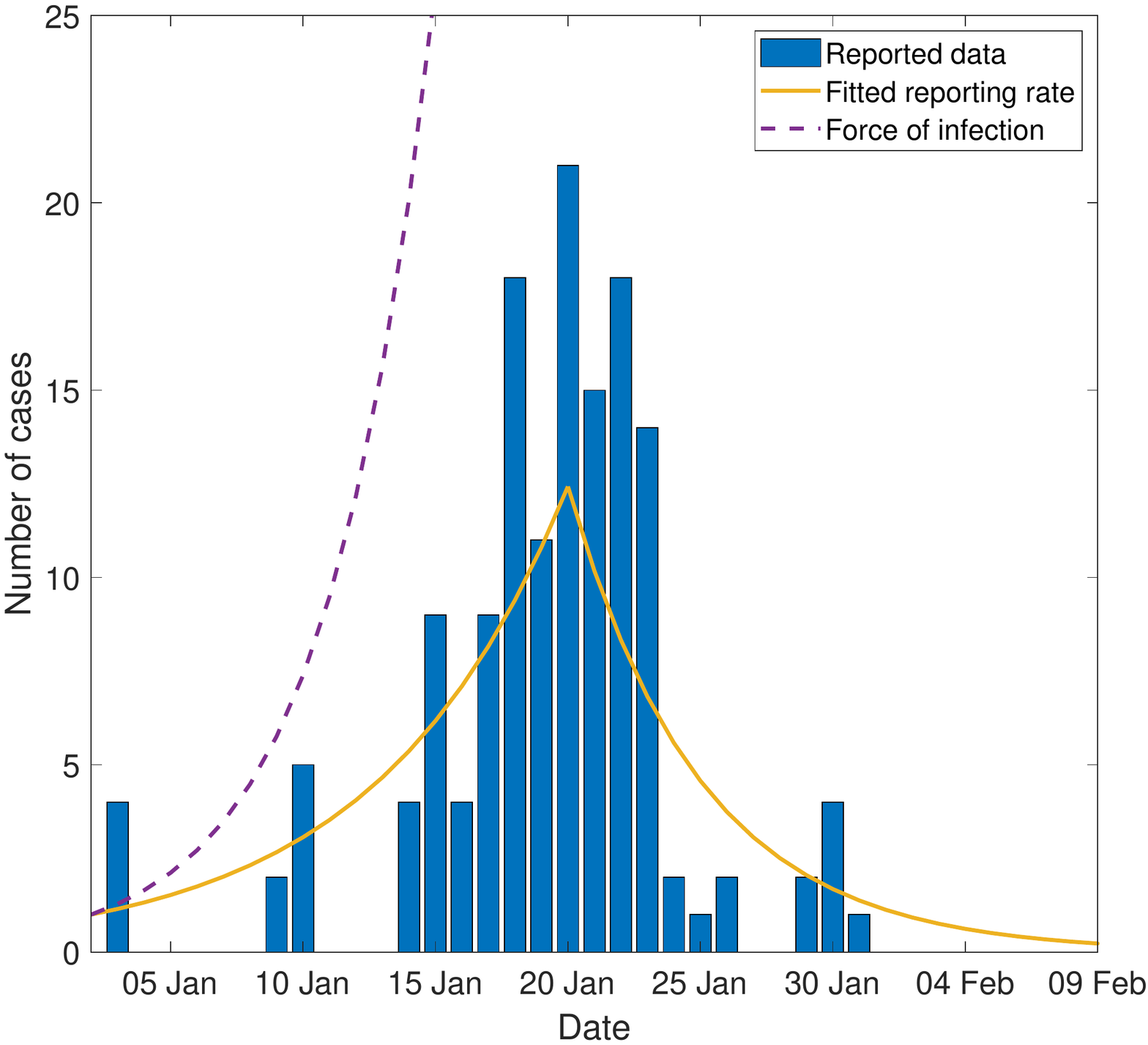}\vspace{-2cm}
        \caption{Illustrating the changing reporting rate for the sample as of $9$ February. The bars indicate the number of new cases corresponding to the date the individual left Wuhan, and the yellow curve indicates the fitted reporting rate for the data in the dataset. The dotted purple line indicates the force of infection with growth rate $r=0.25$.}
        \label{fig:growth}
\end{subfigure}\hspace{1cm}
\centering
\begin{subfigure}[t]{0.45\textwidth}
        \includegraphics[width=1\linewidth]{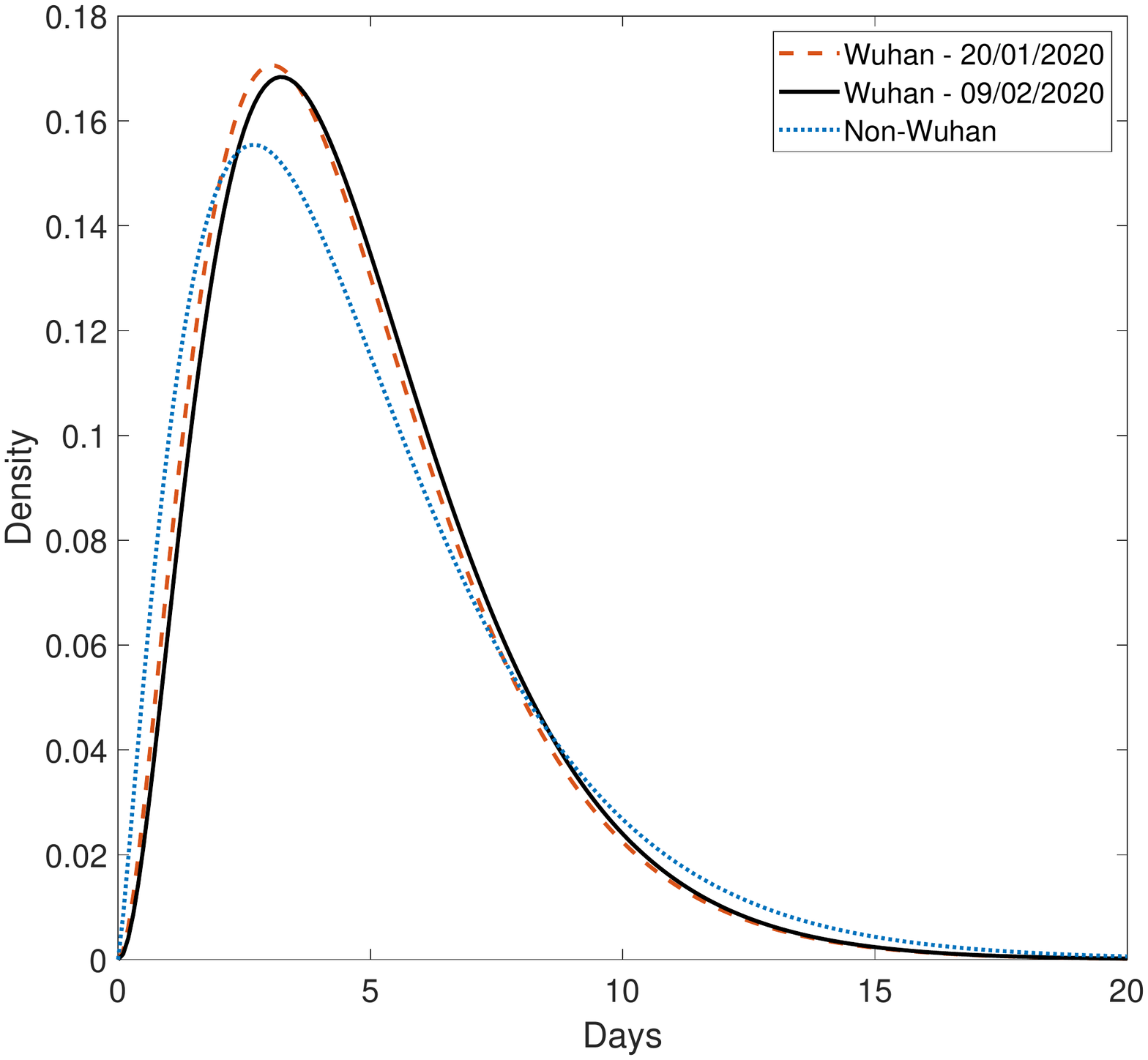}\vspace{-2cm}
        \caption{Figure showing the maximum likelihood distributions for the estimated length of the incubation period in days for the three different data sets (with the truncation correction). All three estimates give rise to very similar distributions.}
        \label{fig:MLE}
\end{subfigure}
\caption{Reporting rate (a) and maximum likelihood distributions (b) for the COVID-19 incubation period.}
\end{figure}

To model the incubation period, we require information regarding when an individual was infected and when they expressed symptoms. Observing exact time of infection is unlikely, but it can be possible to find potential exposure windows. We consider three different data sets. The first two consist of individuals who travelled from Wuhan before expressing symptoms. We can assume these individuals were infected in Wuhan, since at the time of this data, the force of infection was significantly higher in Wuhan than elsewhere. The length of time spent in Wuhan therefore provides a window during which each individual became infected, and for many of these individuals we also have the date of symptom onset. In the early stages, the growth rate in reported cases was constant, and dependent on the epidemic growth rate in Wuhan and the rate at which people left Wuhan. By using travel to estimate the true number of cases, we estimate the exponential growth rate in Wuhan as $r=0.25$ (see Section~\ref{sec:transport}). Therefore, the force of infection on day $i$, $g(i)$, is proportional to $\mathrm{e}^{0.25i}$. After the $23$ of January when significant travel bans were introduced, the rate at which individuals left Wuhan diminished significantly, causing the reporting rate to suddenly drop. Therefore, if the data is truncated after the $23$ of January, the reporting rate must be appropriately adjusted. This is illustrated in Figure~\ref{fig:growth}. The difference between these two datasets is the truncation date, with the first truncated at $20$ January and the second at $9$ February. The third dataset contains cases that were infected through a discrete infection event, such as spending time with a known infected case. In this ``non-Wuhan" dataset, the reporting rate is constant and the force of infection can be assumed constant over each exposure window. The source we use for these three data sets is a publicly available line-list~\cite{Sun2020}. 

Incubation periods, and many other delay distributions, are generally observed to have right skewed distributions. We therefore choose to use a Gamma distribution, though other distributions can also be applied using the proposed methods, such as Weibull and Log-normal. To fit the data, we use maximum likelihood estimation. To adjust for the biases we use a ``forwards'' approach~\cite{Nishiura2010,Tomba2010,Sun1995,Svensson2007}, where we condition on the time of the first event, time of exposure, and find the distribution looking forward to the second event, time of symptom onset. For a data point $\{a_i,b_i,y_i\}$, where infection occurs between $a_i$ and $b_i$, and $y_i$ is the symptom onset date, the likelihood function is given by
\begin{equation*}
L(y_i|a_i,b_i,\theta)=\frac{\int_a^bg(i)f_\theta(y-i)\mathrm{d}i}{\int_a^b\int_0^{T-i}g(i)f_\theta(x)\mathrm{d}x\mathrm{d}i},
\end{equation*}
where $g(\cdot)$ is the density function of the infection date and $f_\theta(\cdot)$ is the density function of the incubation period parameterised by $\theta$. From this, the likelihood function for our dataset $X$ is given by
\begin{equation*}
L(X|\theta)=L(\bigcap_i y_i|\bigcap_i (a_i\cap b_i),\theta)=\prod\limits_i L(y_i|a_i,b_i,\theta).
\end{equation*}
This approach is independent of the reporting rate bias, since the reporting rate depends on the date an individual leaves Wuhan ($b_i$), which is conditioned against (see Appendix~\ref{app:reporting}). We use the mean and standard deviation to characterise the MLE. Since the tail of the incubation period is important when designing quarantine strategies, we then calculate the probability that the incubation period is longer than 14 days and find the minimum day by which 99\% of cases will have expressed symptoms (excluding true asymptomatic cases). We also investigate the reporting date uncertainty mentioned in Section~\ref{sec:data} by considering the different extremes that the data could represent. This is achieved through adding or subtracting a day to all recorded data.

Methods accounting for truncation and growth biases in epidemic data have been discussed widely in the literature~\cite{Kalbfleisch1991,Nishiura2010,Su2012,Taylor2003}, however there are fewer applications to outbreaks~\cite{Farewell2005}. In the context of COVID-19, estimates have considered growing force of infection, for example~\cite{Lauer2020}, and some approaches have considered truncation, for example~\cite{Linton2020}. However, these attempts do not adjust for the reporting rate in the data or use the correct force of infection, causing the incubation period to be overestimated. 

\subsubsection{Truncation}
Here we demonstrate the importance of truncation (Table~\ref{Table:Results1}). We use the data truncated at $20$ January, which has exposure windows between $1$ December and $19$ January. This data set is chosen since it is most sensitive to truncation due to the exponentially growing force of infection and high reporting rate. Without accounting for truncation, the length of the incubation period is significantly underestimated, which could have a large impact on the success of intervention strategies.

\begin{table}[h!]
\caption{Effect of accounting for truncation on the incubation period} 
\begin{center}
{
\begin{tabularx}{\textwidth}{|X| l| c |c|c|c|c|}
\hline
Method & Mean & Standard deviation & 14 day risk & $99 \%$ confidence date & Sample size \\ \hline
Uncorrected & 3.49 & 2.05 & 0.00060 & 10 & 65 \\ \hline
Truncation corrected & 4.69 & 2.78 & 0.0075 & 14 & 65\\ \hline
\end{tabularx}
\label{Table:Results1} 
}
\end{center}
\end{table}
\subsubsection{Different data sets}
To demonstrate the effectiveness of the bias correction method, we compare three different data sets (Table~\ref{Table:Results2}). The similar distributions predicted across these datasets suggests a robust method. Figure~\ref{fig:MLE} compares the full distributions for these three estimates.

\begin{table} [h!]
\caption{Effect of different data sets on the incubation period} 
\begin{center}
{
\begin{tabularx}{\textwidth}{|X| l| c |c|c|c|c|}
\hline
Method & Mean & Standard deviation & 14 day risk & $99 \%$ confidence date & Sample size\\ \hline
Wuhan - 20/01/2020 & 4.69 & 2.78 & 0.0075 & 14 & 65 \\ \hline
Wuhan - 09/02/2020 & 4.84 & 2.79 & 0.0081 & 14 & 162\\ \hline
Non-Wuhan & 4.84 & 3.22 & 0.016 & 16 & 52 \\ \hline
\end{tabularx}
\label{Table:Results2} 
}
\end{center}
\end{table}

\subsubsection{Reporting date uncertainty}
Here we investigate the effect that uncertainty in the reporting date can have on the estimations, using the data truncated at $9$ February (Table~\ref{Table:Results3}). The standard interval is the recorded data, wide intervals are obtained by removing a day from the exposure window lower bound and adding a day to the upper bound, and the narrow interval vice versa. The uncertainty in the reporting date can impact the estimated incubation period, showing that it is important to consider this risk when designing interventions. 

\begin{table} [h!]
\caption{Effect of uncertainty in the reporting date on the incubation period} 
\begin{center}
{
\begin{tabularx}{\textwidth}{|X| l| c |c|c|c|c|}
\hline
Method & Mean & Standard deviation & 14 day risk & $99 \%$ confidence date & Sample size\\ \hline
Standard Intervals & 4.84 & 2.79 & 0.0081 & 14 & 162 \\ \hline
Wide Intervals & 4.21 & 2.56 & 0.0041 & 13 & 162\\ \hline
Narrow Intervals & 5.55 & 2.86 & 0.0112 & 15 & 162\\ \hline
\end{tabularx}
\label{Table:Results3} 
}
\end{center}
\end{table}

\subsubsection{Implications}

When constructing intervention strategies for an epidemic, the incubation period is an important parameter. For example, consider the quarantine strategy deployed in many countries during the early stages of the epidemic, aimed at preventing cases being imported from Wuhan. This strategy quarantined individuals upon their return from Wuhan for 14 days. For such a strategy to be effective, we require most incubation periods to be less than 14 days, so that the majority of infected people would develop symptoms before quarantine ended, enabling them to be further isolated. In this analysis, we show that in the worst-case scenario we would expect 1 in 62 cases to slip through this quarantine, with the best fit predicting 1 in 101 cases. Therefore, the 14 day quarantine period would capture the majority of cases. Throughout the epidemic, this seems to have been reasonably successful, and prevented early seeding of cases in many countries. However, potentially due to complicated travel patterns or asymptomatic transmission, cases have slipped through detection and not been quarantined, which unfortunately has led to the situation observed today. 

In addition to the incubation period, there are many other delay distributions that must be estimated while an epidemic is growing related to infectious diseases that can be estimated using the same technique. These include the generation time, the time between two infection events in a transmission chain; the serial interval, the time between symptom onset of an infector to their infectee; and the onset-to-death delay, the time from symptom onset to death.

%%%%__________________________________________________________________________

%%%%% TRANSPORT___________________________________________________________________
\subsection{Transportation modelling and under-reporting}
\label{sec:transport}

Transportation modelling plays a crucial role in the early stages of an outbreak; an infected individual may travel outside of the region in which they were originally infected and seed further infections across geographical scales which are impossible to contain. Furthermore, as the rate of travelling increases, the number of observed cases within the known ``origin" region decreases, and if exportation is not taken into account this results in an underestimation of the number of cases. These underestimates can be improved by looking at the total number of cases across all known affected regions, but doing so introduces further complications. For example, if an individual has less severe symptoms they may not seek medical assistance, thereby not being recorded as a case at their destination. This underestimation of cases can have significant effects if the traveller is able to infect more people. A new transmission chain can thus be started which remains undetected for some time due to a lacking known connection to the ``origin" region.

In the ``origin" region an individual with mild symptoms may still be tested for an infection due to a higher level of alertness in the local health care system. However, this level of active case-finding may not be present elsewhere, or may not have been allocated a comparable level of resources. Further complications to this model arise from the incubation period of individuals wherein detection is unlikely, and the variations in movement and mixing between people when preventative measures are put in place.

We consider a metapopulation model seeded with an infection in one of the regions, $O$, and investigate how exportation from this region combined with variability in case-finding can alter estimates for the doubling time and the expected portion of the population we expect to identify. This accordingly bounds the proportion of the infected population one would be able to target for personal intervention (e.g. quarantine or treatment). Note that the proportion of identified cases need not necessarily correlate with the proportion of the infected population who exhibit symptoms.

Let us assume that movement from $O$ begins at time $t=t_c$, and occurs with a constant rate $\rho$; this can be thought of as the surge in travel in China during the beginning of the Spring Festival. In the early phase of an epidemic, we can assume the incidence $I(t)=I_0 \mathrm{e}^{rt}$ of cases to be growing exponentially with a rate $r$. The number of cases at time $t$ which were infected a time $\tau$ ago is denoted by $i(t,\tau)$, where the probability of detecting a case that infected a host a time $\tau$ ago is given by $p(\tau)=\Theta(\tau-\tau_{\mathrm{inc}})$ where $\Theta(\cdot)$ is the Heaviside step function and $\tau_{\mathrm{inc}}$ is the incubation period. That is, a case is detected immediately following the end of the incubation period. We assume that the incubation period is characterised by a gamma-distribution with mean $4.84$ and standard deviation $3.22$ (Table~\ref{Table:Results2}), since this estimate is independent of the exponential growth rate (Section~\ref{sec:incubation}).
The number of observed cases in $O$ is given by
\begin{equation*}
C_O(t)=I_0e^{-\rho(t-t_c)\Theta(t-t_c)}\int_0^{t}\frac{1}{\Gamma(k)\theta^k}\tau_{\mathrm{inc}}^{k-1}e^{-\tau_{\mathrm{inc}}/\theta}\int_0^t e^{r(t-\tau)}\Theta(\tau-\tau_{\mathrm{inc}})\mathrm{d}\tau \mathrm{d}\tau_{\mathrm{inc}} ,
\end{equation*}
where we have assumed that recovery of cases is negligible over the time scale of case observations. If we consider travel to $i$ other regions from $O$, the total number of observed cases in all destinations is
\begin{equation*}
C_D(t)=\sum_{i\neq O}C_i(t)=\omega C_O(t)\left(e^{\rho(t-t_c)\Theta(t-t_c)}-1\right),
\end{equation*}
where $\omega$ is the mean case-finding ability across all destinations. In the presence of real-time transition probabilities $p_{ij}$ of moving between two regions, these estimates can be further elaborated.

Historic estimates for Chinese travel data indicate a mean travel rate from Hubei province of $\rho=0.029$ which began on January 10 \cite{travel2,travel1}. Assuming an incubation period of approximately 5 days for the infection, this suggests a rate $r=0.22\pm0.01$ when ignoring travel exportation, in contrast to $r=0.25\pm0.01$ when accounting for $\rho$. This difference may seem small, but it reduces the doubling time by approximately 12 hours. The expected value of $r$ grows linearly with the exportation rate, which has also been observed with real-time travel models \cite{Kraemer2020}. Further models have also been developed which consider travel and exportation of cases in greater detail \cite{Chinazzi2020,Gostic2020}. 

The relationship between the observed cases in our origin and destinations can be used to determine the case-finding ability, though it should be noted that $\omega$ likely varies with time as burdens are increased on public services and the number of cases grow. Early estimates using data from~\cite{CNInc} indicate at most an $80\%$ case-finding ability, suggesting thousands of undetected cases exported to other regions of China, a sufficient quantity to sustain further transmission post-exportation independently of the number of asymptomatic cases present.

The intention of these estimates is not to provide specific values for the doubling time of the spread of COVID-19 in China (as the estimates above use historic travel data and are limited by the availability of data), but to bring attention to the unusual circumstances surrounding changes in contact patterns, and mobility during the Chinese Spring Festival, the largest human migration on Earth~\cite{travel2}. Failing to account for the significant level of dispersion or exportation of cases during these circumstances will significantly skew our estimates.

%%%%_______________________________________________________________________

%%%% GEN SIZE______________________________________________________________
\subsection{Estimating the size of the first generation from the observed number of symptomatic individuals}
\label{sec:gen-size}
In a scenario where a single individual exposes a group, it can be unclear how many people have been infected since they do not immediately develop symptoms. However, knowing the true prevalence in the population is essential to determine the most effective interventions to put in place, and to estimate future burdens on public services. 
Using the probability density function of the incubation period, we consider the efficacy of using the time it takes for people to present with symptoms as a predictor for the size of the infected group. 
This analysis is an effective ready reckoner at early stages of a novel infection, or in close contact environments, and is useful for predicting generation size when a complete data set is not yet available. In this analysis we focus on a scenario where infection time is known. In reality, we may only know an exposure window. For short exposure windows this method can still be valid, but for longer exposure windows it will need extending to account for this added uncertainty.

We assume that the number of individuals who have been exposed to potential infection is known, in which case the number of people who are infected can be assumed to be binomially distributed with an unknown probability $P$ that each individual has been infected.
To determine the distribution of infected individuals, we use the available information regarding the number of individuals who have expressed symptoms. 
This yields two cases. In the first case, we assume that the true number of symptomatic individuals are observed. In the second case, we take the number of observed symptomatic individuals as a lower bound on the true value. 

We wish to determine the probability that the first generation has $e_0$ individuals, $E_0=e_0$, given that $i_\tau$ symptomatic individuals are observed on day $\tau$, $I_\tau=i_\tau$, which is given by
\begin{eqnarray}
    \mathbb{P}\left(E_0=e_o\middle|I_\tau=i_\tau\right)&=&\int_0^1\mathbb{P}\left(E_0=e_0\cap P=p\middle|I_\tau=i_\tau\right)\mathrm{d}p\\
    &=&\int_0^1\mathbb{P}\left(E_o=e_0\middle|I_\tau=i_\tau\cap P=p\right)\times \mathbb{P}\left(P=p\middle|I_\tau = i_\tau\right)\mathrm{d}p\\
    &=&\int_0^1\frac{\mathbb{P}\left(E_0=e_0\cap I_\tau=i_\tau\middle|P=p\right)}{\mathbb{P}\left(I_\tau=i_\tau\middle|P=p\right)}\times \mathbb{P}\left(P=p\middle|I_\tau=i_\tau\right)\mathrm{d}p
\label{eqn:H-1}
\end{eqnarray}
To solve this, we need to determine the distribution of the infection probability $P$ given the number of observed symptomatics. Assuming that $P$ is uniformly distributed, we have
\begin{eqnarray*}
\mathbb{P}\left(P=p\middle|I_\tau=i_\tau\right)&\propto& \mathbb{P}\left(I_\tau=i_\tau\middle|P=p\right)\\
&=&\binom{n}{i_\tau}\left(p F\left(\tau\right)\right)^{i_\tau}\left(1-p F\left(\tau\right)\right)^{n-i_\tau}\\
\mathbb{P}\left(P=p\middle|I_\tau=i_\tau\right) &=& c\times\binom{n}{i_\tau}\left(p F\left(\tau\right)\right)^{i_\tau}\left(1-p F\left(\tau\right)\right)^{n-i_\tau},
\end{eqnarray*}
where $F(\cdot)$ is the cumulative density function for the incubation period, $n$ is the number of initially exposed individuals, and $c$ is a normalising constant such that
\begin{align*}
c = \left(\int_0^1\binom{n}{i_\tau}\left(p F\left(\tau\right)\right)^{i_\tau}\left(1-p F\left(\tau\right)\right)^{n-i_\tau}\mathrm{d}p\right)^{-1}\\
=\frac{\left(\binom{n}{i_\tau}\right)^{-1}\left(i_\tau+1\right)\left(F(\tau)\right)^{-i_\tau}}{_2F_1\left(i_\tau+1,i_\tau-n,i_\tau+2,F\left(\tau\right)\right)}\\
\mathbb{P}\left(P=p\middle|I_\tau=i_\tau\right) = \frac{p^{i_\tau}\left(i_\tau+1\right)\left(1-p F\left(\tau\right)\right)^{n-i_\tau}}{_2F_1\left(i_\tau+1,i_\tau-n,i_\tau+2,F\left(\tau\right)\right)},
\end{align*}
where $_2F_1$ represents the hypergeometric function~\cite{Abramowitz1948}.
Substituting this into Equation~(\ref{eqn:H-1}) gives

\begin{eqnarray*}
      \mathbb{P}\left(E_0=e_o\middle|I_\tau=i_\tau\right) &=&\int_0^1\frac{\binom{n}{e_0}p^{e_0}\left(1-p\right)^{n-e_0}\binom{e_0}{i_\tau}F\left(\tau\right)^{i_\tau}\left(1-F\left(\tau\right)\right)^{e_0-i_\tau}}{\binom{n}{i_\tau}\left(pF\left(\tau\right)\right)^{i_\tau}\left(1-pF\left(\tau\right)\right)^{n-i_\tau}}\\
    &&\times \frac{p^{i_\tau}\left(i_\tau+1\right)\left(1-pF\left(\tau\right)\right)^{n-i_\tau}}{_2F_1\left(i_\tau+1,i_\tau-n,i_\tau+2,F\left(\tau\right)\right)}\mathrm{d}p\\
    &=&\int_0^1p^{e_0}\left(1-p\right)^n\times\left(\frac{1-F\left(\tau\right)}{1-p}\right)^{e_0}\times\left(1-F\left(\tau\right)\right)^{-i_t}\frac{\left(n-i_\tau\right)!\left(i_\tau+1\right)}{\left(n-e_0\right)!\left(e_0-i_\tau\right)!}\\
    &&\times\frac{1}{_2F_1\left(i_\tau+1,i_\tau-n,i_\tau+2,F\left(\tau\right)\right)}\mathrm{d}p,
\end{eqnarray*}
which simplifies to
\begin{eqnarray*}
    \mathbb{P}\left(E_0=e_o\middle|I_\tau=i_\tau\right) &=&\frac{e_0!}{\left  (e_0-i_\tau\right)!}\left(1-F\left(\tau\right)\right)^{e_0}\times\frac{\left(1-F\left(\tau\right)\right)^{-i_\tau}\left(n-i_\tau\right)!\left(i_\tau+1\right)}{\left(n+1\right)!_2F_1\left(i_\tau+1,i_\tau-n,i_\tau+2,F\left(\tau\right)\right)}.
\end{eqnarray*}

This gives a distribution of the generation-size based on the number of observed symptomatic individuals by time $\tau$. We can extend it to investigate a scenario where no symptomatic individuals have been observed by time $\tau$ by using a value of 0 for $I_\tau$:

\[\mathbb{P}\left(E_0=e_0\middle|I_\tau=0\right)=\frac{\left(1-F\left(\tau\right)\right)^{e_0}}{\left(n+1\right) {_2F_1}\left(1,-n,2,F\left(\tau\right)\right)}\]

This can be used to illustrate worst and best case scenarios given $\tau$ time has passed without symptomatic individuals. Additionally, if we consider the probability that $E_0$ = 0, we can find the value of $\tau$ where we can have a 95\% confidence that there will not be a second generation:

\[\mathbb{P}\left(E_0=0\middle|I_\tau=0\right)=\frac{1}{\left(n+1\right) {_2F_1}\left(1,-n,2,F\left(\tau\right)\right)}>0.95\]

This analysis considers the case when the number of observed symptomatic individuals to date is the true number. In practice however, we do not generally observe every symptomatic individual, so the number of observations is only a lower bound on the true number. To address this, rather than considering $I_\tau$ as the total number of people who have developed symptoms by time $\tau$, we can define $\tilde{I}_\tau$ as the minimum number of people who have developed symptoms by time $\tau$. We assume that the probability that $\tilde{I}_\tau$ is equal to $\tilde{i}_\tau$ for a given value of $i_\tau$ is uniform at $\frac{1}{i_\tau+1}$. We can then use the same methods as above to infer a distribution for $P$. Details are provided in Appendix~\ref{app:Hugo}.
\begin{figure}[h]
\begin{center}
\includegraphics[width=0.7\textwidth]{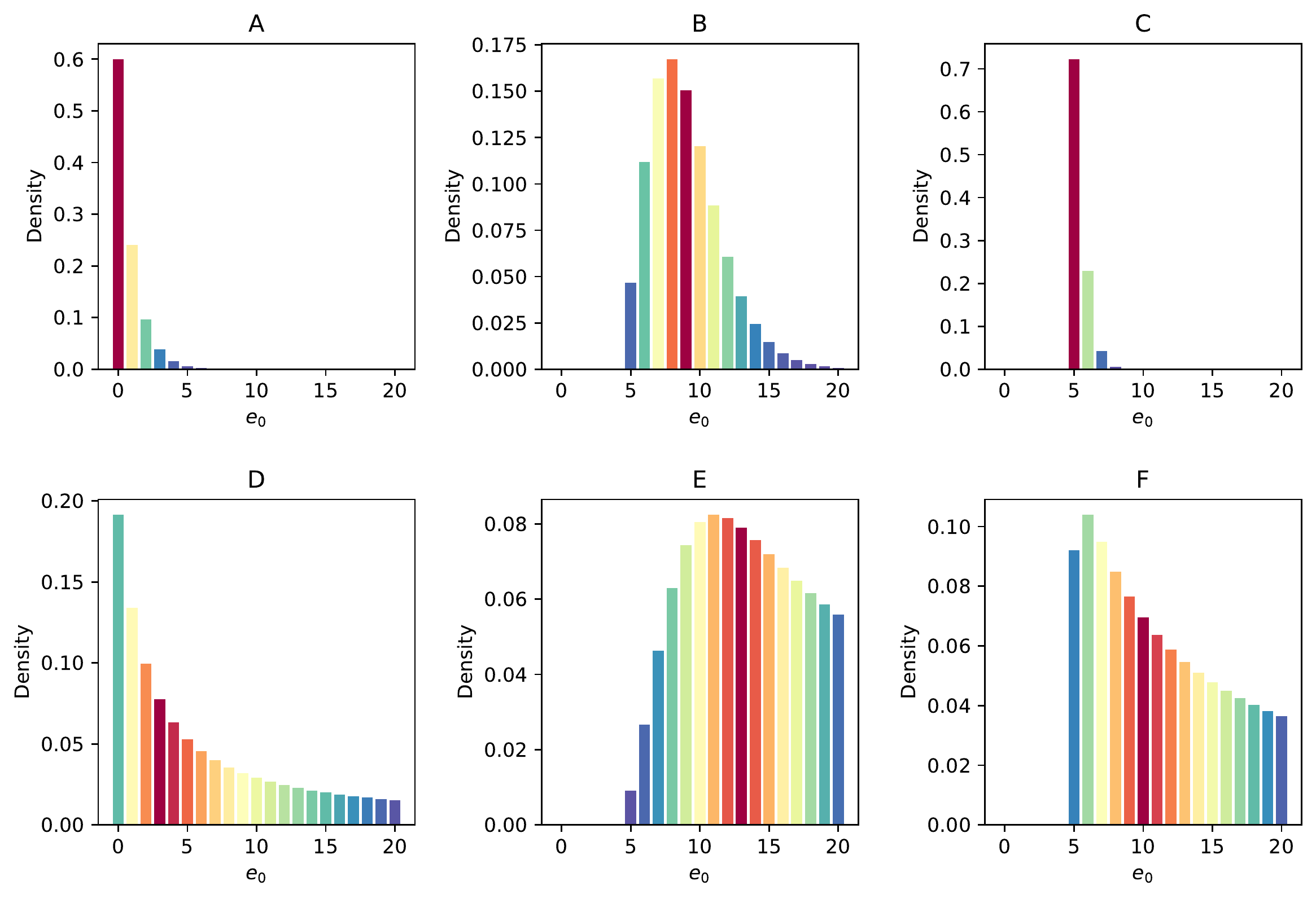}
\caption{Prediction of the size of the first generation, $e_0$, in an infection event in which 20 people were exposed. A, B and C show the density when the number of observed symptomatics is taken to be the true number of symptomatics, D, E and F consider the case where the observed symptomatics is a lower bound on the true symptomatics. A and D consider the case when zero symptomatics are observed after 5 days, B and E when 5 are observed after 5 days, and C and F when 5 are observed after 10 days. The incubation period for the disease has been modelled as a gamma distribution with a mean of 4.84 and standard deviation of 2.79 (Table~\ref{Table:parameters2}).}
\label{fig:Hugo}
\end{center}
\end{figure}

As we can see from Figure~\ref{fig:Hugo}, this method can be used to predict the number of infected individuals in the original exposed group. However, we have also demonstrated the importance of caution when interpreting this data. If there is uncertainty surrounding the presentation of symptomatic patients, using $\tilde{I}_{\tau}$ as a lower bound is a robust method to ensure the size of the generation is not underestimated.

%%%%________________________________________________________

%%%% PARAMETERS____________________________________________
\subsection{COVID-19: Indicative parameters}
Here, we provide a table with some of the key parameters for COVID-19. From Table~\ref{Table:parameters1}, it can be seen that parameter estimates vary widely across different reports. This can be due to the biases in the data not being correctly accounted for, such as the growth rate of the epidemic, travel patterns and time scale over which data is gathered. 

In addition to the incubation period distribution, we use the method from Section~\ref{sec:incubation} to estimate a variety of delays from UK data provided by Public Health England through a data sharing agreement. This data takes the form of line-lists reporting date of onset, hospitalisation, ICU admission, and recovery or death. These estimates are given in Table~\ref{Table:parameters2}, along with the other estimates from Sections~\ref{sec:incubation} and~\ref{sec:transport}.

\begin{table}[h!]
\fontsize{10}{12}
\caption{Indicative parameter estimates from the literature.} 
\begin{center}
{
\begin{tabularx}{\textwidth}{|X| l| c |c|c|c|c|c|}
\hline
Parameter & Mean & Standard deviation & Confidence interval & Sample size & Source\\ \hline
Serial interval & 3.96 & 4.75 & 3.53 - 4.39 & 468 & \cite{Du2020} \\ \hline
Serial interval & 7.5 & 3.4 & 5.3 - 19 & 6 &\cite{LiQ2020} \\ \hline
R0 & 2.57 & - & 2.37 - 2.78 & - & \cite{Chinazzi2020}  \\ \hline
R0 & 3.11 & - & 2.39 - 4.13 & - & \cite{Read2020} \\ \hline
R0 & 4.5 & - & 4.4 - 4.6 & - & \cite{LiuT2020}   \\ \hline
Case ascertainment & 0.244 & - &  0.127 - 0.358 & - & \cite{Chinazzi2020}    \\ \hline
Incubation period & 4.8 & - & 2.2 - 7.4 & 16 & \cite{LiuT2020} \\ \hline
Incubation period & 6.4 & 2.3 & 5.6 - 7.7 & 88 & \cite{Backer2020}  \\ \hline
Doubling time & 4.2 & - & 3.8 - 4.7 & - &\cite{Chinazzi2020} \\ \hline
Doubling time & 2.1 & - & - & - & \cite{Read2020} \\ \hline
Doubling time & 2.4 & - & - & - & \cite{LiuT2020} \\ \hline
Onset to death & 20.2 & 11.6 & 15.1 - 29.5 & 34 & \cite{Linton2020} \\ \hline
Onset to confirmation & 4.8 & 3.03 & - & 38 & \cite{Kraemer2020} \\ \hline
Onset to hospital & 5 & - & - & - &\cite{Ferguson2020} \\ \hline 
Hospitalisation to death & 13 & - & 8.7 - 20.9 & - & \cite{Linton2020} \\ \hline
Under reporting & 0.138 & - & - & - & \cite{Lin2020} \\ \hline
\end{tabularx}
\label{Table:parameters1} 
}
\end{center}
\end{table}

\begin{table}[h!]
\fontsize{10}{12}
\caption{Indicative parameter estimates using methods proposed in this paper.} 
\begin{center}
{
\begin{tabularx}{\textwidth}{|X| l| c |c|c|c|c|}
\hline
Parameter & Mean & Standard deviation &  Type & Sample size \\ \hline
Incubation period & 4.84 (4.43 - 5.27) & 2.79 & Gamma distribution & N=162\\ \hline
Onset to hospital (UK) & 5.25 (4.60 - 5.97) & 3.32 & Gamma distribution & N=90 \\ \hline
Doubling time & 2.77 (2.67-6.87) & - &  - & -  \\ \hline
Growth rate & 0.25 (0.24-0.26) & - &  - & - \\ \hline
\end{tabularx}
\label{Table:parameters2} 
}
\end{center}
\end{table}

%%%%_____________________________________________________________

\section{Modelling intervention strategies}\label{sec:models}

%%%% COMPLIANCE_________________________________________________

\subsection{Adherence}
When designing intervention strategies, we need to consider how adherence may alter their effectiveness. This is important, since highly effective interventions may not be adhered to if they present great individual cost to a population. In this case, a theoretically less effective intervention may perform better, if it has sufficient reduction in individual-level cost. In this section, we illustrate the potential impacts of adherence on the effectiveness of interventions using a toy model. 

\begin{figure}[h!]
\centering
\includegraphics[width=12cm]{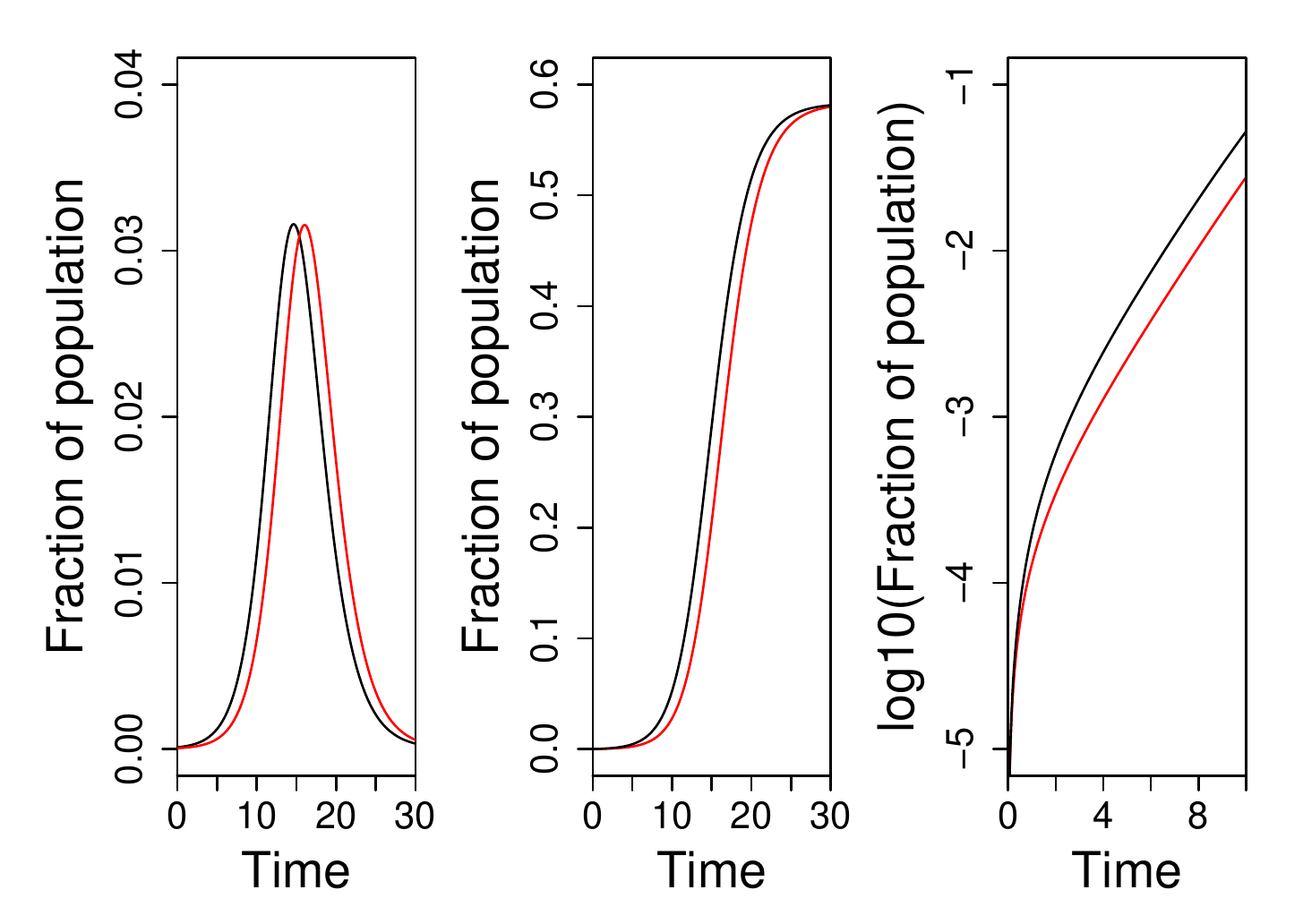}
\caption{Comparing the different definitions of adherence. The left panel shows the people in $I$ state from $SIR$ model (red) and from $I_A$ state in $SI_AI_BR$ model (black), with the same overall mean level of adherence. The resulting recovered curves are in the middle panel, with the right panel showing the recovered cases on log scale.}
\label{fig:adherence}
\end{figure}

Consider a standard SIR model, and denote by $S(t)$ and $R(t)$, respectively, the  susceptible and recovered/immune fractions of the population at time $t$. We can write $S$ in terms of $R$ such that
\[S(t) =S(0) \exp \left( -R_0 R(t) \right)\]
and let $t\to \infty$ to get the final size formula
\[1-R({\infty}) =S(0) \exp \left( -R_0 R(\infty) \right),\]
where $R(\infty)$ is the fraction of cases at end of outbreak.
This gives a ready reckoner for the eventual attack rate if interventions are not put in place or come in too late to be effective. To illustrate, if we have $R_0=3$ (and $S(0)\approx 1$), then $R(\infty) =0.94$. 
If an intervention is put in place that reduces (with full adherence) $R_0<1$ then the outbreak will be controlled. Indeed, let us assume that $R_0$ is reduced to zero by the intervention: for example, assume that social distancing is perfect and the number of contacts of a fully-adherent individual is zero. If only 50\% of people adhere to the intervention then the average number of contacts is effectively reduced by a half and logically $R_0^\dagger = R_0/2=1.5$ (the $\dagger$ representing quantities post intervention) and $R(\infty)^\dagger =0.58$ in this case.
However, this assumes that adherence is an independent random process at each contact. This suggests that for each contact an individual would ordinarily make, they ``toss a coin" to decide whether to isolate or not. In reality, individuals are more likely to show polarity, where some individuals reduce all their contacts and follow the measures and a proportion of individuals choose to not adhere to the intervention. If there was distinct polarity in the population such that 50\% adhered perfectly and 50\% ignored policy, then a toy model can be created with two infectious groups, $I_A$ and $I_B$, that behave differently. In this case
\begin{eqnarray*}
\dot{S} &=& -(R_A I_A+R_B I_B)S, \\
\dot{I}_A&=&\phi(R_A I_A+R_B I_B)S-I_A,\\
\dot{I}_B&=&(1-\phi)(R_A I_A+R_B I_B)S-I_B,
\end{eqnarray*}
where a dot over a variable represents its time derivative. Such an epidemic model, where the two groups have the same susceptibility but different infectivity, has the same final size as an epidemic in a single-type model with the same $R_0$ (e.g.\ see \cite{andreasen2011final}).  However, they have different durations as can be seen in Figure~\ref{fig:adherence}, where $\phi=1/2$, $R_B=0$ and $R_A=3$.
This shows that the assumptions about the nature of adherence predict the same growth rate and final size, but that the more polarised adherence has faster early growth and therefore an earlier peak.

This issue of independent versus polarised adherence is related to the idea of all-or-nothing  versus leaky vaccination~\cite{Goldstein2009, Consequences2014}, where you either vaccinate a fraction of the population with 100\% efficacy or vaccinate 100\% of the population with reduced efficacy~\cite{House2011}.
Note however that vaccination reduces your susceptibility (whether only or also), rather than only your infectivity as in the model discussed above, and variation in susceptibility does reduce the final size, with imperfect coverage with a perfect vaccine (all-or-nothing) leading to a lower final size than full coverage with a leaky vaccine (all individuals having the same mean susceptibility).

%%%%_____________________________________________________________

%%%% NURSING____________________________________________________
\subsection{Care home model}
The ongoing COVID-19 outbreak is known to have higher mortality rates amongst the elderly, the immunocompromised and those with respiratory and health complications \cite{Guan2020,WuJT2020,WuZ2020,Yang2020,Zhou2020}. In this section, we model the introduction of an infectious disease into care homes, in order to obtain estimates of the final size of the epidemic in the vulnerable population as well as predictions for the number of hospitalisations and fatalities.

Modelling of care homes in the UK is conducted against the backdrop of a wider epidemic in the general population, which we here assume to be following SEIR dynamics with a basic reproduction number $R_0$ that might be different from the within-care home reproduction number $R_C$.

\begin{figure}
\begin{center}
\includegraphics[]{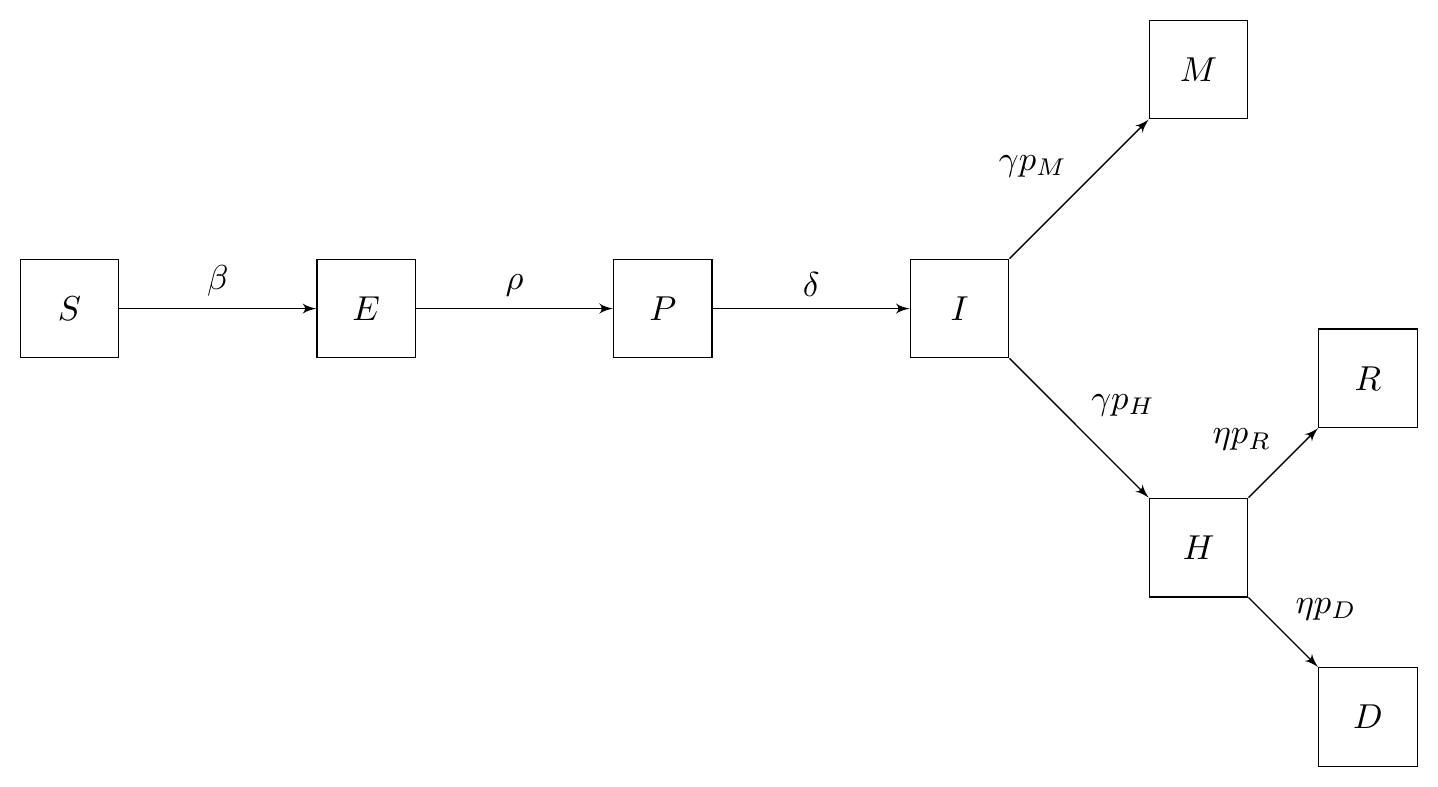}
\caption{ Compartmental model for disease dynamics within a care-home. We extend a deterministic SEIR model to include compartments for prodromal (infectious) cases (P), mildly symptomatic cases that recover without requiring hospitalisation (M), cases that do require hospitalisation and are removed from the care home (H), cases that die in hospital (D) and cases that recover in hospital (R).}\label{fig:J-3}
\end{center}
\end{figure}

Care homes are assumed to be closed populations, with the infection entering each of them independently with a certain probability. Infection is seeded only once, and within-care home outbreaks then evolve independently from, and do not contribute to, other care home outbreaks and the epidemic in the background population. To keep track of hospitalisations, we model the within-care home infection dynamics using a compartmental model that, in addition to SEIR model, has compartments for mildly symptomatic prodromal cases (P), who show no symptoms but are capable of transmitting the virus, those who recover from the disease after mild symptoms that did not require hospitalisation (M), those who have severe symptoms and are admitted to hospital (H), those who recover after hospitalisation (R), and those that die (D). This is illustrated in Figure~\ref{fig:J-3}.

The stochastic component of the model, i.e.\ the random introduction of the infection in care homes, is modelled using the Sellke construction \cite{Andersson2000}. Each care home $i$ is given an individual, random threshold of resistance, $Q_i$, which is drawn from an $\mathrm{U}(0,1)$ distribution. At time $t$, we then calculate the infection pressure ${\mathrm{IP}}(t)$ from the background epidemic so that care home $i$ becomes infected at time $T_i$, where $T_i = {\mathrm{inf}}\{t | {\mathrm{IP}}(t) > Q_i\}$. 
The infection pressure up to time $t$ for a median sized care home is the integral from 0 to $t$ of the force-of-infection (FOI) applied to the care home coming from all infectious sources, multiplied by a probability $p$. This probability represents the probability of the infection being introduced to a median-sized care home. For other care homes, we allow this probability to be proportional to its size, under the assumption that larger care homes employ more staff and are therefore at higher risk of introduction.
When the infection pressure becomes higher than an individual care home's resilience threshold, that care home begins its own deterministic infection dynamics with a single initial infected case.

We run this model on data for the entire care home population in the UK, so that there are approximately 15,000 care homes with a total population of approximately 450,000 residents~\cite{cqc}. 
In this model we only consider the vulnerable population within care homes. 
We assume $R_0 = 1.5$ in the background epidemic, a relatively low value that somehow accounts for a certain degree of control, and an $R_C = 3$ to allow relatively explosive epidemics in care homes due to potentially more frail individuals, difficulty in isolation and staff inadvertently passing the infection from one case to the next. The other parameters in the baseline scenario are reported in Table \ref{table:J-1}.

\begin{figure}
        \centering
\begin{subfigure}[t]{0.45\textwidth}
       \includegraphics[width=1\linewidth]{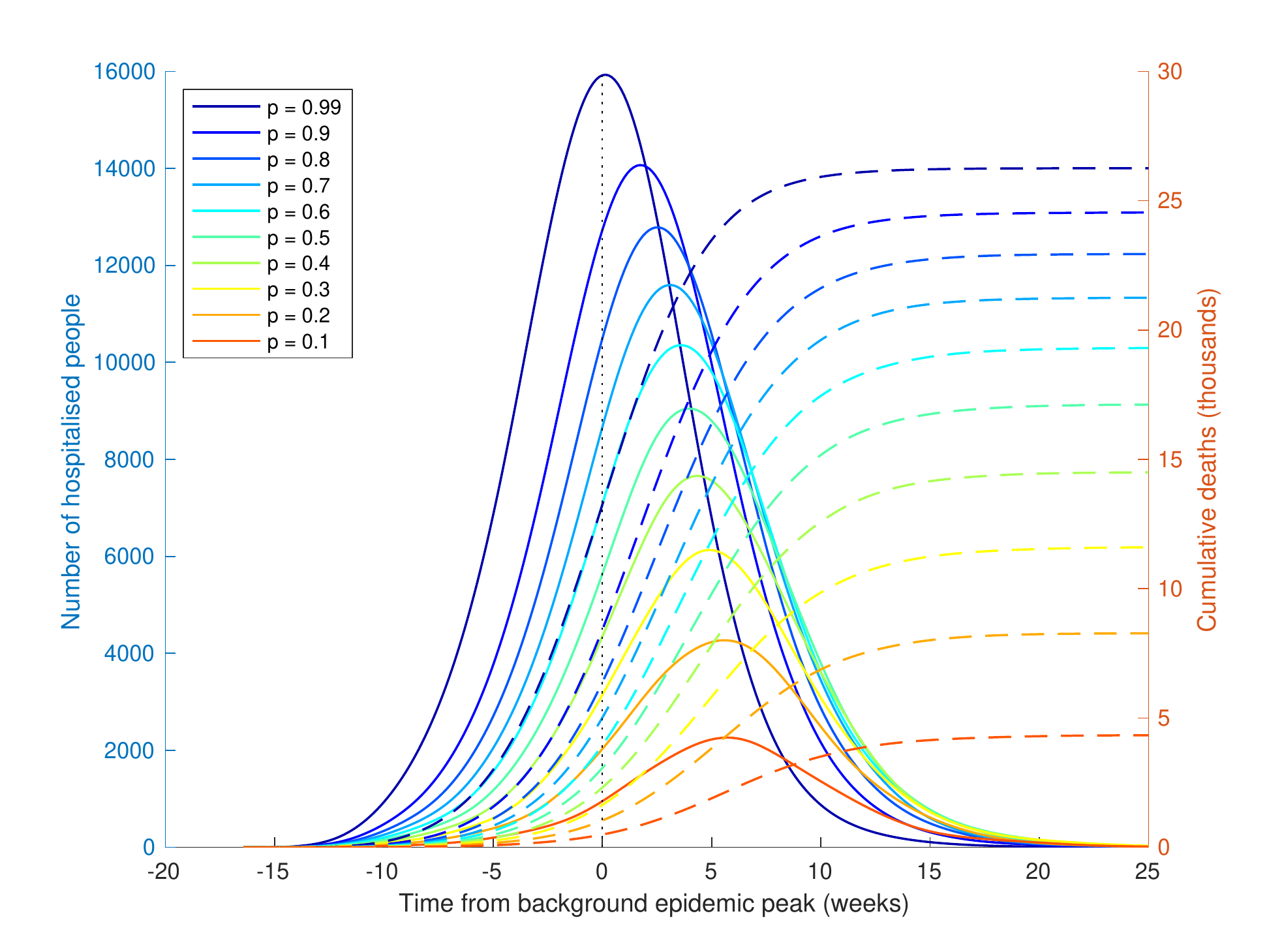}
        \caption{Hospital prevalence (solid lines, left axis) and cumulative number of deaths (dashed lines, right axis). The $x$-axis shows time in weeks, with 0 (vertical dotted line) denoting the peak of the background epidemic. Colours refer to different values of the probability $p$ that a median-sized care home experiences an introduction.}
        \label{fig:J-1}
\end{subfigure}\hspace{1cm}
\centering
\begin{subfigure}[t]{0.45\textwidth}
        \includegraphics[width=1\linewidth]{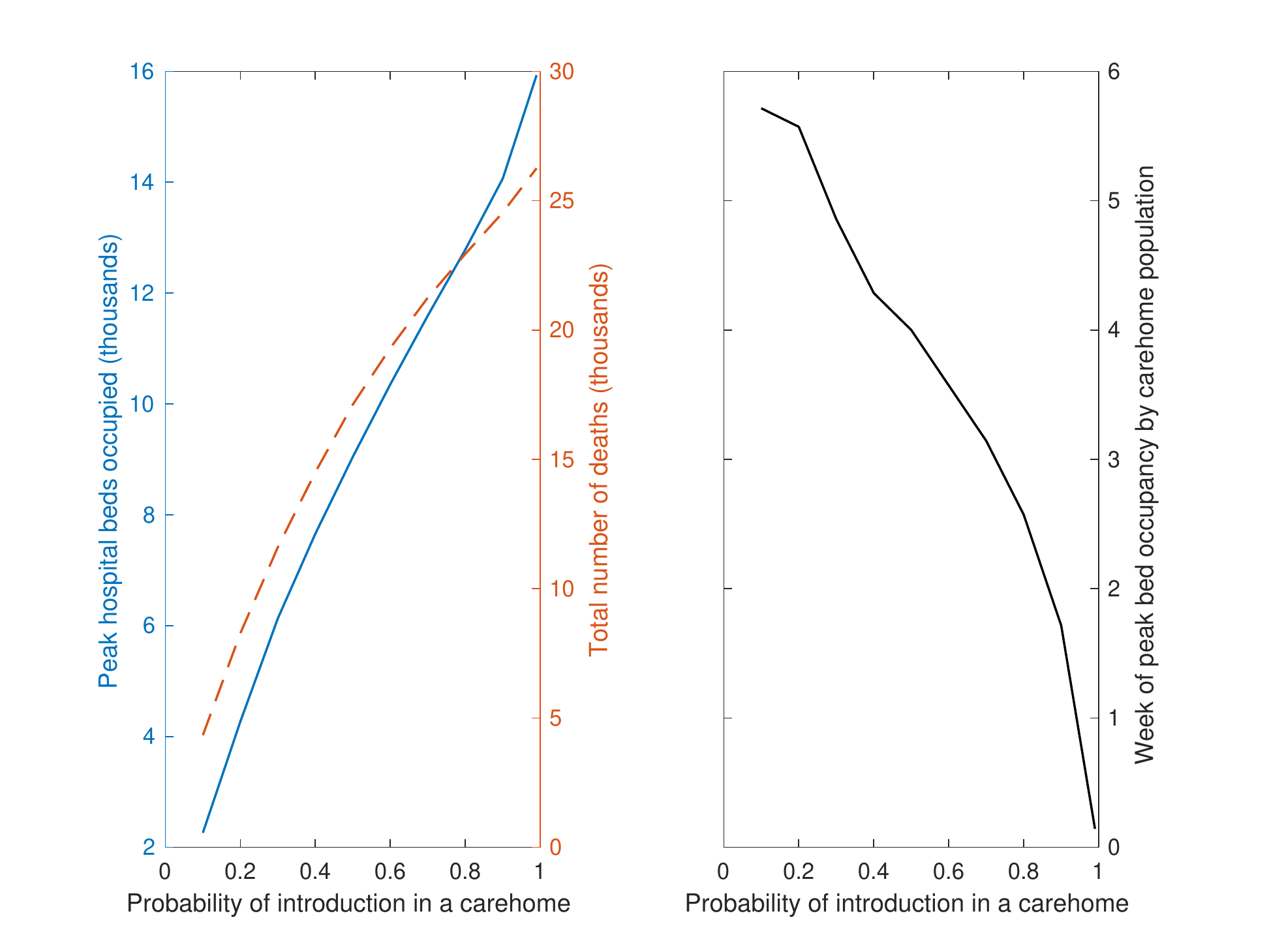}
        \caption{Left plot shows the predicted height of the peak in the number of hospital beds occupied (solid blue line, left axis) and total number of deaths (dashed red line, right axis), as a function of the probability $p$ of introduction in a median-sized care home. Right plot shows the timing of the peak, which always appears to occur at a similar time or later, compared to the background epidemic. The peak arrives earlier for larger values of $p$, as high $p$ corresponds to multiple simultaneous introductions early on in the background epidemic.}
        \label{fig:J-2}
\end{subfigure}
\caption{Hospitalisation prevalence (a) and hospitalisation peak (b) for the care home model.}
\end{figure}

Figure~\ref{fig:J-1} shows number of hospital beds occupied and the cumulative number of deaths for the parameter values chosen and for different values of $p$. Time is shown in weeks, where week zero represents the peak of the external/background epidemic. Figure~\ref{fig:J-2} summarises the first, showing the impact of reducing $p$ on the demand for hospital beds and on the final number of deaths. It also shows the impact that changing $p$ has on the timing of the peak. 

Reducing $p$ corresponds to increasing protection of the vulnerable population in care homes by preventing introduction of infection (e.g.\ screening, testing and promoting hygiene among staff, etc.), a policy sometimes termed `cocooning' or `shielding'.
The results show that reducing $p$ from 0.99 to 0.1 corresponds to a reduction of around 22,000 (83\%) in the number of deaths and around 131,500 in the number of hospitalisations (83\%). Strategies aimed at reducing the probability of introduction into a care home, such as reducing the number of visitors or increased monitoring and protection of care home staff, are therefore predicted to have a large impact on the number of cases in vulnerable care home populations. 

There are a variety of assumptions underpinning this model. Firstly, the background epidemic ignores structure and assumes homogeneous mixing. This is likely to make the peak more pronounced, so presents a worst case scenario for the demand on hospital beds. The assumption of $R_0$ for the background epidemic only affects the shape and duration of the background epidemic, since the tuneable parameter $p$ controls the risk of introduction to the care home.  That is, if $R_0$ is small, a large $p$ still presents a high force of infection into the care homes. Therefore, for a fixed $p$, we expect that changing $R_0$ does not affect the total number of deaths, but it changes the peak hospitalisation incidence because a faster and more explosive background epidemic makes epidemics in care homes more synchronised. In fact, when testing the impact of a longer, flatter background epidemic, for example obtained by simulating three slightly desynchronised background SEIR epidemics, results have lower peaks and a much more variable timing (not shown). Assuming each care home is independent might not be realistic, since it is likely that staff are shared between multiple homes, in which case they can act as vectors of transmission between homes. However, current outbreaks are already quite synchronised (the within care home outbreaks occur at similar times), so the effect of this assumption is likely to be minimal. The final major assumption is that the epidemic within the care homes is deterministic. This removes the probability of random extinction and random delays, and should obviously be relaxed with a stochastic model, given half of care homes have size smaller than 25. However, the extinction probability is very low with $R_C=3$, so this stochastic effect is unlikely to have a large impact. Random delays, instead, may change the shape and timing of the epidemic, which could potentially reduce the peak burden. Therefore, this model represents a worst case scenario.

\begin{table}
\caption{Information and values for each parameter in our within care home model.}\label{table:J-1}
\begin{tabularx}{\textwidth}{ |X|c|c|c| }
\hline
Parameter & Value & Details \\
\hline
$R_0$ & 1.5 & Basic reproductive ratio (for external epidemic) \\
\hline
$\beta$ & $R_0\times\gamma$ & Infectiousness\\
\hline
$\rho$ & $1/5$ & Reciprocal of period between exposure and asymptomatic infectiousness\\
\hline
$\delta$ & $1/2$ & Reciprocal of period between asymptomatic infectiousness and onset of symptoms \\
\hline
$\gamma$ & $1/4$ & Reciprocal of infectious period \\
\hline
$p_L$ & 0.64 & Proportion of vulnerable infectious cases who recover without severe symptoms\\
\hline
$p_H$ & 0.36 & Proportion of vulnerable infectious cases who are hospitalised\\
\hline
$\eta$ & $1/14$ & Reciprocal of period of hospitalisation\\
\hline
$p_R$ & 5/6 & Proportion of hospitalised cases that recover \\
\hline
$p_D$ & 1/6 & Proportion of hospitalised cases that die\\
\hline
\end{tabularx}
\end{table}

%%%%_____________________________________________________________

%%%% HOUSHOLD_____________________________________________________
\subsection{Household isolation modelling}
\label{sec:households}
In the absence of cure or vaccine for COVID-19, governments worldwide must rely on non-pharmaceutical interventions (NPIs) to control the outbreak \cite{OxCGRT}. A natural such intervention is to ask individuals who express symptoms similar to COVID-19 to isolate
themselves, but variants to such individual isolation might include policies sometimes referred to as household isolation, household quarantine and mixed isolation. In this section, we
investigate how such strategies affect the spread of the epidemic when bearing in mind that adherence to each intervention may differ.

Individual isolation relies on individuals staying in isolation when they
express symptoms, thereby stopping transmission. However, there is potential asymptomatic or prodromal
transmission before they go into isolation. Additionally, isolation strategies
generally ask infected individuals to remain at home, which presents an infection
risk to the other members of their household, who may go on to spread the infection. 

The term `household isolation' refers to a policy where, upon detection of symptoms within a household, all
individuals within the household go into isolation for a fixed duration of time.
This strategy reduces the risk that other household members, if they are infected within the household, transmit in the community when pre-symptomatic (and hence before they self-isolate themselves) or if asymptomatic but still infectious. 

A blanket policy invoking a fixed duration of household isolation might cover the full epidemic in a small household. However, a larger household might present multiple generations of infection, potentially extending the within-household outbreak beyond the fixed duration of the household isolation policy.
To address this issue, `household quarantine' is another potential strategy. Upon
detection of symptoms, the entire household is isolated until a fixed duration of
time after the last symptomatic case within the household expresses symptoms.
This ensures that there are no symptomatic cases evading intervention but applies quite drastic measures to the household. 

A fourth strategy, that reduces the cost relative to household quarantine, is mixed isolation. Here, upon detection of symptoms the entire household is isolated for a fixed length of time. Any subsequent cases within the household then undergo individual isolation as described above. This reduces the risk of cases not being isolated whilst allowing recovered individuals to return to work. There is however still some remaining risk that infected individuals may not yet express symptoms after the end of the isolation period, but this risk can be controlled through the duration of each isolation.

Although there is now a rich theoretical literature on households models~\cite{BALL201563,Ball2011,ball1997}, the mainstream methodological tools in this research area present important limitations that make them not directly applicable to studying these control policies. First, exact theoretical or asymptotic results in these models are mostly restricted to time-integrated quantities, i.e.\ those quantities that do not depend on the detailed temporal shape at which the infectivity is spread by an individual: these are $R_0$ (or any other reproduction number~\cite{BALL2016108,PELLIS201285}, e.g.\ the household reproduction number $R_\ast$), the probability of a large epidemic, and the epidemic final size \cite{Andersson2000}. For this reason, the vast majority of the literature relies on the standard stochastic SIR model \cite{Andersson2000}, despite its unrealistic infectivity profile. Even if more recent work has expanded beyond time integrated quantities, for example considering the real-time growth rate~\cite{BALL2016108,Pellis2011a}, if the interest is on tracking the dynamics of infection spread, a model based on full temporal representation of between- and within-household dynamics \cite{House:2008} appears necessary. 

A second limitation of standard household models is the key assumption of constant parameter values. This appears essential for any form of analytical progress. However, in the context of the interventions discussed above, a reduction in transmission between households, as well as a potential increase in within the household, require parameters to change over time.

To overcome these limitations, we consider two approaches. The first approach fully captures both within and between-household dynamics with a master-equation formalism, i.e.\ by relying on a Markovian within-household dynamics and keeping track of the expected number of households in each possible state of their internal dynamics. The second approach has a greater emphasis on within-household dynamics, and is fundamentally an independent-households, individual-based, stochastic simulation. The more limited mathematical tractability is the price to pay for an increased flexibility, as the within-household Markov assumption is relaxed and exact distributions for delays between events, typically informed by the data, can be explicitly inputted. Although both approaches can account for increased within-household transmission as isolation and quarantine are imposed, we only consider this for the second method here. This aspect allows us to study the increased risk of infection a vulnerable individual in the household would experience following the implementation of a control policy.

To model the households in the UK, we construct a realistic distribution of household sizes. 
We take this demographic data from the 2001 Census \cite{Census:2001}. More recent information, though less specific on large household sizes, shows that sizes of smaller households are largely unchanged over time \cite{FH:2019}.

\subsubsection{Population and household transmission}
\label{sec:Householdindep}

In this section, we investigate the above intervention strategies under the assumption that a fraction of households adhere $100\%$ with an intervention and the remaining households ignore the intervention. To model the interventions, we implement a
dynamical household model that explicitly represents the small sizes of
households.

The dynamics of the outbreak are simulated using an SEPIR model. This model
assumes that there are five possible states in which an individual can be.
These are, susceptible, latent, mildly symptomatic prodrome, symptomatic
infectious and removed. Following \cite{Cauchemez:2004}, we assume that within-household transmission scales with the inverse of the household size to a specified power $\eta$.
Such a model can be used to investigate how the pathogen spreads through and between households.

The methodology involved is the use of self-consistent differential equations, first written down by Ball \cite{Ball:1999}. More recent developments, including numerical
methods for these equations, include
\cite{House:2008,Ross:2009,Black:2017,Kinyanjui:2018}. Important features
of this approach include allowing for a small, finite size of each household
in which random effects are important and each pair can only participate in one
infection event.

\textbf{Model}

Let $Q_{n,s,e,p,i}(t)$ be the proportion of households in the population at
time $t$ of size $n$, with $s$ susceptibles, $e$ exposed, $p$ prodromal, and
$i$ symptomatic infectious individuals. The number of recovered individuals
will be $n-s-e-p-i$. In the absence
of household-based interventions, we have
\ba
\ddt{}Q_{n,s,e,p,i} & = - \left(s r_{s\rightarrow e}(t,\mathbf{Q}) + e
r_{e\rightarrow p} + p r_{p\rightarrow i} + i r_{i\rightarrow \emptyset} 
+ n^{-\eta}s p \tau_p + n^{-\eta}s i \tau_i \right) Q_{n,s,e,p,i}
\\ & \qquad
+ (s+1) r_{s\rightarrow e}(t,\mathbf{Q}) Q_{n,s+1,e-1,p,i}
+ (e+1) r_{e\rightarrow p} Q_{n,s,e+1,p-1,i}
\\ & \qquad
+ (p+1) r_{p\rightarrow i} Q_{n,s,e,p+1,i-1}
+ (i+1) r_{i\rightarrow \emptyset} Q_{n,s,e,p,i+1}
\\ & \qquad
+ n^{-\eta}(s+1) p \tau_p Q_{n,s+1,e-1,p,i}
+ n^{-\eta}(s+1) i \tau_i Q_{n,s+1,e-1,p,i}
\text{ ,}\nonumber
\ea
where we take any $Q$ with logically impossible indices just to equal $0$, $r_{a\rightarrow b}$ is the rate from state $a$ to $b$, and $\tau_a$ is the transmission rate from an individual in state $a$. The transmission into households is given by
\be
r_{s\rightarrow e}(t,\mathbf{Q}) = \Lambda(t) +
\sum_{n=1}^{n_{\mathrm{max}}}
\sum_{s=0}^{n}
\sum_{e=0}^{(n-s)}
\sum_{p=0}^{(n-s-e)}
\sum_{i=0}^{(n-s-e-p)}
\left(p \beta_p(t) + i \beta_i(t) \right) Q_{n,s,e,p,i}
\text{ .}\nonumber
\ee
Here $\Lambda$ represents infections imported from outside the population of
households, and the other terms represent between-household transmissions. We
take a `global' intervention as part of the baseline, in particular, we can
model phenomena such a school closures that hold during a set of times
$\mathcal{T}$ as
\be
\beta_x(t) = \begin{cases} (1-\varepsilon) \beta_x(0) & \text{ if } t \in \mathcal{T}
\text{ ,} \\
\beta_x(0) & \text{ otherwise,} \end{cases}\nonumber
\ee
for $x \in \{p, i\}$. We call $\varepsilon$ the \textit{global reduction}. We
will generally drop this $t$-indexing for simplicity, and will also consider
only a household isolation strategy (though the other strategies can be considered similarly, with an example of how other strategies could be captured in this model framework given in Appendix \ref{app:exhh}).
Suppose a fraction $\alpha_{\mathrm{W}}$ of households isolates when there is
at least one symptomatic case in the household. These households do not
experience new infections, meaning that the dynamics become
\ba
\ddt{}Q_{n,s,e,p,i} & = - \left(
\left(1-\alpha_{\mathrm{W}} \indic{i>0} \right)s r_{s\rightarrow e}(t,\mathbf{Q})
+ e r_{e\rightarrow p} + p r_{p\rightarrow i} + i r_{i\rightarrow \emptyset} 
+ n^{-\eta}s p \tau_p + n^{-\eta}s i \tau_i \right) Q_{n,s,e,p,i}
\\ & \qquad
+ \left(1-\alpha_{\mathrm{W}} \indic{i>0} \right)
(s+1) r_{s\rightarrow e}(t,\mathbf{Q}) Q_{n,s+1,e-1,p,i}
+ (e+1) r_{e\rightarrow p} Q_{n,s,e+1,p-1,i}
\\ & \qquad
+ (p+1) r_{p\rightarrow i} Q_{n,s,e,p+1,i-1}
+ (i+1) r_{i\rightarrow \emptyset} Q_{n,s,e,p,i+1}
\\ & \qquad
+ n^{-\eta}(s+1) p \tau_p Q_{n,s+1,e-1,p,i}
+ n^{-\eta}(s+1) i \tau_i Q_{n,s+1,e-1,p,i}
\text{ ,}\nonumber
\ea
and also do not transmit outside, meaning that the rate of between-household
transmission becomes
\be
r_{s\rightarrow e}(t,\mathbf{Q}) = \Lambda(t) +
\sum_{n=1}^{n_{\mathrm{max}}}
\sum_{s=0}^{n}
\sum_{e=0}^{(n-s)}
\sum_{p=0}^{(n-s-e)}
\sum_{i=0}^{(n-s-e-p)}
\left(1-\alpha_{\mathrm{W}} \indic{i>0} \right) \left(p \beta_p + i \beta_i \right)
Q_{n,s,e,p,i}
\text{ .}\nonumber
\ee

\textbf{Parameterisation}

\begin{figure}[h!]
\begin{center}
\includegraphics[width=0.9\textwidth]{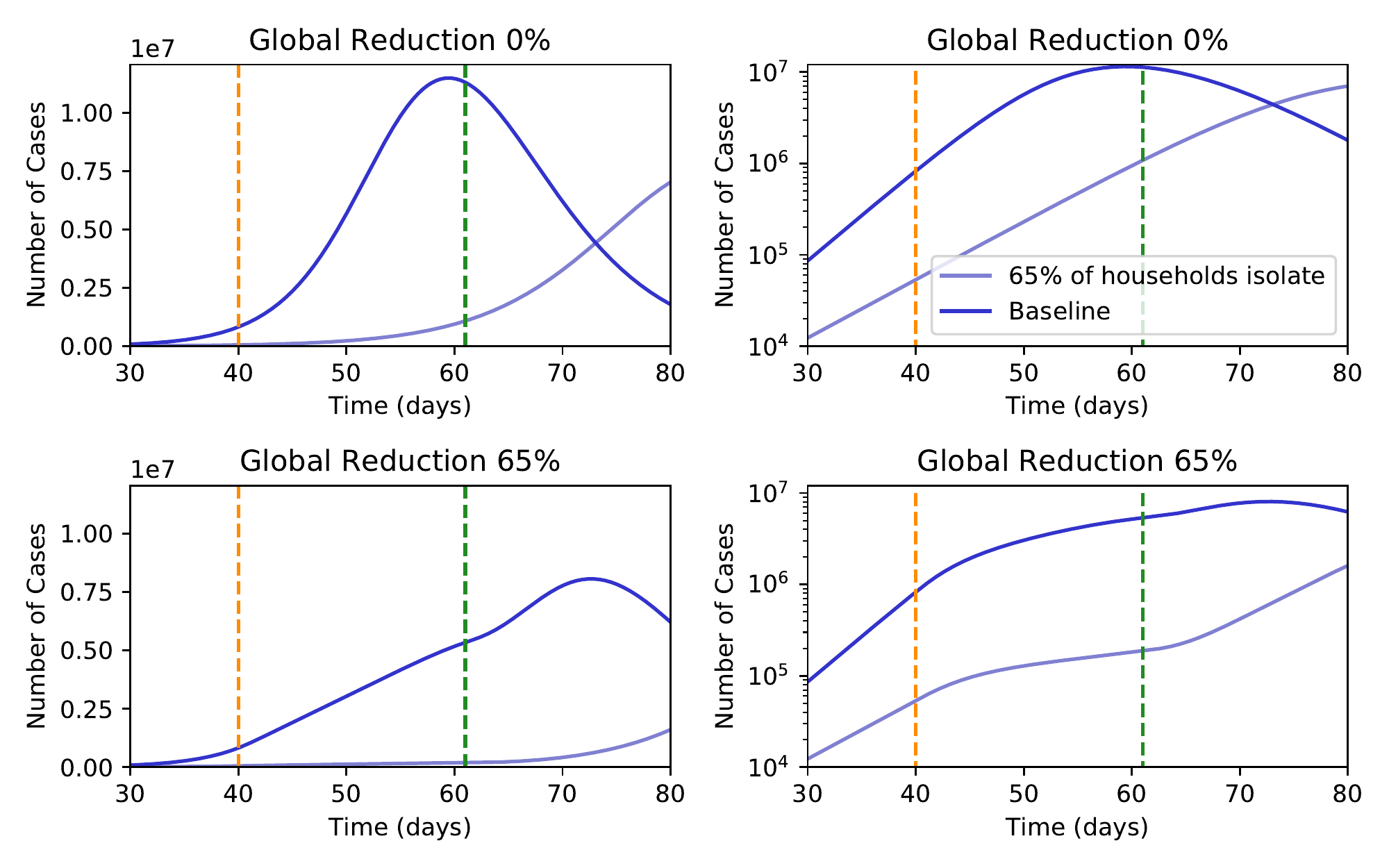}
\caption{Investigating the impact of household intervention with $\alpha_W = 65\%$ on the number of cases, for two levels of global intervention. The top two figures have no global intervention, and the bottom two have  an $\epsilon = 65\%$ reduction in global transmission, e.g. school closure or partial lockdown, lasting for 21 days (between the vertical lines). The left-most figures have a linear y-axis. The right-most figures show that same results on a logarithmic y-axis.}\label{fig:T-1}
\end{center}
\end{figure}

\begin{figure}[h!]
\begin{center}
\includegraphics[width=0.9\textwidth]{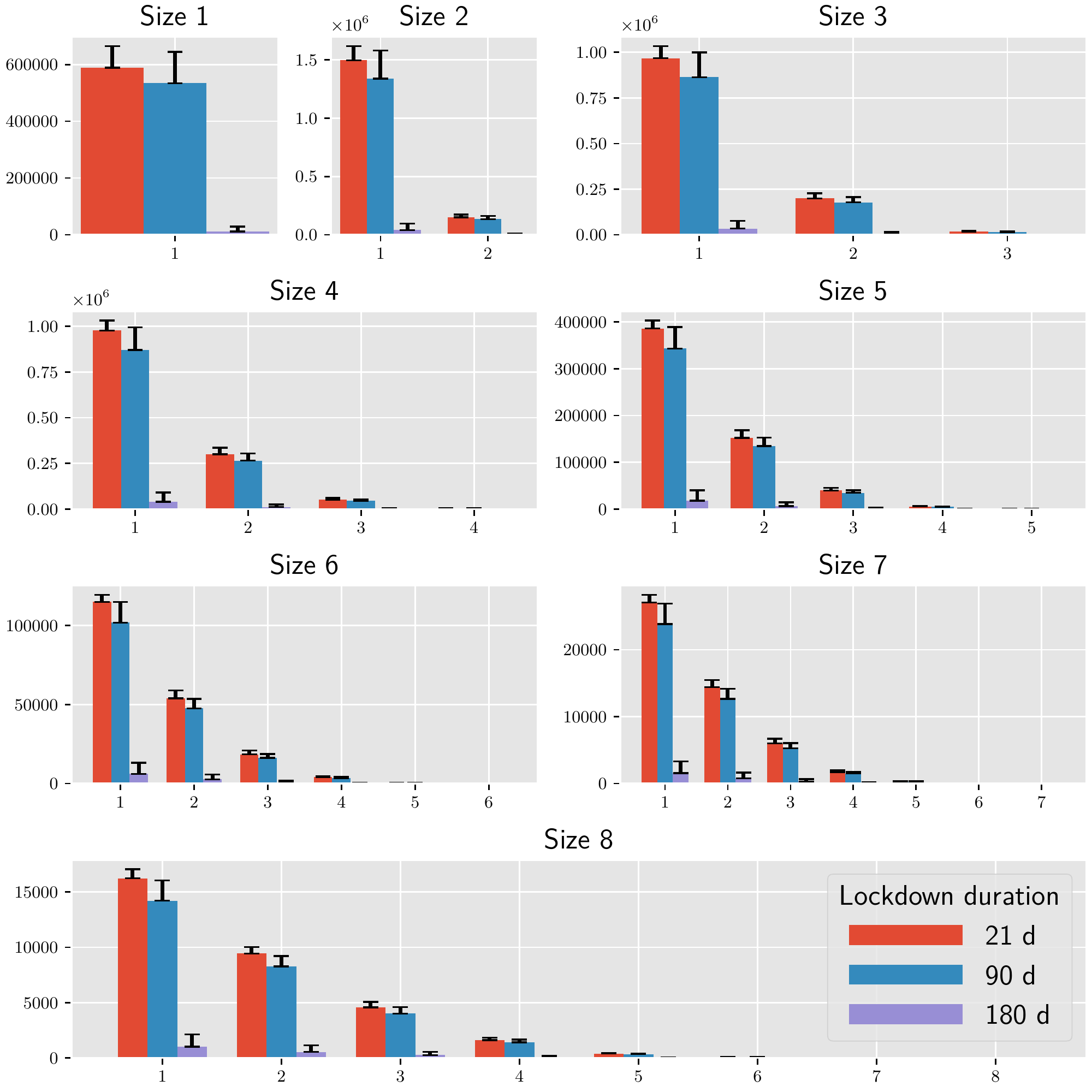}
\caption{Histograms representing the number of infectious cases (x-axis) at different household sizes, against a background of household isolation and for three levels of global transmission reduction. The global reduction takes the form of a lockdown reducing contacts by 65\% for 21, 90 and 180 days. The number of cases is a cumulative measure of probing the state of each household every two weeks over the 180 day period. The error bar is one standard deviation constructed by varying isolation adherence ($\alpha_W$), global reduction ($\epsilon$) and secondary probability of attack. Parameters are in the main text.}\label{fig:T-2}
\end{center}
\end{figure}

Using the methods in \cite{Black:2017,Ross:2009}, it is possible to fit household models of this kind to the overall growth rate, $r$, which we take to correspond to a doubling time of three days. Natural history parameters can then be set directly based on reasonable estimates: $r_{e\rightarrow p}$ to the inverse of the latent period; $r_{p\rightarrow i}$ to the inverse of the prodromal period; $r_{i\rightarrow \emptyset}$ to the inverse of the symptomatic period. Shaw~\cite{Shaw2016} analyses various household datasets for respiratory pathogens and estimates values for $\eta$ close to $1$, so this is taken to be $0.8$. The remaining degrees of freedom are relative infectiousness of the prodrome (taken as a third) and the probability of transmitting within a pair, which we can take as a typical value given by Shaw~\cite{Shaw2016}. For the numerical results in figures~\ref{fig:T-1} and~\ref{fig:T-2}, the baseline natural history parameters are chosen to be $r_{e\rightarrow p}=1/5$, $r_{p\rightarrow i}=1/3$, $r_{i\rightarrow \emptyset}=1/4$.

\textbf{Summary}

Using the given parameter values for our baseline scenario, we consider a combination of household isolation (which follows all-or-nothing adherence) with global
reduction in transmission (which follows leaky adherence) for three weeks and show the results in
Figure~\ref{fig:T-1}. The distribution of infectious individuals varies with household size, which is shown in Figure~\ref{fig:T-2} for different durations of global intervention. Applying household isolation at 65\% adherence ($\alpha_W=0.65$) manages to reduce the spread of infection, but appears insufficient in this model and with baseline parameters for controlling the outbreak in the long-term, unless other intervention strategies that reduce the global transmission $\epsilon$ are adopted at the same time. Alternatively, different levels of adherence can be considered to determine if and when control may be achieved purely through household-based interventions.. For the model proposed in the next section, we look into the effectiveness of increasing adherence.


\subsubsection{Non-Markovian models with enhanced within-household transmission}
\label{sec:within}
The model described above has the advantage of being able to track the dynamics within the household as well as the overall epidemic in the population in a relatively efficient manner. We now discuss a different framework that loses part of the capability in keeping track of the overall epidemic, but offers further flexibility both in the impact of policies on the within-household dynamics and in the distributions between events in the infectious life of an individual. We use this model to investigate the relative effectiveness of the different control policies. We also consider allowing recovered individuals to leave the household, even in the context of household isolation or household quarantine. This has no impact on the transmission dynamics, but reduces the individuals' life disruption and potential economic cost of any policy implemented.

This model assumes that there is no reintroduction within households so each household can only be isolated or quarantined once. The assumption that only one household member is infected from outside is approximately satisfied if we assume homogeneous mixing between households and a large number of households, which are all fully susceptible at the start of the epidemic. However, the reality of heterogeneous mixing makes reintroduction a likely possibility even early on in the epidemic.
This model, therefore, lacks an explicit description of the social network structure beyond the household. For simplicity, we assume that within households all individuals are identical in terms of their disease dynamics, although the method might be extended to allow for different age/risk groups with different disease dynamics.
We assume that the level of within-household transmission in a household of size $n$ scales proportionally to $1/(n-1)$, though we acknowledge that true transmission is slightly more complex~\cite{Cauchemez2009}.

\noindent\textbf{Model}

We consider independent households of size $n=1,2,\dots,8$, for each of which $n_e$ stochastic simulations of the within-household epidemic are performed based on the Sellke construction~\cite{sellke_1983,Andersson2000}.
Given all infectious contacts outside lead to an actual infection because we are in the early phase of the epidemic and there is no depletion of susceptibles, each case infects, on average, $R_g$ new cases outside. Inside the household, a case would infect on average $R_h$ cases in an infinitely large household, but not all infectious contacts lead to real infections, given local saturation effects: in a household of size $n$, each infectious individual makes on average $R_h/(n-1)$ infectious contacts with each other specific individual throughout the infectious period, but only the first one will result in an infection, and only if the individual was susceptible at the time of contact. 

Each individual is given an indicator function of whether they are symptomatic or not (individuals show symptoms independently of each other with probability $p_s$) and a resilience threshold. This last quantity is drawn from an exponential distribution with mean 1, and represents the overall infection pressure this individual is able to withstand before they get infected. The infection pressure up to time $\tau$ is the integral from 0 to $\tau$ of the force-of-infection (FOI) applied to this individual coming from all infectious sources. 

At the beginning of the within-household epidemic, a single initial case is assumed. Time is discretised with a predefined time step $dt = 0.1$ days. At any time step, the current infectivity of all infectives in that time step is summed over, keeping track differently of the infectivity spread outside and inside the household. An overall measure of the accumulated infectivity within the household is updated at each time step and when this crosses the resilience threshold of a susceptible individual, they acquire the infection.

\begin{figure}[h!]
\begin{center}
\includegraphics[width=1\textwidth]{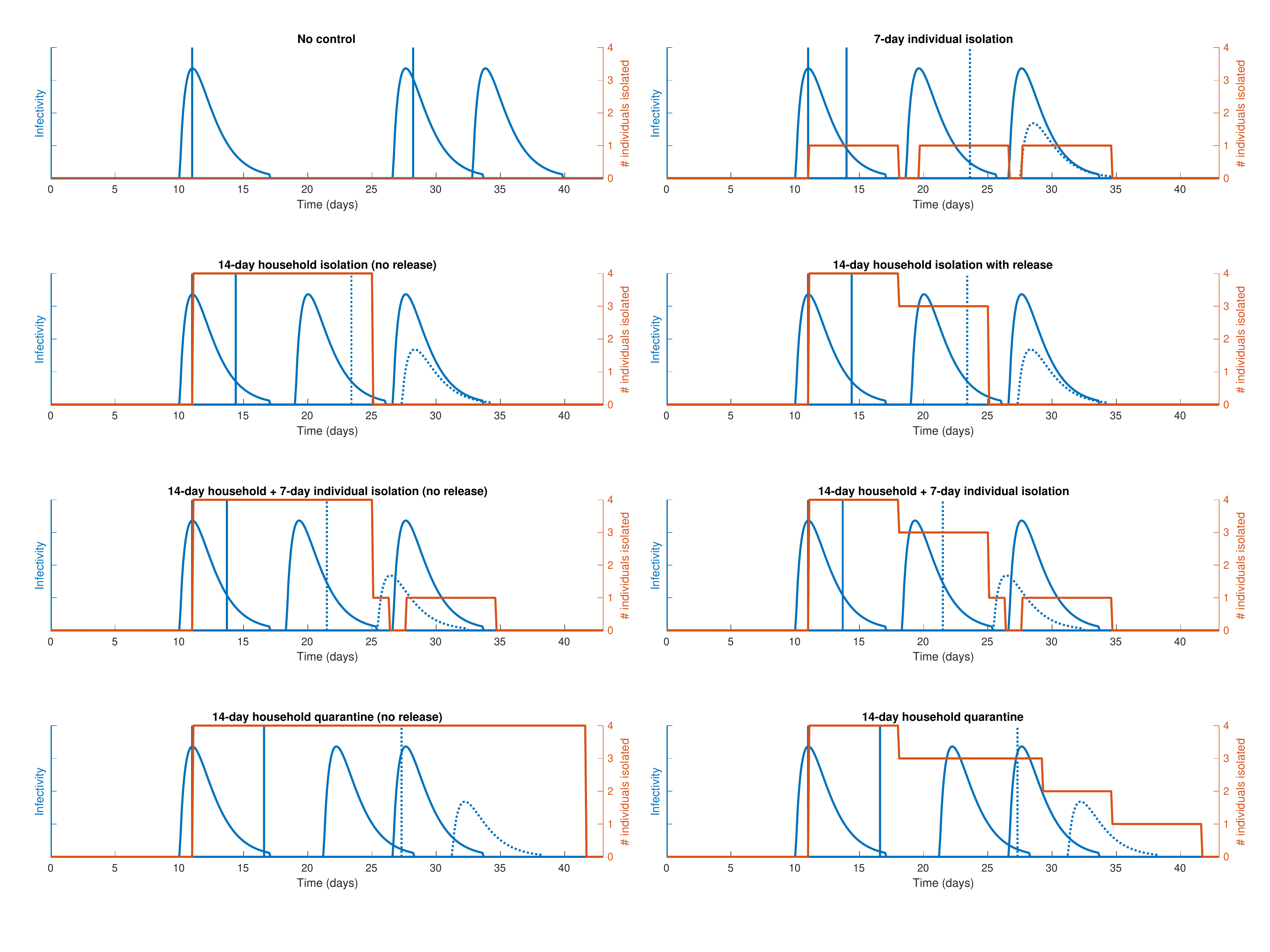}
\caption{Impact of various control policies on a single epidemic realisation in a household of size 4. The blue solid lines represent the infectivity of symptomatic individuals. Vertical solid lines represent the times of infection of symptomatic cases. Dashed curves and vertical lines represent the infectivity and time of infection of asymptomatic cases, which are assumed to be half as infectious as symptomatic ones. The total number of cases under isolation (possibly compounded, e.g. both household and individual isolation) is shown in red (right axis). All random numbers involved in the realisation of the stochastic epidemic are drawn at the start, before the impact of each control policy is implemented. Row 1 shows no isolation and individual and isolation. Rows 2, 3 and 4 show, respectively, household isolation, mixed isolation and household quarantine. The difference between the columns is that the basic policy on the left is ``upgraded'' to the more cost-effective version on the right that allows recovered individuals to leave the house as they cannot transmit outside anymore. When no control is implemented, the primary case (individual A, say, infected at time 0) infects another individual (B, say) around time 11. After a long latent period (i.e. incubation minus prodromal), B becomes infectious and infects a further individual (C, say). The last individual (D, say) escapes infection. When different intervention strategies are in place, within-household infectivity is increased. This can result in individual C becoming infected earlier in the outbreak and individual D no longer escaping infection, both due to the increased force of infection. In this simulation, the dynamics for individual B do not change since they are infected before A becomes symptomatic. Individual D is infected earliest under mixed isolation, because within-household transmission is higher than household isolation alone, due to increased adherence from individual isolation also being in place. Adherence levels to household quarantine are lower than those of household isolation, due to the higher demand of full quarantine, thus leading to less enhanced within-household transmission. The more severe the intervention, the better it captures the infectious periods of infected individuals within the household.  
}
\label{fig:L-within}\end{center}
\end{figure}

We assume an individual spends half of their time outside and half inside the household. When self-isolation starts, the assumed adherence $a_i$ represents the fraction of the time spent outside that is shifted from outside to within the household. Therefore, for perfect adherence, from the moment symptoms occur, the individual stops transmitting outside but their infectivity within the household grows by 100\%. We also explore variations in this compensatory behaviour, so that the time of an individual is split in a more flexible proportion than $1:1$. The same argument applies to other control policies, with adherence levels $a_h$ for household isolation and $a_q$ for household quarantine. When multiple control policies are in place at the same time, their effect is assumed to be multiplicative: if an individual has symptoms and the household isolates, the outside transmission rate from that individual is reduced from baseline by a multiplicative factor $(1-a_i)\times(1-a_h)$ and the within-household transmission rate is the baseline value plus a fraction $1-((1-a_i)\times(1-a_h))$ of the baseline value. Therefore, implementing a control policy that reduces transmission outside might lead to more infections in the household (see Figure~\ref{fig:L-within} and the associated accumulated infection pressure in Figure \ref{fig:L-supp}).

We denote by $\beta_g^n(\tau)$ the average global infectivity profile of a household of size $n$, i.e. the time-point average of the rates at which new cases outside are generated by any case infected in all simulated epidemics in a household of size $n$. During the exponentially growing phase, any global infection starts a new within-household epidemic. Furthermore, larger households are more likely to be infected because they have more members. Therefore, if $h_n$ is the probability that a randomly selected household has size $n$,
$$\pi_n = \frac{n h_n}{\sum_m{m h_m}}$$
gives the probability that the household of a randomly selected individual is of size $n$. This is called the size-biased distribution. The global infectivity profile of the average household infected during the exponentially growing phase is then:
\begin{equation*}
\beta_g(\tau) = \sum_n \pi_n \beta_g^n(\tau)
\label{eq:avHinfprof}
\end{equation*}
and the area under this curve is known in the literature as the household reproduction number, and is typically denoted by$R_\ast$~\cite{ball1997,Ball2016,Fraser2007}. If enough transmission is prevented, so that $R_\ast < 1$, the epidemic is controlled. 

\noindent\textbf{Parameterisation}

At baseline we take $R_g = 2.5$ and $R_h = 2.5$, which gives a real-time growth rate $r$ of about $0.245$. Each individual, irrespective of whether they will be infected or not, is given (independently of each other) a duration of: incubation period (randomly drawn from a Gamma distribution with mean $\mu_E = 4.84$ and standard deviation $\sigma_E = 2.79$ days), prodromal period ($d_p = 1.5$ days), and delay from onset of symptoms to when the individual detects the symptoms and enters isolation ($d_d = 0.5$ days), and the period of self-isolation ($d_i = 7$ days). At the end of the incubation period, a symptomatic individual starts showing symptoms, which allow the triggering of control policies (after the delay $d_d$). Asymptomatic individuals do not trigger any policy. After a latent period (defined as the incubation minus the prodromal period) and irrespective of symptoms, any infected case starts an infectious period with an infectivity that changes over time following the probability density function of a Gamma distribution (mean $\mu_F = 2.2$ and standard deviation $\sigma_F = 1.64$ days). Asymptomatic cases are assumed half as infectious as symptomatic ones (relative infectivity $f_a = 0.5$).

\begin{figure}
        \centering
\begin{subfigure}[t]{0.45\textwidth}
        \includegraphics[width=1\linewidth]{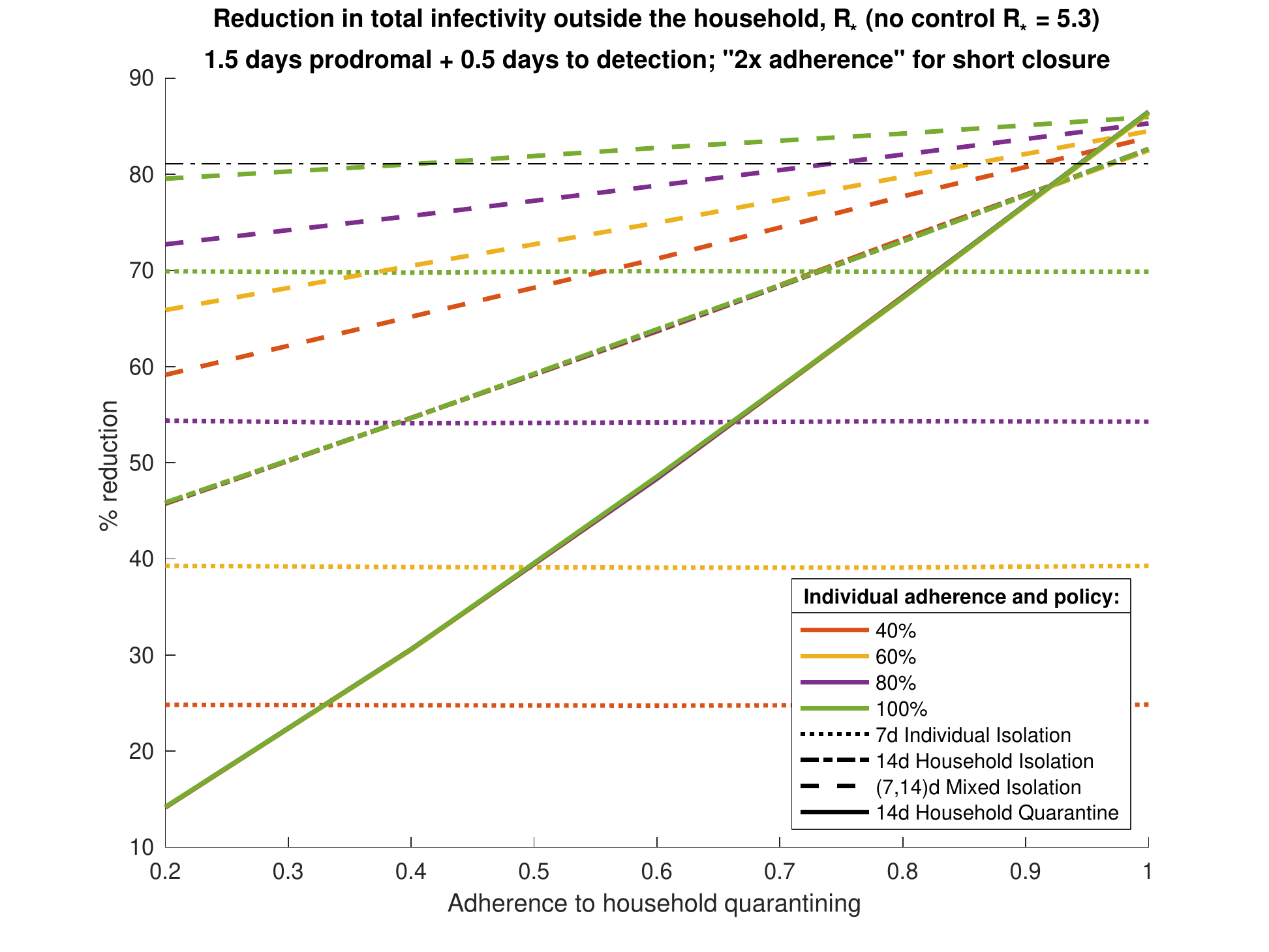}
        \caption{Percentage reduction in $R_\ast$.}
        \label{fig:L-2}
\end{subfigure}\hspace{1cm}
\centering
\begin{subfigure}[t]{0.45\textwidth}
        \includegraphics[width=1\linewidth]{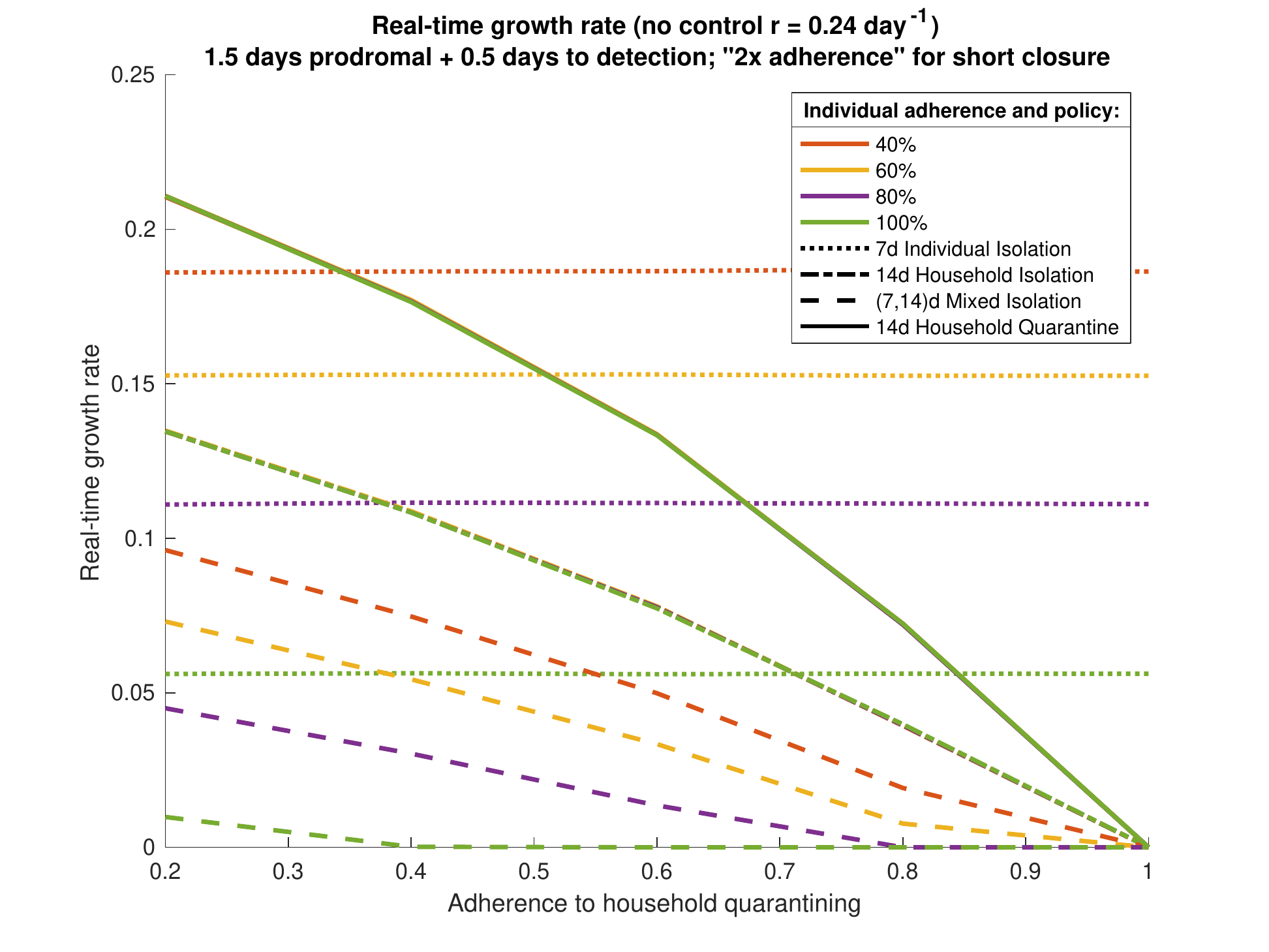}
        \caption{Reduction in the real time growth rate.\vspace{1cm}}
        \label{fig:L-1}
\end{subfigure}
        \centering
\begin{subfigure}[t!]{0.45\textwidth}
        \vspace{-1cm}\includegraphics[width=1\linewidth]{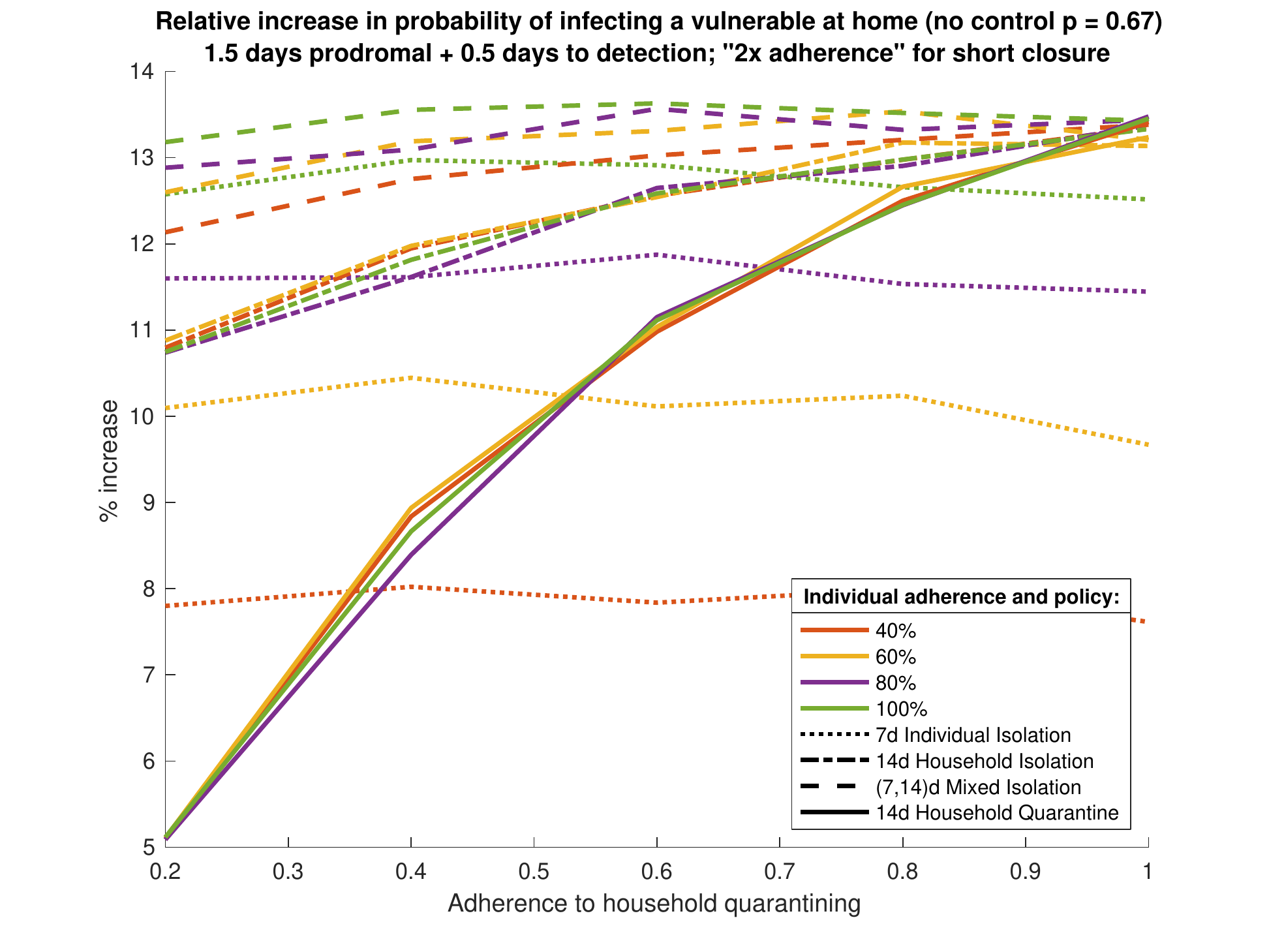}
        \caption{Increase in risk of infection an initially susceptible vulnerable person experiences in the household.}
        \label{fig:L-3}
\end{subfigure}\hspace{1cm}
\begin{subfigure}[t!]{0.45\textwidth}
      \vspace{-1cm}\hspace{-0.5cm}\includegraphics[height=0.8\linewidth,width=1.1\linewidth]{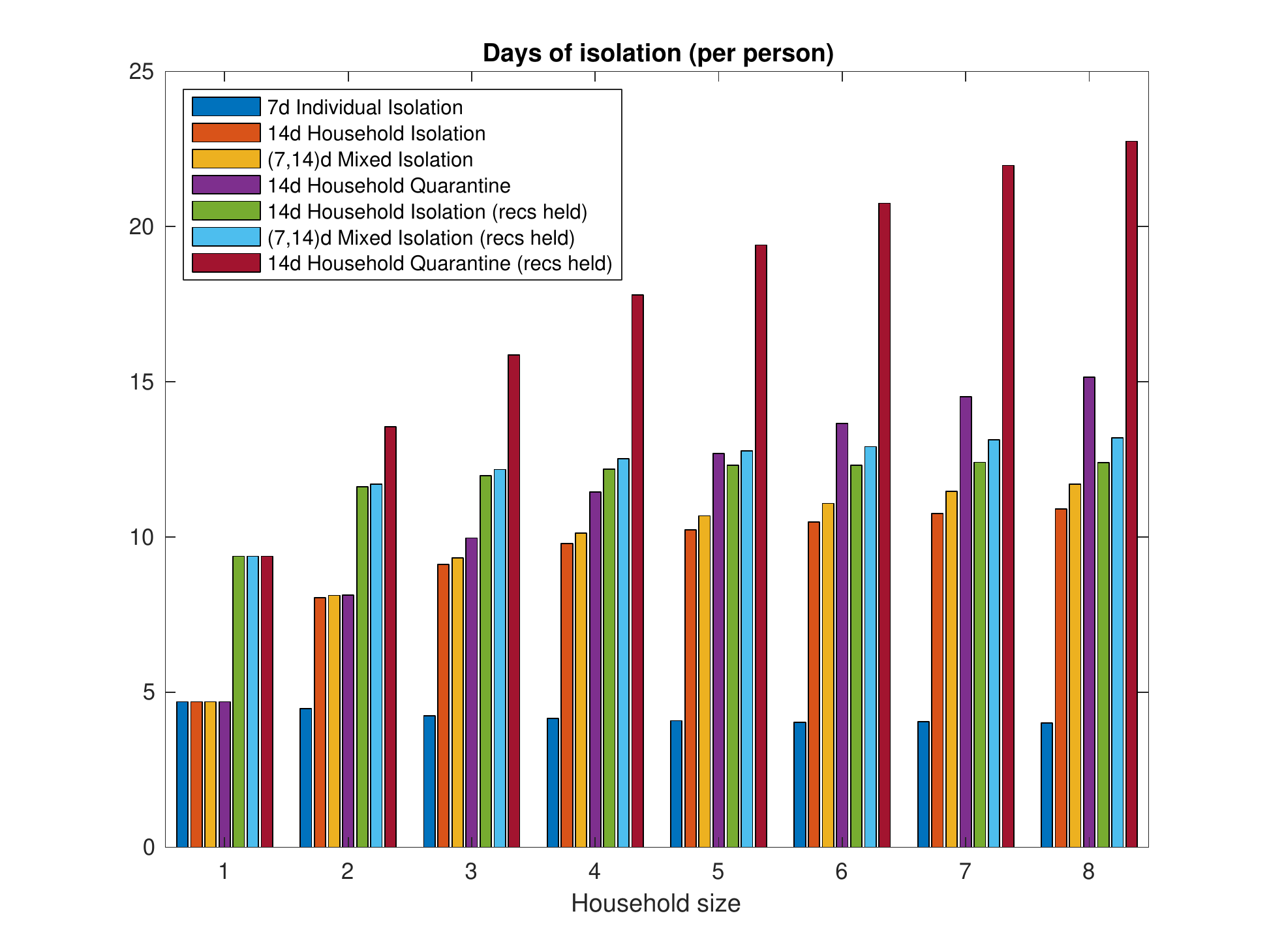}
        \caption{Average number of days of isolation a person experiences in households of different sizes.}
        \label{fig:L-4}
\end{subfigure}
\caption{Impact of different control policies and levels of adherence on transmission, infection risk, and time in isolation.\\
(a) Percentage reduction in $R_\ast$, defined as the total amount of community transmission spread by an average household early in the epidemic, which equals 3.1 for baseline parameters in the absence of control; (b) real-time growth rate, which is assumed to be 0 (rather than negative) when the infection is controlled; (b) increase in risk of infection an initially susceptible vulnerable person experiences in the household; and (d) the average number of days of isolation a person experiences on average in households of different sizes, computed for each size as the average total person-days in isolation divided by the number of individuals in the household. In (a)-(c), line styles refer to different control policies and colours to different levels of adherence to individual isolation. The $x$-axis gives the adherence to household quarantine, and household isolation (which is less demanding) is assumed to be ``twice as high'', meaning it is at the midpoint between that of household quarantine and 1 (e.g.\ 0.6 for an $x$ value of 0.2, 0.9 for an $x$ value of 0.8, etc.). The black dash-dotted line in (a) gives the amount  needed to control the spread by achieving $R_\ast=1$. Notice how: the effect of individual isolation is independent of adherence to household quarantine (dotted lines); the effect of household isolation is independent of adherence to individual isolation (overlapping dash-dotted lines); mixed isolation is always superior to household isolation; household quarantine is only optimal at really high levels of adherence (for these baseline parameters, generally, beyond the level needed to achieve control), but quickly becomes suboptimal to mixed isolation as adherence is reduced. When a sufficiently large reduction in $R_\ast$ is achieved in (a), the growth rate drops to 0 in (b). The increased risk an initially susceptible vulnerable person is infected at home (c) does not reflect this effect, as it represents the increased risk conditional on a introduction: if the infection were controlled in the community, the overall risk of a vulnerable person getting infected would vanish as the risk of introduction in the household vanishes. For these plots, $n_e = 10 000$ simulations are performed for each household size. Nevertheless, a large amount of stochastic noise is still visible in (c). In (d), the same control policies are considered, but the household-based ones are considered both in their na\"{i}ve form (where recovered individuals remain isolated), and in their upgraded version where recovered individuals are free to leave the house: they are identical in terms of transmission but the na\"{i}ve versions are significantly more costly in terms of person-days of isolation. In a household of size 1 (no within-household transmission), the days in isolation would be exactly 7 or 14 if all cases were symptomatic (here $p_s = 2/3$); similarly, in all households, individual isolation would total exactly 7 days if all cases were symptomatic and all individuals in the household were ultimately infected.}
\end{figure}

\noindent\textbf{Summary}

Under the baseline scenario control can in principle be achieved, but only for high levels of adherence, which might be difficult to enforce for a prolonged length of time. More importantly, the model's conclusions are highly sensitive to variations in parameter choices, which are uncertain. Parameters that present problems here are the delay from symptom onset to isolation (with control failing for 1 day detection delay unless adherence is essentially perfect), proportion of asymptomatic infections (any chance of control lost at 50\%) and the strength of asymptomatic transmission. The short delay before symptomatic individuals isolate may be unrealistic unless the susceptible population is very well-informed about symptoms that call for isolation, and so likely does not apply in very early stages of an outbreak. Overall, in the face of the many uncertainties, household-based interventions triggered purely by symptoms appear useful to slow the spread but need to be complemented by other policies.

Comparing the different strategies (Figure~\ref{fig:L-1}), household quarantine can be optimal (as one might expect), but this requires high adherence levels. As adherence drops, this strategy becomes suboptimal to mixed isolation. Mixed isolation is significantly better than household isolation on its own and requires little extra social cost, so should not cause adherence to drop (relative to household isolation adherence levels). The difference between the two strategies comes down to the transmission slipping through after the 14 day household isolation. The cheapest strategy, when considering working age adults, is individual isolation (Figure~\ref{fig:L-4}), but the effect is limited compared to the other models and cannot achieve control in the baseline scenario even with $100\%$ adherence.

Overall, the mixed isolation strategy appears to be most cost-effective. However, this is dependent on the assumption that adherence is better for 14 day isolation rather than a very long quarantine. It can be observed that household-based interventions are more effective than individual isolations, demonstrating the importance of these strategies in designing intervention policy. Figure~\ref{fig:L-within} shows how the different isolation strategies contain the infectious periods of individuals within the household and also indicates the number of individuals being isolated within the household. 

To study the impact such an increased within-household transmission has on the chance that a vulnerable individual is infected in the household, we randomly choose one non-primary case in the household as the vulnerable one and count how many of the $n_e$ epidemics result in this individual being infected under the different control policies (Figure~\ref{fig:L-3}). Under these interventions, the risk of a vulnerable individual getting infected within-household, conditional on the infection entering it in the first place, is in the range $15 - 25\%$. 

%%%%_____________________________________________________________

%%%%_EXTINCTION___________________________________________________
\subsection{Extinction probabilities}
\label{sec:extinction}
Social distancing, isolation and lockdowns act to mitigate the spread of an infectious disease and reduce the number of cases. However, such interventions, particularly widespread lockdowns, cannot be maintained indefinitely and must be lifted at some point. For the disease to be controlled, these interventions can be implemented until pharmaceutical interventions are developed, such as a vaccine, or until the case numbers are low enough that the disease may go extinct.
Here, we consider the situation where interventions are lifted just before extinction, when the number of cases has reached a low but non-zero initial value $n_0$: at this point, the number of cases might rebound or might go extinct by random chance despite an $R_0>1$. We use a time-inhomogeneous birth-death chain model \cite{Kendall1948} to investigate the probability of extinction in this context. The $n_0$ ``initial'' cases give rise to new cases at a time-dependent rate $\beta(t)$ and recover at rate $\delta(t)$. Letting $Z(t)$ denote the random variable that gives the number of cases at time $t$, we are first interested in obtaining an expression for the probability generating function
\begin{equation*} 
Q(t, s) = {\mathbb{E}}[s^{Z(t)}] = \sum_{n=0}^\infty \mathbb{P}(Z(t) = n)s^n.
\end{equation*}
It follows that $Q(t, s)$ satisfies the differential equation
\begin{equation}
\frac{\partial Q}{\partial t} = (\delta(t)- (\delta(t) + \beta(t))s +\beta(t)s^2)\frac{\partial Q}{\partial s}, \label{Qeq}
\end{equation} 
subject to the initial condition
\begin{equation*}
Q(0, s) = s.
\end{equation*}
Solving for $Q$ and setting $s=0$ gives the probability that, at time $t$, the number of cases has reached zero and the disease has become extinct. We denote this probability by $q(t)$ \cite{Alexander2012}, which is given by
\begin{equation}
q(t) = 1 - \Big(\int_0^t (\beta(t_1){{\mathrm{e}}^{I(t_1)})dt_1 + {\mathrm{e}}^{I(t)} } \Big)^{-1}.
\end{equation}
The above case considers a closed population. Since the virus has spread worldwide, for any population of interest, immigration of infected individuals cannot be ignored. To capture this, we model the case where immigration from external sources is introduced into the system at a rate $\eta(t)$, and are similarly interested in the random variable $Y(t)$, which denotes the number of cases at time $t$. The corresponding generating function, $R(t, s)$, for this random variable satisfies
\begin{equation}
\frac{\partial R}{\partial t} = (\delta(t)- (\delta(t) + \beta(t))s +\beta(t)s^2)\frac{\partial R}{\partial s} + \eta(t)(s-1)R.  \label{Req}
\end{equation} 
Again, solving for $R(t, s)$ and setting $s=0$ gives the probability, $r(t)$, that there are no cases of infected individuals left, at which time a new case can only arise through immigration from an external source. This probability is given by
\begin{equation}
r(t) = {\mathrm{exp}}\Big(-(1-q(t)) \int_0^t \eta(t_1) dt_1)  \Big).\label{Rsol}
\end{equation}

\begin{figure}[h!]
\begin{center}
\includegraphics[width=0.6\textwidth]{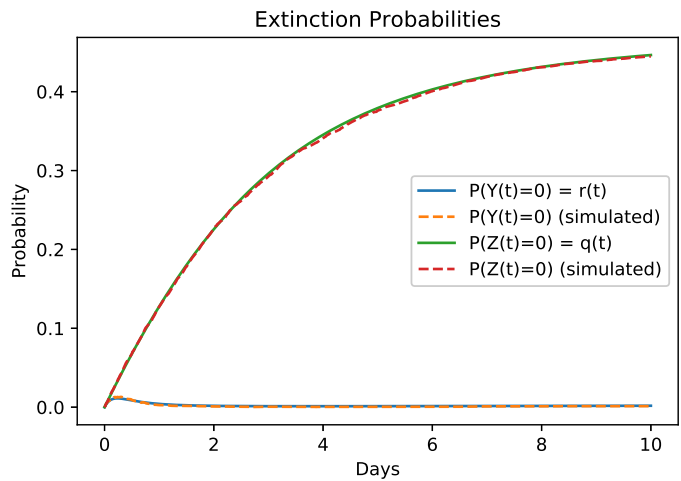}
\caption{Extinction probabilities, both analytic and simulated, for the choices of $\beta(t), \delta(t)$ and $\eta(t)$ described in the main text. The simulated extinction probabilities were calculated from 10,000 simulations of a birth-death chain both with and without immigration, for which the code can be obtained via the supplementary material. }\label{fig:extinction}
\end{center}
\end{figure}

We simulate data based on one initial case $n_0 =1$, though this may easily be extended to any number of initial cases. We run simulations both with and without immigration, choosing $\beta(t) = 3/(7(1+5{\mathrm{e}}^{-t}))$ and $\delta(t) = 1/7$ for all $t$, so that an effective reproduction number given by $\beta(t)/\delta(t)$ grows gradually from 0.5 to 3 after interventions are released, and choosing immigration rate $\eta(t) = W_0{\mathrm{e}}^{-t}$, where $W_0$ is the initial (constant) rate of importation of cases before any controls on immigration are put into effect. We set $W_0=5$ imported cases per day. With these choices of parameters, the resulting extinction probabilities are given in Figure~\ref{fig:extinction}. Note that we are assuming the immigration rate is decreasing to 0, so if the infection is controlled internally for long enough, an overall ultimate extinction is possible in this model. For these parameter choices, the final probability of extinction, defined as ${\mathrm{lim}}_{t \to \infty}q(t)$ (without immigration) and ${\mathrm{lim}}_{t \to \infty}r(t)$ (with immigration) are approximately 0.446 and 0.002, respectively. It should be noted that $q(t)$ concerns the best case scenario with only one initial case. Increasing the number of initial cases $n_0$ scales the probability of extinction by $q(t)^{n_0}$. These probabilities suggest that, without widespread immunity, stochastic extinction might be aided by social distancing but is heavily compromised by immigration. Border controls, therefore, if of limited use when transmission is self-sustaining, become key when the number of cases is low. Note that we have assumed an importation function $\eta(t)$ that goes to 0 for large $t$, in line with a pandemic that goes extinct in other geographical regions. However, the presence of an animal reservoir might lead to an importation function that is non-zero over longer time scales, thus effectively making ultimate extinction impossible unless the effective reproduction number is kept below one by a systematic and permanent intervention (e.g.\ technology-based change in behaviour) or herd immunity.

%%%%_____________________________________________________________

%%%% CONTACT_____________________________________________________
\subsection{Contact tracing and household isolation}
Contact tracing is a complementary control policy to isolation or quarantine. When a case is discovered, attempts are made to identify and isolate individuals who may have been infected. In doing so, some of the secondary cases will be discovered and isolated early in their infection, decreasing their effective infectious period. If contact tracing is successful, it can greatly reduce the effective reproduction number of the infection, and in combination with other interventions may drive an epidemic extinct, as was seen in the case of SARS~\cite{Wilder2020}.

Contact tracing in itself presents numerous challenges, which are exacerbated by its success relying not only on the effectiveness of the tracing process but also the underlying transmission characteristics.
For COVID-19, some of these challenges include mild symptoms which cause infections not to be reported, pre-symptomatic transmission which occurs before a case is reported, and short generation times~\cite{Ganyani2020} which can cause the epidemic to outrun contact tracing. Additionally, contact tracing is only feasible for smaller case numbers, because each case generates multiple contacts to follow up, so the tracing workload expands dramatically, and an increasing number of chains remain unobserved. This makes it a viable strategy in the early days of an outbreak, or, if containment has failed, following a period of severe interventions, such as a lockdown. Combining contact tracing with isolation is being considered by many countries as part of a test, trace and isolate strategy to be implemented once lockdowns or comparable measures are lifted, provided these lockdowns succeed at driving case numbers sufficiently low. In this section, we develop a household-level contact tracing model for an emerging outbreak, since we do not wish to make assumptions about immunity or depletion of susceptibles. These assumptions can be added to the model as the availability of data into immunity improves. We are interested in the likelihood that the contact tracing process is overwhelmed by large case numbers and the likelihood that, combined with isolation, it can drive the disease to extinction.

The early days of an outbreak can be modelled using a branching process, where generations of infections produce infectious offspring. Contact tracing processes can be incorporated as a superinfection along the tree generated by the branching process~\cite{Ball2015}. When a node is ‘superinfected’ by the contact tracing process, it is isolated.

We model the infection spreading through a fully susceptible population of individuals, segmented into households of different sizes according to the 2019 ONS survey~\cite{FH:2019}, and progress through discrete time steps of 1 day. As such, our branching process is at the household level, coupled with localised within-household epidemics. This allows us to model contact tracing strategies that isolate whole households, which may contain several undetected infections. It also enables a wider range of contact tracing strategies to be modelled, each with different intervention scope and costs.

Each day, individuals (or nodes) make contacts to a random set of individuals; divided into local contacts to members of the same household, and global contacts to members of other households. The number of individuals contacted in a day is distributed using an overdispersed negative binomial distribution and parameterised using estimates from the POLYMOD social contact survey~\cite{Mossong2008}, stratified by household size. Since the probability that a contact causes infection cannot be directly observed, we use improper hazard rates that give rise to the 5 day COVID-19 generation time~\cite{Ferretti2020} and $R_0 = 3$.

For contact tracing to begin, an infection must be diagnosed, which we assume occurs 70\% of the time among infected individuals due to flaws in reporting or very mild symptoms in those infected. We assume a Gamma distributed incubation period with mean 4.84 (Table~\ref{Table:parameters2}) and a Geometric reporting delay from symptom onset with mean 4.8 days~\cite{Kraemer2020}. Intuition suggests that if $R_0=3$ then tracing two thirds of contacts will control the epidemic. However, in practice transmission may occur before tracing, so this will not reduce the number of infectious contacts by two thirds. To demonstrate this, we assume that contact tracing successfully traces two thirds of contacts. Trained professionals have to trace all reported contacts from the last 14 days, so we assume that the contact tracing delay follows a Geometric distribution with a mean of 2 days. Individuals are considered recovered 21 days after infection, as the chances that they are still transmitting then are negligible.

Though our general framework can be modified extensively, we assume the following contact tracing strategy. When an individual reports infection, their household is immediately isolated. Contact tracing attempts are then made for all households connected to one of the individuals in this household, whether symptomatic or not. When a connected household is identified (after the contact tracing delay), all individuals within the household are immediately placed under observation. If any of the individuals in the observed households develop symptoms, then the household becomes isolated and the contact tracing process continues to connected households. When a household is isolated, we assume all individuals are isolated with 100\% adherence, and cannot transmit the virus within or outside the household. The assumption that isolation prevents local infections is unrealistic, but does not change the overall behaviour of the process as there are no more global infections. This strategy imposes high individual-level cost, since by isolating all individuals within a household, it isolates individuals who have not had direct contact with an infected individual. In practice, such a strategy may have poor adherence. Figure~\ref{fig:M-1} shows an example contact tracing network.

\subsubsection{Hitting times of contact tracing capacities}
When choosing contact tracing strategies, a balance must be struck between the effectiveness of a strategy and the resources that it requires. Some strategies are only feasible when there are few infections, since the resources required can grow rapidly depending on the dynamics of the outbreak and the contact tracing process.

To define the capacity of the contact tracing process, we consider the ability of a public health agency to observe the condition of those asked to self-isolate, due to their recent exposure to an infected individual. The health agency must remain in contact for the duration of the 14 day self-isolation period, so that if any individual under isolation develops symptoms and then tests positive, the contact tracing process can be initiated on this node. We will define the capacity of the contact tracing process to be the number of people that can be placed under observation and assume two possible capacities: 800 and 8000. We assume that when a node is contact traced, they are asked to report their global contacts for the last 14 days. All global contacts are assumed to be to a new person since we are in the early stages of an outbreak.

\begin{table} [h!]
\caption{Contact tracing capacity hitting probability and hitting time distribution} 
\begin{center}
\begin{tabularx}{0.5\textwidth}{| X|l| c |}
\hline
Quantity & Results \\ \hline
Mean (time 800 capacity reached) & 13.9 days  \\ \hline
Mean (time 8000 capacity reached) & 22.5 days \\ \hline
Hitting probability (800) & 81.2\%  \\ \hline
Hitting probability (8000) & 76.8\%  \\ \hline
\end{tabularx}
\label{Table:M-1} 
%}
\end{center}
\end{table}

\begin{figure}
        \centering
\begin{subfigure}[t]{0.45\textwidth}
\includegraphics[width=1\linewidth, height=0.2\textheight]{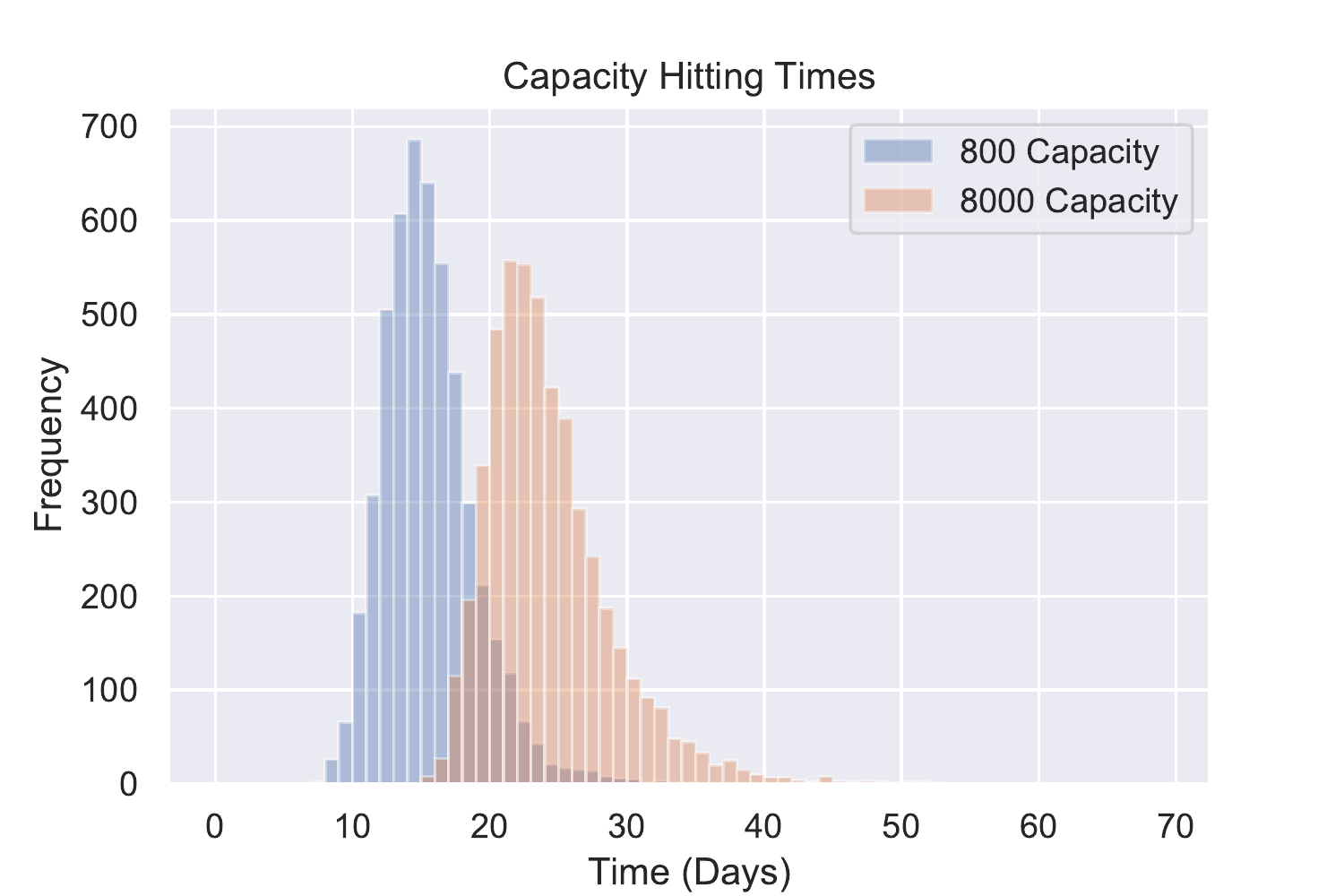}
\caption{Distribution of hitting times for 800 (blue) and 8000  (orange) contacts traced. The distributions appear to follow Gumbel distributions.}
\label{fig:M-2}
\end{subfigure}\hspace{1cm}
\centering
\begin{subfigure}[t]{0.45\textwidth}
\includegraphics[width=1\linewidth, height=0.2\textheight]{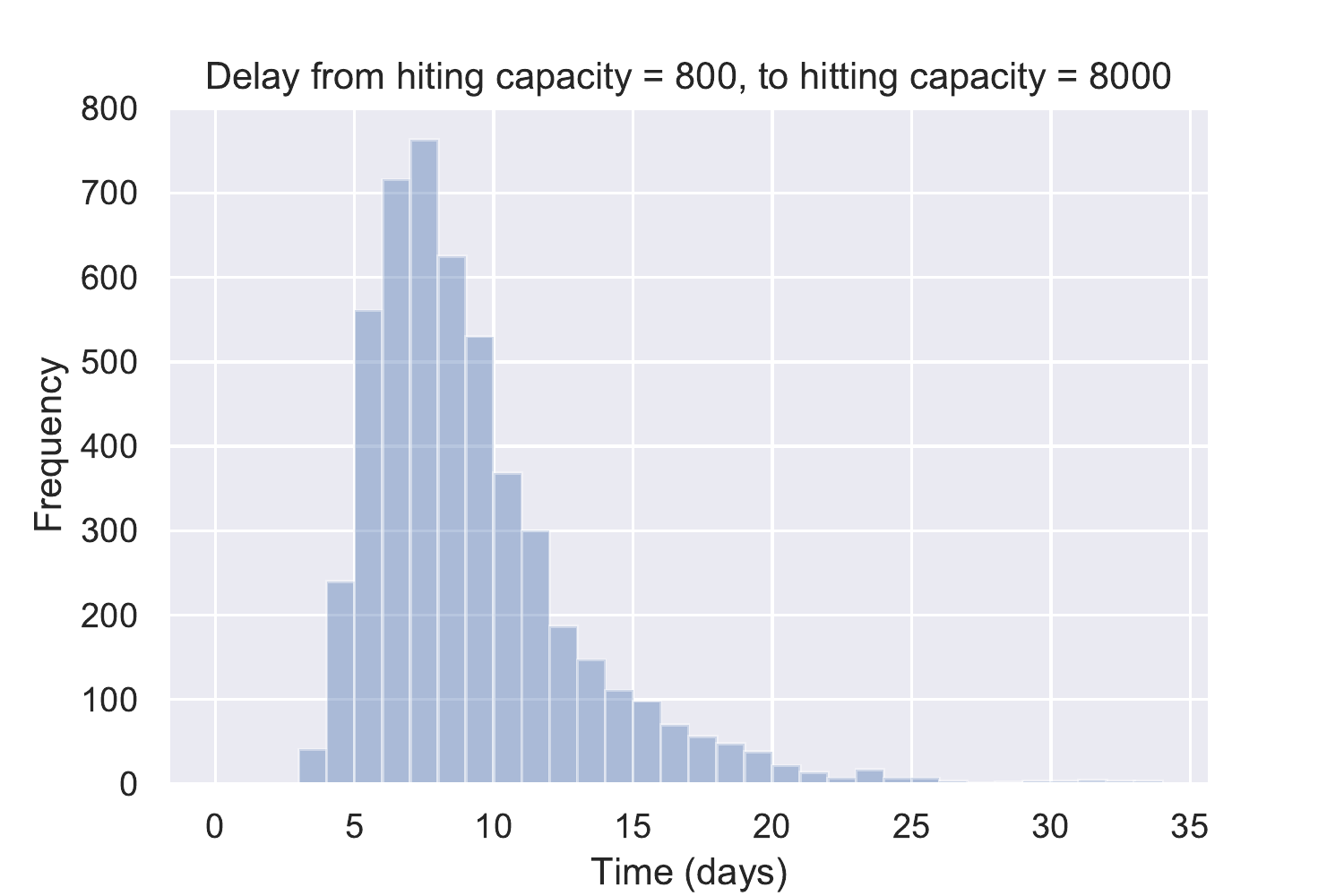}
\caption{The delay between hitting 800 contacts traced and 8000. This distributions appears to follow a Gumbel distribution.}
\label{fig:M-3}
\end{subfigure}
\caption{Capacity hitting times for the contact tracing model.}
\end{figure}

We carried out 6507 simulations of the contact tracing process for 150 days. Contact tracing capacity was reached in 5000 simulations, and in 180 the epidemic neither went extinct nor was the 8000 capacity reached. In the remaining simulations, the epidemic went extinct. Figure~\ref{fig:M-2} and Table~\ref{Table:M-1} show that increasing the contact tracing capacity tenfold less than doubles the time until that capacity is reached. However, it does increase the odds of driving the epidemic to extinction without hitting the capacity by about 10\% (Table~\ref{Table:M-1}). Different contact tracing strategies will strain different aspects of the health agency. A strategy that generates large amounts of work is only feasible if there are few active infections. The optimal strategy will need to compromise and may need to change depending on the number of active infections, which cannot be directly observed.

\subsubsection{Extinction Time}
When there is a small number of cases in a single country, it may be possible to drive the pathogen to extinction. This small case number could correspond to the start of an outbreak or removing of severe interventions. We consider the latter case, but conservatively assume a fully susceptible population. 

\begin{figure}
        \centering
\begin{subfigure}[t]{0.45\textwidth}
\includegraphics[width=1\linewidth, height=0.2\textheight]{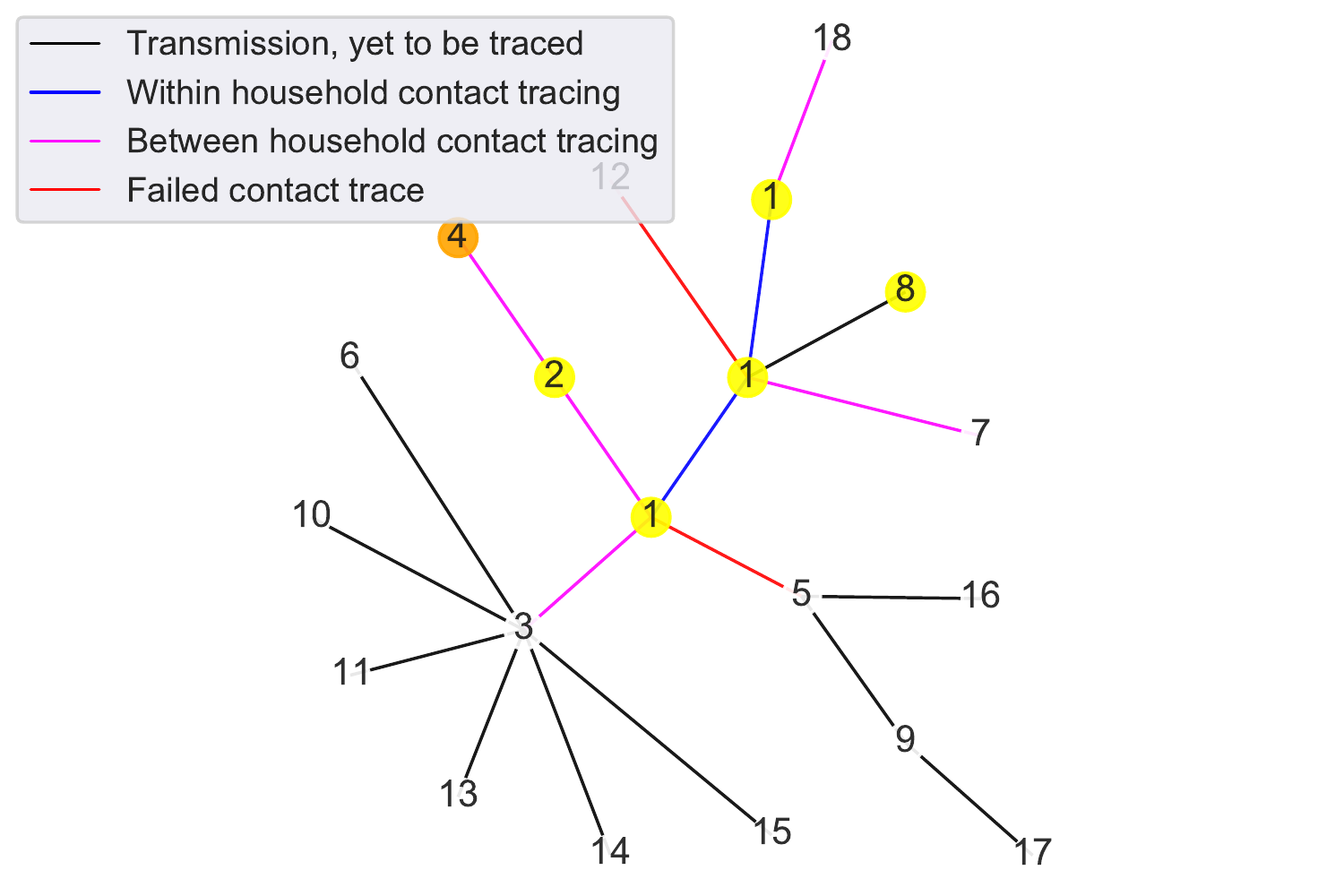}
\caption{A visualization of the branching process model after 20 days. Node numbers represent the household that they belong to. Nodes 3, 7 and 18 will be contact traced but the process has not reached them yet, due to contact tracing delays. Node 4 has been contact traced, but has not been quarantined in the plotted scenario, as it does not display symptoms.}
\label{fig:M-1}
\end{subfigure}\hspace{1cm}
\centering
\begin{subfigure}[t]{0.45\textwidth}
\includegraphics[width=1\linewidth, height=0.2\textheight]{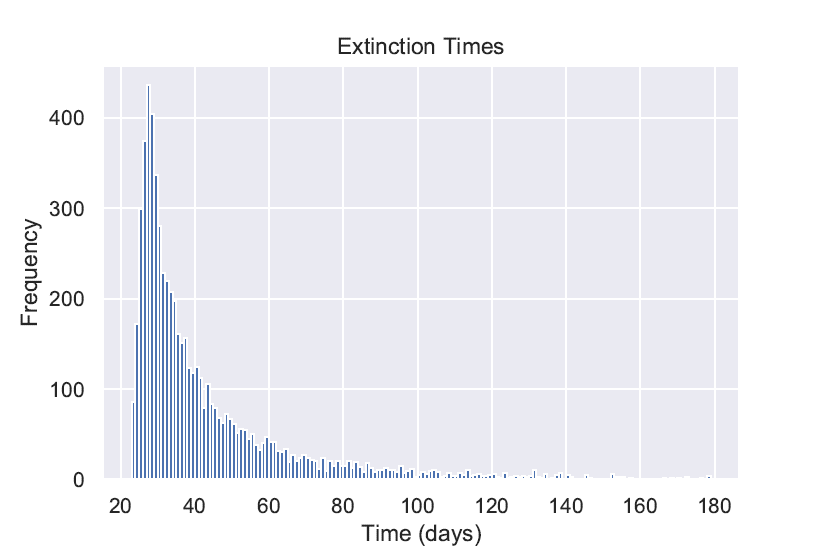}
\caption{Distribution of extinction times for epidemics that transmit the infection at least once. The distribution takes the shape of a generalised extreme value distribution, and there are several epidemics that still have not gone extinct after interventions have been in effect for long periods of time, which exemplifies how difficult it can be to make an epidemic extinct.}
\label{fig:M-4}
\end{subfigure}
\caption{Example of the contact tracing process (a) and the extinction times distribution (b) for the contact tracing model.}
\end{figure}

We assume that social distancing is enforced on day 0 and reduces global contacts by 70\%. 
Since we are interested in extinction, we will no longer consider the contact tracing capacity. Under these baseline parameter assumptions and 10000 simulations, the combined force of this contact tracing strategy and isolation is enough to drive the epidemic extinct (Figure~\ref{fig:M-4}), but measures will need to be in place for months in some cases. If the infection is ever re-imported, then the process would begin again, since herd immunity is not achieved. Note that the minimum extinction time is 21 days due to this being the time after which an infected individual is labelled recovered.

Additionally, this model only considers extinction under the assumption that no cases are imported. In Section~\ref{sec:extinction}, we have shown that importation of cases significantly reduces the extinction probability. This suggests that extinction may no longer be guaranteed, and the time to extinction will be significantly increased. This analysis has focused on a single contact tracing strategy using indicative parameters for COVID-19. The proposed model can be extended to more strategies and region specific parameters to inform the design of control policies. Also, as is shown in Table~\ref{Table:M-1}, contact tracing capacity is likely to be reached, which may prevent extinction from being achieved. This complication is compounded by the issues of loss of immunity or the presence of an animal reservoir discussed in Section \ref{sec:extinction}.

%%%%_____________________________________________________________

%%%% DISCUSSION_____________________________________________________
\section{Discussion}
In this manuscript we have presented a range of mathematical tools to tackle infectious disease outbreaks. In particular, these tools address various technical questions posed by the authors to support the ongoing public health response to COVID-19. This toolkit considers both estimation efforts for key parameters, and investigative efforts (often numerical simulations) in gauging the effectiveness of various intervention or control measures. Joint consideration of estimation and simulation efforts is critical. Parameter estimates are obtained using a certain set of assumptions regarding the data, and investigations or simulations utilising these estimates should ensure that their underlying assumptions are consistent. These challenges in model construction and applicability of statistical methods are compounded by the limitations of the data with which decisions must be made. Some of the biases present in the data can be addressed with an improved data collection methodology - often challenging in the context of a fast-moving outbreak -  but many are also inherent to the nature of early outbreak data~\cite{Britton2019,Lesko2020}. The consequent lack of intuitive insight from this data underscores the need for careful parametric estimates, especially considering the large variability in predicted outcomes resulting from small differences in parameters. Even with robust estimates for some parameters, many other parameters are challenging to estimate using the available data. Therefore, models need to address this variability and uncertainty in order to inform public health policy.

We have presented methods to address biases arising from a growing force of infection, changes in the reporting rate, truncated data samples and a varying travel rate. We use these methods to account for these biases when estimating delay distributions, such as the incubation period and hospitalisation-to-death delay, and the growth rate/doubling time (results given in Table~\ref{Table:parameters2}).  These biases can have significant impact when estimating key parameters: the mean incubation period estimates for COVID-19 range from 3.48 days without correcting for truncation to 4.84 days with the correction, and the doubling time in Hubei province decreases from 3.15 days without correcting for travel to 2.77 days. These differences can significantly alter our understanding of the outbreak, and could have a large impact on policy and public health. For instance, underestimating the incubation period may lead to quarantine strategies failing to identify infected individuals if the quarantine length is too short. Overestimating the doubling time (or underestimating the growth rate) will underestimate the risk posed to the host population - both in terms of final size of the epidemic and the rate at which it spreads, which can have significant public health impacts as discussed in~\cite{pellis2020challenges}.

It is important to note that the above-mentioned biases, and consequent impact of implementing the methods correcting for their presence, may vary across setting. 
As an example, the potential underestimation of the COVID-19 growth rate is exacerbated by an overlap in early outbreaks with a period of significant travel and movement in China, and would be less detrimental if first observed in other populations such as Italy. Also, for the incubation period, we have shown two different types of data; one from Wuhan and one from discrete infection events. In the Wuhan data set, truncation and force of infection biases are very important, whereas in the other data set, there is no force of infection bias since the infection events are observed. 

When an outbreak occurs in an enclosed group, such as a large gathering, we may wish to know how many individuals are likely to be infected. We developed a statistical method to estimate the first generation size based on the number of symptomatic individuals, taking care to account for the uncertainty in this quantity. This ready reckoner can inform testing of large groups to help control the disease spread, but does not apply to later generations or the possible interventions enacted on the population.

Building on these enclosed population scenarios, we have developed a set of models that investigate public control measures or interventions on enclosed populations, such as households and care homes. These structured descriptions improve the population risk profiles relative to assumptions of homogeneous mixing. 
A complementary aspect to a structured population when modelling interventions is adherence. Motivated by vaccination modelling, we consider leaky adherence, where every individual/household chooses to adhere or not whenever an event occurs, and all-or-nothing adherence, where some households adhere every time and some never adhere. We observed that in a homogeneous population, although the two types of adherence predict the same growth rate and final size, the timing of the peak and the early growth can be faster under all-or-nothing adherence. This insight, combined with lessons from the vaccination literature, suggests that efforts should focus on ensuring complete adherence in individuals/households with some level of pre-existing adherence, rather than pushing non-adherent individuals/households to change behaviour. 

Dedicated modelling of disease spread in care homes is essential due to the documented history of co-morbidity of their residents during pandemics~\cite{Guan2020,WuJT2020,WuZ2020,Yang2020,Zhou2020}.
We do so by regarding care homes as closed populations that are subjected to a force of infection from an external epidemic. We develop a tool for analysing the risk posed to this population by determining the peak size of the epidemic within the care homes and the number of deaths. 
Applying this model to COVID-19, we find that by ``cocooning" the care homes, i.e.\ shielding them to reduce the chance of introduction from the external outbreak, we can significantly reduce the size of the peak and therefore reduce the number of deaths. However, assessing the necessary level of shielding requires accurate characterisation of the external force of infection, and underestimating this may invalidate shielding efforts. A limitation to the proposed model is the deterministic within care home epidemic. However, since the average size of care homes is relatively large and we assume a high $R_0$ within care homes, the deterministic assumption is unlikely to significantly alter the conclusions.

When modelling households however, we are concerned with much smaller population sizes. Therefore, it is important to consider stochastic effects within each household, combined with between-household dynamics. We consider two different household models: one which contains features of both within- and between-household transmission, where small-scale transmission can be linked to the epidemic on a population level, and another which facilitates more detail in the within-household transmission and delay distributions, but with reduced correspondence to the population-wide transmission. With the first model, a 65\% adherence to household isolation appears insufficient to control the epidemic without severe global reductions in transmission. Coupled with a short term global reduction, the epidemic can be controlled, but upon lifting the global intervention, which could take the form of a lockdown, household isolation is insufficient to maintain control. For the second model, we look into changing the strength of adherence, and the impact this can have on achieving control. Indicative but reasonable parameter values suggest that the COVID-19 outbreak can potentially be controlled using household isolation strategies, provided the level of adherence is sufficiently high. However, such a high level of adherence may be difficult to maintain in the long-term and this modicum of control is anyway highly sensitive to the chosen parameters. We further investigated the efficacy of various isolation or quarantine measures. A policy of individual isolation struggles to curtail the epidemic for any adherence. Instead, mixed isolation, whereby first the whole household isolates and any individual infected during isolation goes on to self isolate after household isolation is lifted, appears to be the most cost-effective strategy. 

Countries have put into place strict social distancing and lockdown intervention to suppress or regain control of epidemics that threaten to overwhelm the health system and cause massive mortality, but they cannot be sustained in the long term without growing social and economic costs. We have shown however, that the probability of the epidemic becoming extinct once these policies are lifted, even when very few cases remain, is very small. We therefore consider a contact tracing intervention as a potential strategy for managing the COVID-19 outbreak, once severe lockdown interventions are lifted. We developed a household-level contact tracing model to explore the feasibility of combining these strategies to control the epidemic. Firstly, we noted that by using knowledge of household structure, we can reduce the burden on the contact tracing process by isolating household and removing them from the contact tracing process once an infected member has been identified. Secondly, we investigated how contact tracing combined with household isolation may drive the disease to extinction, finding that aggressive contact tracing coupled with household isolation can drive the epidemic to extinction under the indicative parameters assumed, when starting with a single infection. However, the time until extinction can be impractically long, which risks the contact tracing capacity being overwhelmed, suggesting such a strategy may be infeasible in practice. A less aggressive strategy could be implemented that would be less likely to overwhelm local health agencies. Whilst it may not lead to extinction, this can still be beneficial at mitigating and controlling the spread of an outbreak as part of a test, trace and isolate strategy.

There are many complexities when modelling an outbreak of a novel infectious disease. To address some of these, we have described a variety of techniques to serve as part of a generally applicable toolkit.
However, our proposed models, and many other models, are subject to important limitations which must be considered prior to their application. Key among these is the lack of heterogeneous population mixing, such as through age-stratification~\cite{Pellis2020} and different risk-groups~\cite{Valdano2019a}, and spatio-temporal variations~\cite{Lau2017}, all of which influence modelling estimates and predictions. 
Nevertheless, the relative simplicity of the presented models allows for the development of qualitative intuition regarding the efficacy of various intervention methods, whilst providing tractable theoretical frameworks which can be further developed and better inform policy-makers.

%%%%____________________________________________________________

%%%%_APPENDIX____________________________________________________
\begin{appendices}
\renewcommand{\thefigure}{S\arabic{figure}}
\setcounter{figure}{0} 
\section{Independence of the incubation period likelihood function and the reporting rate}
\label{app:reporting}
To estimate the incubation period distribution, we need to find the distribution that maximises the probability of observing the sampled data. However, the sampled data does not directly record incubation period, and instead contains infection exposure window and the symptom onset date. Additionally, the sample does not contain all individuals, and therefore there is a reporting rate that must be incorporated into the likelihood function. If the reporting rate is constant, it can be ignored. However, in the data coming out the Wuhan, the reporting rate varies significantly, since individuals are no longer exported from Wuhan after travel restrictions. Since the reporting rate depends on individuals leaving Wuhan, the main factor effecting the probability that a case is included in the data set is the date an individual leaves Wuhan. Therefore, the reporting rate depends on the days an individual spends in Wuhan and is independent of their symptom onset date. We need the likelihood function that an individual was infected between days $a$ and $b$, had symptom onset on day $y$, and was included in the data set. 

We will condition against the infection window, $a<I<b$, the case being included in the data set, $x \in D$, and symptom onset occurring before the truncation date, $Y<T$, and determine the probability that for such an individual we observe the given symptom onset date, $Y=y$. That is, we need to find $P(Y=y|\{a<I<b\}\cap\{Y\leq T\}\cap \{x \in D\})$. This can be rearranged to
\begin{align}
\frac{P(\{Y=y\}\cap\{a<I<b\}\cap\{Y\leq T\}\cap \{x \in D\})}{P(\{a<I<b\}\cap\{Y\leq T\}\cap \{x \in D\})}&=\frac{P(\{Y=y\}\cap \{x \in D\}|\{a<I<b\})}{P(\{Y\leq T\}\cap \{x \in D\}|\{a<I<b\})} \nonumber \\ \nonumber
&=\frac{P(\{Y=y\}|\{a<I<b\}) P(\{x \in D\}|\{a<I<b\})}{P(\{Y\leq T\}|\{a<I<b\})P(\{x \in D\}|\{a<I<b\})} \\ \nonumber
&=\frac{P(\{Y=y\}|\{a<I<b\})}{P(\{Y\leq T\}|\{a<I<b\})} \\ \nonumber
&=\frac{P(\{Y=y\}\cap\{a<I<b\})}{P(\{Y\leq T\}\cap\{a<I<b\})}\\ \nonumber
&=\frac{\int_a^bP(\{Y=y\}\cap\{I=i\})di}{\int_a^bP(\{Y\leq T\}\cap\{I=i\})di}=\frac{\int_a^bg(i)f_X(C-i)di}{\int_a^bg(i)\int_0^{T-i} f_X(x)dxdi}.
\end{align}
Therefore, the likelihood function $P(Y=y|a<I<b, Y \leq T)$ is independent of the reporting rate for the data coming out of Wuhan in the early days of the outbreak. 

\section{Estimating the generation size based using a lower bound on the number of symptomatic individuals}
\label{app:Hugo}
In the analysis in Section~\ref{sec:gen-size}, it is assumed that every person who has developed symptoms by time $\tau$ is known to the observer. However, depending on the disease, symptoms can be subjective. One person may not notice something another person may visit hospital for. Additionally, one person may not want to come forward with symptoms if they are worried about the repercussions of coming forward (for example being isolated against their own will).

To address this, rather than considering $I_{\tau}$ as the total number of people who have developed symptoms by time $\tau$, we can consider $\tilde{I}_{\tau}$ as the number of people who have presented with symptoms by time $\tau$. We do no know the true value of $I_{\tau}$, but we know that it cannot be below $\tilde{I}_{\tau}$. We assume that the probability of $\tilde{I}_{\tau}$ being $\tilde{i}_{\tau}$ for a given value of $i_{\tau}$ is uniform at $\frac{1}{i_{\tau}+1}$. We can then use the same methods as above to infer a distribution for $P$.

\begin{equation*}
  \mathbb{P}\left(\tilde{I}_\tau=\tilde{i}_\tau\middle|I_\tau=i_\tau\right)=\left\{
  \begin{array}{@{}ll@{}}
    \frac{1}{i_\tau + 1}, & \text{if }0\leq \tilde{i}_\tau\leq i_\tau \\
    0, & \text{otherwise}
  \end{array}\right.
\end{equation*} 

\subsection{Inferring a distribution for $P$ given $\tilde{I}_{\tau}=\tilde{i}_{\tau}$}
We again assume an uninformative uniform prior distribution for $P$ and as a result the posterior distribution for $P$ is proportional to the likelihood that $\tilde{I}_{\tau}=\tilde{i}_{\tau}$ given $P=p$:

\begin{align*}
    \mathbb{P}\left(P=p\middle|\tilde{I}_{\tau}=\tilde{i}_{\tau}\right)\propto \mathbb{P}\left(\tilde{I}_{\tau}=\tilde{i}_{\tau}\middle|P=p\right)\\
    =\sum_{i_{\tau}=\tilde{i}_{\tau}}^n\mathbb{P}\left(I_{\tau}=i_{\tau}\middle|P=p\right)\times\frac{1}{i_{\tau}+1}\\
    =\sum_{i_{\tau}=\tilde{i}_{\tau}}^n\binom{n}{i_{\tau}}\left(pF\left(\tau\right)\right)^{i_{\tau}}\left(1-p F\left(\tau\right)\right)^{n-i_{\tau}}\frac{1}{i_{\tau}+1}\\
    =\binom{n}{\tilde{i}_{\tau}}\frac{\left(pF\left(\tau\right)\right)^{\tilde{i}_{\tau}}}{\tilde{i}_{\tau}+1}\left(1-pF\left(\tau\right)\right)^{n-\tilde{i}_{\tau}}{_2F_1}\left(1,\tilde{i}_{\tau}-n,\tilde{i}_{\tau}+2,\frac{-p F\left(\tau\right)}{1-pF\left(\tau\right)}\right)\\
    \\
    \mathbb{P}\left(P=p\middle|\tilde{I}_{\tau}=\tilde{i}_{\tau}\right) = c\times\\
    \binom{n}{\tilde{i}_{\tau}}\frac{\left(p F\left(\tau\right)\right)^{\tilde{i}_{\tau}}}{\tilde{i}_{\tau}+1}\left(1-p F\left(\tau\right)\right)^{n-\tilde{i}_{\tau}}{_2F_1}\left(1,\tilde{i}_{\tau}-n,\tilde{i}_{\tau}+2,\frac{-pF\left(\tau\right)}{1-pF\left(\tau\right)}\right)
\end{align*}
where $c$ is a normalising constant such that
\begin{align*}
    c = \left(\int_0^1\binom{n}{\tilde{i}_{\tau}}\frac{\left(pF\left(\tau\right)\right)^{\tilde{i}_{\tau}}}{\tilde{i}_{\tau}+1}\left(1-p F\left(\tau\right)\right)^{n-\tilde{i}_{\tau}} {_2F_1}\left(1,i_l-n,i_l+2,\frac{-p F\left(\tau\right)}{1 - p F\left(\tau\right)}\right)dp\right)^{-1}\\
    =\left(\binom{n}{i_l}\frac{F\left(\tau\right)^{\tilde{i}_{\tau}}}{\tilde{i}_{\tau}+1}\int_0^1p^{\tilde{i}_{\tau}}\left(1-pF\left(\tau\right)\right)^{n-\tilde{i}_{\tau}} {_2F_1}\left(1,\tilde{i}_{\tau}-n,\tilde{i}_{\tau}+2,\frac{-pF\left(\tau\right)}{1-pF\left(\tau\right)}\right)dp\right)^{-1}\\
    \mathbb{P}\left(P=p\middle|\tilde{I}_\tau=\tilde{i}_\tau\right)=p^{\tilde{i}_{\tau}}\left(1-pF\left(\tau\right)\right)^{n-\tilde{i}_{\tau}}{_2F_1}\left(1,\tilde{i}_{\tau}-n,\tilde{i}_{\tau}+2,\frac{-pF\left(\tau\right)}{1-pF\left(\tau\right)}\right)\\
    \times\left(\int_0^1y^{\tilde{i}_{\tau}}\left(1-yF\left(\tau\right)\right)^{n-\tilde{i}_{\tau}} {_2F_1}\left(1,\tilde{i}_{\tau}-n,\tilde{i}_{\tau}+2,\frac{-yF\left(\tau\right)}{1-yF\left(\tau\right)}\right)dy\right)^{-1}
\end{align*}
However, this integral cannot be found analytically and instead must be calculated numerically.
\subsection{Inferring a distribution for $E_0$ given $\tilde{I}_\tau=\tilde{i}_\tau$}\label{blargh}
We can use the same probability analysis in Section~\ref{sec:gen-size} to write the following probability formula:
\begin{align*}
    \mathbb{P}\left(E_0=e_o\middle|\tilde{I}_\tau=\tilde{i}_\tau\right) = \int_0^1\frac{\mathbb{P}\left(E_0=e_o\cap \tilde{I}_\tau=\tilde{i}_\tau\middle|P=p\right)}{\mathbb{P}\left(\tilde{I}_\tau=\tilde{i}_\tau\middle|P=p\right)}\times \mathbb{P}\left(P=p\middle|\tilde{I}_\tau=\tilde{i}_\tau\right)dp\\
    =\int_0^1\frac{\binom{n}{e_0}p^{e_0}\left(1-p\right)^{n-e_0}\sum_{i=\tilde{i}_\tau}^{e_0}\binom{e_0}{i}F\left(\tau\right)^i\left(1-F\left(\tau\right)\right)^{e_0-i}\times\frac{1}{i+1}}{\binom{n}{\tilde{i}_{\tau}}\frac{\left(pF\left(\tau\right)\right)^{\tilde{i}_{\tau}}}{\tilde{i}_{\tau}+1}\left(1-pF\left(\tau\right)\right)^{n-\tilde{i}_{\tau}} {_2F_1}\left(1,\tilde{i}_{\tau}-n,\tilde{i}_{\tau}+2,\frac{-p F\left(\tau\right)}{1-pF\left(\tau\right)}\right)}\\
    \times p^{\tilde{i}_{\tau}}\left(1-pF\left(\tau\right)\right)^{n-\tilde{i}_{\tau}} {_2F_1}\left(1,\tilde{i}_{\tau}-n,\tilde{i}_{\tau}+2,\frac{-pF\left(\tau\right)}{1-pF\left(\tau\right)}\right)\\
    \times\left(\int_0^1y^{\tilde{i}_{\tau}}\left(1-yF\left(\tau\right)\right)^{n-\tilde{i}_{\tau}} {_2F_1}\left(1,\tilde{i}_{\tau}-n,\tilde{i}_{\tau}+2,\frac{-yF\left(\tau\right)}{1-yF\left(\tau\right)}\right)dy\right)^{-1}dp\\
    =\int_0^1\frac{\left(\frac{p\left(1-F\left(\tau\right)\right)}{1-p}\right)^{e_0}\left(\frac{1}{1-F\left(\tau\right)}\right)^{\tilde{i}_\tau}\left(1-p\right)^n\left(n-\tilde{i}_\tau\right)! {_2F_1}\left(1,\tilde{i}_\tau-e_0,\tilde{i}_\tau+2,\frac{-F\left(\tau\right)}{1-F\left(\tau\right)}\right)}{\left(e_0-\tilde{i}_\tau\right)!\left(n-e_0\right)!}\\
    \times\left(\int_0^1y^{\tilde{i}_{\tau}}\left(1-yF\left(\tau\right)\right)^{n-\tilde{i}_{\tau}} {_2F_1}\left(1,\tilde{i}_{\tau}-n,\tilde{i}_{\tau}+2,\frac{-yF\left(\tau\right)}{1-yF\left(\tau\right)}\right)dy\right)^{-1}dp\\
\end{align*}

\section{Population and household transmission: Individual isolation strategy}
\label{app:exhh}

Individual isolation does not intimately involve the household and so we assume
that a fraction $\alpha_{\mathrm{I}}$ of symptomatic cases self-isolates and
ceases transmission outside the household, meaning that we take the baseline
but with
\be
r_{s\rightarrow e}(t,\mathbf{Q}) = \Lambda(t) +
\sum_{n=1}^{n_{\mathrm{max}}}
\sum_{s=0}^{n}
\sum_{e=0}^{(n-s)}
\sum_{p=0}^{(n-s-e)}
\sum_{i=0}^{(n-s-e-p)}
\left(p \beta_p + (1-\alpha_{\mathrm{I}}) i \beta_i \right) Q_{n,s,e,p,i}
\text{ .}
\ee
To capture the essential features of household quarantine, we need to add
states to the dynamical variables.
Let $Q_{n,s,e,p,i,\mathbf{f}}(t)$ be the proportion of households in the
population at time $t$ of size $n$, with $s$ susceptibles, $e$ exposed, $p$
prodromal, and $i$ symptomatic infectious individuals, and with vector of
`flags' $\mathbf{f}$ representing implementation of more complex interventions.

We now suppose that a fraction $\alpha_{\mathrm{S}}$ of households start to
isolate when there is at least one symptomatic case in the household, and stop
isolation 14 days after the absence of symptoms in the household. This is
modelled by having a flag $f=0$ if the household is not isolating and $f=1$ if
it is. The dynamics become
\ba
\ddt{}Q_{n,s,e,p,i,f} & = - \left(\indic{f=0}s r_{s\rightarrow e}(t,\mathbf{Q}) + e
r_{e\rightarrow p} + p r_{p\rightarrow i} + i r_{i\rightarrow \emptyset} 
+ n^{-\eta}s p \tau_p  + n^{-\eta}s i  \tau_i 
+ \indic{f=1} \sigma \right) Q_{n,s,e,p,i,f}
\\ & \qquad
+ \indic{f=0}(s+1) r_{s\rightarrow e}(t,\mathbf{Q}) Q_{n,s+1,e-1,p,i,f}
+ (e+1) r_{e\rightarrow p} Q_{n,s,e+1,p-1,i,f}
\\ & \qquad
+ (1-\indic{f=1 \& i=1 \& s+e+p=n-1} \alpha_{\mathrm{S}})
(p+1) r_{p\rightarrow i} Q_{n,s,e,p+1,i-1,f}
+ (i+1) r_{i\rightarrow \emptyset} Q_{n,s,e,p,i+1,f}
\\ & \qquad
+ n^{-\eta}(s+1) p \tau_p Q_{n,s+1,e-1,p,i,f}
+ n^{-\eta}(s+1) i \tau_i Q_{n,s+1,e-1,p,i,f}
\\ & \qquad
+ \indic{f=0} \sigma Q_{n,s,e,p,i,1}
+ \indic{f=1 \& i=1 \& s+e+p=n-1} \alpha_{\mathrm{S}}
(p+1)r_{p\rightarrow i} Q_{n,s,e,p+1,i-1,0}
\text{ ,}
\ea
with between-household term
\be
r_{s\rightarrow e}(t,\mathbf{Q}) = \Lambda(t) +
\sum_{n=1}^{n_{\mathrm{max}}}
\sum_{s=0}^{n}
\sum_{e=0}^{(n-s)}
\sum_{p=0}^{(n-s-e)}
\sum_{i=0}^{(n-s-e-p)}
\left(p \beta_p + i \beta_i \right) Q_{n,s,e,p,i,0}
\text{ .}
\ee

\section{Non-Markovian household model of within-household transmission}
\begin{figure}
\centering
\includegraphics[width=0.8\textwidth]{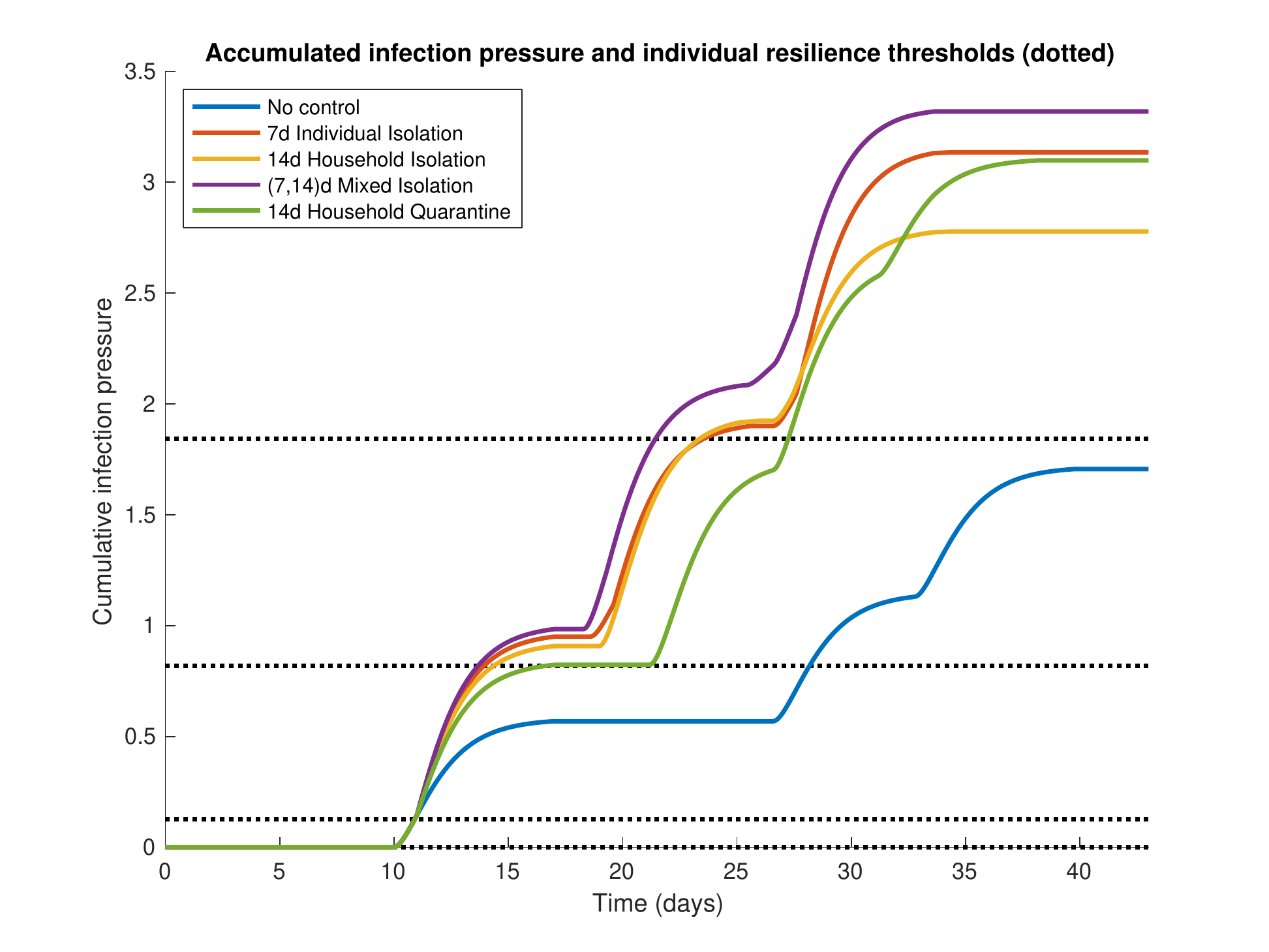}
\caption{Accumulated infection pressure in the simulation presented in Figure \ref{fig:L-within} for different control policies. Horizontal dotted lines represent individuals' resilience thresholds. As time progresses, the accumulated infection pressures (coloured lines) increase and when they cross the resilience thresholds, the corresponding individual acquires infection. Notice that: in the absence of control, one individual escapes infection; with household isolation only, the infection pressure reaches a relatively low endpoint because of the last symptomatic individual slipping through and not transmitting much in the household; with mixed isolation, infection pressure is higher due to combined adherence; and with household quarantine, the infection pressure builds up more slowly at the beginning due to lower adherence.}
\label{fig:L-supp}
\end{figure}

The non-Markovian household model of Section \ref{sec:within} relies on the Sellke construction. In relation to Figure \ref{fig:L-within}, we report in Figure \ref{fig:L-supp} the infection pressure that accumulates for the different control policies.

\end{appendices}
%%%%_____________________________________________________________

\section*{Acknowledgements}
%The authors thank Robert Sawko at IBM Research for discussions on numerical implementation of household models. 
PD and TF are supported by the NIHR Manchester Biomedical Research Centre. LP, HS and CO are funded by the Wellcome Trust and the Royal Society (grant 202562/Z/16/Z). EF is funded by the MRC (grant MR/S020462/1). MF is supported by The Alan Turing Institute under the EPSRC grant EP/N510129/1. TH is supported by the Royal Society (Grant Number INF/R2/180067) and Alan Turing Institute for Data Science and Artificial Intelligence. IH is supported by the National Institute for Health Research Health Protection Research Unit (NIHR HPRU) in Emergency Preparedness and Response and the National Institute for Health Research Policy Research Programme in Operational Research (OPERA) and Alan Turing Institute for Data Science and Artificial Intelligence.

\section*{Author contributions}
CO and HS compiled the manuscript. All authors were involved in the research and revising of the manuscript. 

\section*{Competing interests}
The authors declare no competing interests.

\section*{Data and materials}
All data and code used in this analysis is provided in the Github repository \texttt{https://github.com/thomasallanh ouse/covid19-stochastics}, with the exception of UK specific data. This data is provided by Public Health England under a data sharing agreement and we are unable to share this data.

\bibliography{refs5}

\begin{thebibliography}{10}

\bibitem{CNInc}
Recorded daily case updates of covid-19 cases.
\newblock \url{http://3g.dxy.cn/newh5/view/pneumonia}.
\newblock Data scraped daily from 21-1-2020 to 3-2-2020.

\bibitem{Abramowitz1948}
M.~Abramowitz and I.~A. Stegun.
\newblock {\em Handbook of mathematical functions with formulas, graphs, and
  mathematical tables}, volume~55.
\newblock US Government printing office, 1948.

\bibitem{Alexander2012}
H.~K. Alexander and S.~Bonhoeffer.
\newblock Pre-existence and emergence of drug resistance in a generalized model
  of intra-host viral dynamics.
\newblock {\em Epidemics}, 4(4):187--202, 2012.

\bibitem{Andersson2000}
H.~Andersson and T.~Britton.
\newblock {Stochastic epidemics in dynamic populations: quasi-stationarity and
  extinction}.
\newblock {\em Journal of Mathematical Biology}, 41(6):559--580, 2000.

\bibitem{andreasen2011final}
V.~Andreasen.
\newblock The final size of an epidemic and its relation to the basic
  reproduction number.
\newblock {\em Bulletin of Mathematical Biology}, 73(10):2305--2321, 2011.

\bibitem{Backer2020}
J.~A. Backer, D.~Klinkenberg, and J.~Wallinga.
\newblock {Incubation period of 2019 novel coronavirus (2019-nCoV) infections
  among travellers from Wuhan, China, 20--28 January 2020}.
\newblock {\em Eurosurveillance}, 25(5), 2020.

\bibitem{Ball:1999}
F.~Ball.
\newblock Stochastic and deterministic models for {SIS} epidemics among a
  population partitioned into households.
\newblock {\em Mathematical Biosciences}, 156(1):41--67, 1999.

\bibitem{BALL201563}
F.~Ball, T.~Britton, T.~House, V.~Isham, D.~Mollison, L.~Pellis, and G.~Scalia
  Tomba.
\newblock Seven challenges for metapopulation models of epidemics, including
  households models.
\newblock {\em Epidemics}, 10:63 -- 67, 2015.
\newblock Challenges in Modelling Infectious Disease Dynamics.

\bibitem{Ball2011}
F.~Ball, T.~Britton, and D.~Sirl.
\newblock {Household epidemic models with varying infection response}.
\newblock {\em Journal of Mathematical Biology}, 63(2):309--337, 2011.

\bibitem{ball1997}
F.~Ball, D.~Mollison, and G.~Scalia-Tomba.
\newblock Epidemics with two levels of mixing.
\newblock {\em The Annals of Applied Probability}.

\bibitem{BALL2016108}
F.~Ball, L.~Pellis, and P.~Trapman.
\newblock Reproduction numbers for epidemic models with households and other
  social structures {II}: Comparisons and implications for vaccination.
\newblock {\em Mathematical Biosciences}, 274:108 -- 139, 2016.

\bibitem{Ball2016}
F.~Ball, L.~Pellis, and P.~Trapman.
\newblock {Reproduction numbers for epidemic models with households and other
  social structures II: comparisons and implications for vaccination}.
\newblock {\em Mathematical Biosciences}, 274:108--139, 2016.

\bibitem{Ball2015}
F.~G. Ball, E.~S. Knock, and P.~D. O’Neill.
\newblock Stochastic epidemic models featuring contact tracing with delays.
\newblock {\em Mathematical Biosciences}, 266:23--35, 2015.

\bibitem{Biggerstaff2020}
M.~Biggerstaff, F.~S. Dahlgren, J.~Fitzner, D.~George, A.~Hammond, I.~Hall,
  D.~Haw, N.~Imai, M.~A. Johansson, S.~Kramer, et~al.
\newblock Coordinating the real-time use of global influenza activity data for
  better public health planning.
\newblock {\em Influenza and Other Respiratory Viruses}, 14(2):105--110, 2020.

\bibitem{Black:2017}
A.~J. Black, N.~Geard, J.~M. McCaw, J.~McVernon, and J.~V. Ross.
\newblock Characterising pandemic severity and transmissibility from data
  collected during first few hundred studies.
\newblock {\em Epidemics}, 19:61--73, 2017.

\bibitem{Britton2019}
T.~Britton and G.~Scalia~Tomba.
\newblock Estimation in emerging epidemics: Biases and remedies.
\newblock {\em Journal of the Royal Society Interface}, 16(150):20180670, 2019.

\bibitem{Butler2014}
D.~Butler.
\newblock Models overestimate {E}bola cases.
\newblock {\em Nature}, 515(7525):18, 2014.

\bibitem{Cauchemez:2004}
S.~Cauchemez, F.~Carrat, C.~Viboud, A.~J. Valleron, and P.~Y. Bo\"{e}lle.
\newblock A {Bayesian MCMC} approach to study transmission of influenza:
  application to household longitudinal data.
\newblock {\em Statistics in Medicine}, 23(22):3469--3487, 2004.

\bibitem{Cauchemez2009}
S.~Cauchemez, C.~A. Donnelly, C.~Reed, A.~C. Ghani, C.~Fraser, C.~K. Kent,
  L.~Finelli, and N.~M. Ferguson.
\newblock {\em New England Journal of Medicine}.

\bibitem{Chew2010}
C.~Chew and G.~Eysenbach.
\newblock {Pandemics in the age of Twitter: content analysis of Tweets during
  the 2009 H1N1 outbreak}.
\newblock {\em PloS One}, 5(11), 2010.

\bibitem{Chinazzi2020}
M.~Chinazzi, J.~T. Davis, M.~Ajelli, C.~Gioannini, M.~Litvinova, S.~Merler,
  A.~P. y~Piontti, K.~Mu, L.~Rossi, K.~Sun, et~al.
\newblock The effect of travel restrictions on the spread of the 2019 novel
  coronavirus {(COVID-19)} outbreak.
\newblock {\em Science}, 2020.

\bibitem{cqc}
Care~Quality Commission.
\newblock {CQC care directory - with ratings (1 April 2020)}, 2020.

\bibitem{SPIM}
{Department of Health and Social Care}.
\newblock {SPI-M} modelling summary for pandemic influenza, Edition: Nov 2018.

\bibitem{Du2020}
Z.~Du, X.~Xu, Y.~Wu, L.~Wang, B.~J. Cowling, and L.~A. Meyers.
\newblock The serial interval of {COVID-19} from publicly reported confirmed
  cases.
\newblock {\em medRxiv}, 2020.

\bibitem{Fang2020}
L.~Fang, G.~Karakiulakis, and M.~Roth.
\newblock {Are patients with hypertension and diabetes mellitus at increased
  risk for COVID-19 infection?}
\newblock {\em The Lancet Respiratory Medicine}, 8(4):e21, 2020.

\bibitem{Farewell2005}
V.~T. Farewell, A.~M. Herzberg, K.~W. James, L.~M. Ho, and G.~M. Leung.
\newblock {SARS} incubation and quarantine times: when is an exposed individual
  known to be disease free?
\newblock {\em Statistics in Medicine}, 24(22):3431--3445, 2005.

\bibitem{Ferguson2020}
N.~M. Ferguson, D.~Laydon, G.~Nedjati-Gilani, N.~Imai, K.~Ainslie, M.~Baguelin,
  S.~Bhatia, A.~Boonyasiri, Z.~Cucunub{\'a}, G.~Cuomo-Dannenburg, et~al.
\newblock Impact of non-pharmaceutical interventions {(NPIs)} to reduce
  {COVID-19} mortality and healthcare demand.

\bibitem{Ferretti2020}
L.~Ferretti, C.~Wymant, M.~Kendall, L.~Zhao, A.~Nurtay, L.~Abeler-D{\"o}rner,
  M.~Parker, D.~Bonsall, and C.~Fraser.
\newblock {Quantifying SARS-CoV-2 transmission suggests epidemic control with
  digital contact tracing}.
\newblock {\em Science}, 2020.

\bibitem{Fraser2007}
C.~Fraser.
\newblock Estimating individual and household reproduction numbers in an
  emerging epidemic.
\newblock {\em PloS One}, 2(8), 2007.

\bibitem{Funk2009}
S.~Funk, E.~Gilad, C.~Watkins, and V.~A.~A. Jansen.
\newblock The spread of awareness and its impact on epidemic outbreaks.
\newblock {\em Proceedings of the National Academy of Sciences},
  106(16):6872--6877, 2009.

\bibitem{Ganyani2020}
P.~Ganyani, C.~Kremer, D.~Chen, A.~Torneri, C.~Faes, J.~Wallinga, and N.~Hens.
\newblock Estimating the generation interval for {COVID-19} based on symptom
  onset data.
\newblock {\em medRxiv}, 2020.

\bibitem{Goldstein2009}
E.~Goldstein, K.~Paur, C.~Fraser, E.~Kenah, J.~Wallinga, and M.~Lipsitch.
\newblock Reproductive numbers, epidemic spread and control in a community of
  households.
\newblock {\em Mathematical Biosciences}, 221(1):11--25, 2009.

\bibitem{Gostic2020}
K.~M. Gostic, A.~C.~R. Gomez, R.~O. Mummah, A.~J. Kucharski, and J.~O.
  Lloyd-Smith.
\newblock {Estimated effectiveness of symptom and risk screening to prevent the
  spread of COVID-19}.
\newblock {\em eLife}, 9:1--18, 2020.

\bibitem{Guan2020}
W.~Guan, Z.~Ni, Y.~Hu, W.~Liang, C.~Ou, J.~He, L.~Liu, H.~Shan, C.~Lei,
  D.~S.~C. Hui, et~al.
\newblock Clinical characteristics of coronavirus disease 2019 in {C}hina.
\newblock {\em New England Journal of Medicine}, 2020.

\bibitem{OxCGRT}
Thomas Hale et~al.
\newblock {Oxford COVID-19 Government Response Tracker (OxCGRT)}, 2020.
\newblock
  https://www.bsg.ox.ac.uk/research/research-projects/oxford-covid-19-government-response-tracker.

\bibitem{House:2008}
T.~House and M.~J. Keeling.
\newblock Deterministic epidemic models with explicit household structure.
\newblock {\em Mathematical Biosciences}, 213(1):29--39, 2008.

\bibitem{House2011}
T.~House and M.~J. Keeling.
\newblock Epidemic prediction and control in clustered populations.
\newblock {\em Journal of Theoretical Biology}, 272(1):1--7, 2011.

\bibitem{bbcvaccine}
S.~Jack.
\newblock {Coronavirus: GSK and Sanofi join forces to create vaccine}, 2020.

\bibitem{Kalbfleisch1991}
J.~D. Kalbfleisch and J.~F. Lawless.
\newblock Regression models for right truncated data with applications to
  {AIDS} incubation times and reporting lags.
\newblock {\em Statistica Sinica}, pages 19--32, 1991.

\bibitem{travel2}
W.~Keju.
\newblock China braces for world's biggest travel rush around {S}pring
  {F}estival, 2019.

\bibitem{Kendall1948}
D.~G. Kendall et~al.
\newblock On the generalized ``birth-and-death" process.
\newblock {\em The Annals of Mathematical Statistics}, 19(1):1--15, 1948.

\bibitem{Kermack1927}
W.~O. Kermack and A.~G. McKendrick.
\newblock A contribution to the mathematical theory of epidemics.
\newblock {\em Proceedings of the Royal Society of London. Series A, Containing
  papers of a mathematical and physical character}, 115(772):700--721, 1927.

\bibitem{Kinyanjui:2018}
T.~Kinyanjui, J.~Middleton, S.~G\"{u}ttel, J.~Cassell, J.~Ross, and T.~House.
\newblock Scabies in residential care homes: Modelling, inference and
  interventions for well-connected population sub-units.
\newblock {\em PLOS Computational Biology}, 14(3):e1006046, 2018.

\bibitem{Kraemer2020}
M.~U.~G. Kraemer, C.~Yang, B.~Gutierrez, C.~Wu, B.~Klein, D.~M. Pigott,
  L.~du~Plessis, N.~R. Faria, R.~Li, W.~P. Hanage, et~al.
\newblock The effect of human mobility and control measures on the covid-19
  epidemic in china.
\newblock {\em Science}, 368(6490):493--497, 2020.

\bibitem{Lau2017}
M.~S.Y. Lau, B.~D. Dalziel, S.~Funk, A.~McClelland, A.~Tiffany, S.~Riley,
  C.~J.~E. Metcalf, and B.~T. Grenfell.
\newblock {Spatial and temporal dynamics of superspreading events in the
  2014-2015 West Africa Ebola epidemic}.
\newblock {\em Proceedings of the National Academy of Sciences of the United
  States of America}, 114(9):2337--2342, 2017.

\bibitem{Lauer2020}
S.~A. Lauer, K.~H. Grantz, Q.~Bi, F.~K. Jones, Q.~Zheng, H.~Meredith, A.~S.
  Azman, N.~G. Reich, and J.~Lessler.
\newblock The incubation period of {2019-nCoV} from publicly reported confirmed
  cases: estimation and application.
\newblock {\em medRxiv}, 2020.

\bibitem{Lesko2020}
C.~R. Lesko, A.~P. Keil, and J.~K. Edwards.
\newblock {The Epidemiologic Toolbox: Identifying, Honing, and Using the Right
  Tools for the Job}.
\newblock {\em American Journal of Epidemiology}.

\bibitem{LiQ2020}
Q.~Li, X.~Guan, P.~Wu, X.~Wang, L.~Zhou, Y.~Tong, R.~Ren, K.~S.~M. Leung,
  E.~H.~Y. Lau, J.~Y. Wong, et~al.
\newblock Early transmission dynamics in {W}uhan, {C}hina, of novel
  coronavirus--infected pneumonia.
\newblock {\em New England Journal of Medicine}, 2020.

\bibitem{LiR2020}
R.~Li, S.~Pei, B.~Chen, Y.~Song, T.~Zhang, W.~Yang, and J.~Shaman.
\newblock Substantial undocumented infection facilitates the rapid
  dissemination of novel coronavirus {(SARS-CoV2)}.
\newblock {\em Science}, 2020.

\bibitem{Lin2020}
H.~Lin, W.~Liu, H.~Gao, J.~Nie, and Q.~Fan.
\newblock Trends in transmissibility of 2019 novel coronavirus-infected
  pneumonia in {W}uhan and 29 provinces in {C}hina.
\newblock {\em Available at SSRN 3544821}, 2020.

\bibitem{Linton2020}
N.~M. Linton, T.~Kobayashi, Y.~Yang, K.~Hayashi, A.~R. Akhmetzhanov, S.~Jung,
  B.~Yuan, R.~Kinoshita, and H.~Nishiura.
\newblock Incubation period and other epidemiological characteristics of 2019
  novel coronavirus infections with right truncation: A statistical analysis of
  publicly available case data.
\newblock {\em Journal of Clinical Medicine}, 9(2):538, 2020.

\bibitem{LiuT2020}
T.~Liu, J.~Hu, J.~Xiao, G.~He, M.~Kang, Z.~Rong, L.~Lin, H.~Zhong, Q.~Huang,
  A.~Deng, et~al.
\newblock Time-varying transmission dynamics of novel coronavirus pneumonia in
  {C}hina.
\newblock {\em bioRxiv}, 2020.

\bibitem{Liu2020a}
Y.~Liu, A.~A. Gayle, A.~Wilder-Smith, and J.~Rockl{\"{o}}v.
\newblock {The reproductive number of COVID-19 is higher compared to SARS
  coronavirus}.
\newblock {\em Journal of Travel Medicine}, 2020.

\bibitem{Consequences2014}
F.~M.~G. Magpantay, M.~A. Riolo, M.~Domenech~de Cell\`{e}s, A.~A. King, and
  P.~Rohani.
\newblock {Epidemiological Consequences of Imperfect}.
\newblock {\em SIAM Journal on Applied Mathematics}, 74(6):1810--1830, 2014.

\bibitem{Mahasem308}
E.~Mahase.
\newblock China coronavirus: what do we know so far?
\newblock {\em BMJ}, 368, 2020.

\bibitem{Majumder2020b}
M.~S. Majumder and K.~D. Mandl.
\newblock {Early in the epidemic: impact of preprints on global discourse about
  COVID-19 transmissibility}.
\newblock {\em The Lancet Global Health}.

\bibitem{Mizumoto2020}
K.~Mizumoto, K.~Kagaya, A.~Zarebski, and G.~Chowell.
\newblock Estimating the asymptomatic proportion of coronavirus disease 2019
  {(COVID-19)} cases on board the {Diamond Princess cruise ship, Yokohama,
  Japan}, 2020.
\newblock {\em Eurosurveillance}, 25(10):2000180, 2020.

\bibitem{travel1}
H.~Morris.
\newblock The largest human migration on the planet: {W}hat happens in {C}hina
  when 1.4bn go on holiday at the same time.
\newblock
  https://www.telegraph.co.uk/travel/news/chinese-new-year-chunyun-in-numbers/.

\bibitem{Mossong2008}
J.~Mossong, N.~Hens, M.~Jit, P.~Beutels, K.~Auranen, R.~Mikolajczyk,
  M.~Massari, S.~Salmaso, G.~Scalia Tomba, J.~Wallinga, et~al.
\newblock Social contacts and mixing patterns relevant to the spread of
  infectious diseases.
\newblock {\em PLoS Medicine}, 5(3), 2008.

\bibitem{Nishiura2010}
H.~Nishiura.
\newblock Time variations in the generation time of an infectious disease:
  implications for sampling to appropriately quantify transmission potential.
\newblock {\em Mathematical Biosciences \& Engineering}, 7(4):851--869, 2010.

\bibitem{Nishiura2020}
H.~Nishiura, T.~Kobayashi, T.~Miyama, A.~Suzuki, S.~Jung, K.~Hayashi,
  R.~Kinoshita, Y.~Yang, B.~Yuan, A.~R. Akhmetzhanov, et~al.
\newblock Estimation of the asymptomatic ratio of novel coronavirus infections
  {(COVID-19)}.
\newblock {\em medRxiv}, 2020.

\bibitem{FH:2019}
{Office for National Statistics}.
\newblock Families and households, Edition: 15 November 2019.
\newblock
  https://www.ons.gov.uk/peoplepopulationandcommunity/birthsdeathsandmarriages/families/datasets
  \\ /familiesandhouseholdsfamiliesandhouseholds.

\bibitem{Census:2001}
{Office for National Statistics}.
\newblock 2001 census aggregate data, Edition: May 2011.

\bibitem{whositrep}
World~Health Organisation.
\newblock Coronavirus disease 2019 {(COVID-19). Situation Report - 69.}, 2020.

\bibitem{PELLIS201285}
L.~Pellis, F.~Ball, and P.~Trapman.
\newblock Reproduction numbers for epidemic models with households and other
  social structures. {I}. definition and calculation of ${R}_0$.
\newblock {\em Mathematical Biosciences}, 235(1):85 -- 97, 2012.

\bibitem{Pellis2020}
L.~Pellis, S.~Cauchemez, N.~M. Ferguson, and C.~Fraser.
\newblock {Systematic selection between age and household structure for models
  aimed at emerging epidemic predictions}.
\newblock {\em Nature Communications}, 11(1), 2020.

\bibitem{Pellis2011a}
L.~Pellis, N.~M. Ferguson, and C.~Fraser.
\newblock {Epidemic growth rate and household reproduction number in
  communities of households, schools and workplaces}.
\newblock {\em Journal of Mathematical Biology}, 63(4):691--734, 2011.

\bibitem{pellis2020challenges}
L.~Pellis, F.~Scarabel, H.~B. Stage, C.~E. Overton, L.~H.~K. Chappell, K.~A.
  Lythgoe, E.~Fearon, E.~Bennett, J.~Curran-Sebastian, R.~Das, M.~Fyles,
  H.~Lewkowicz, X.~Pang, B.~Vekaria, L.~Webb, T.~House, and I.~Hall.
\newblock {Challenges in control of Covid-19: short doubling time and long
  delay to effect of interventions}, 2020.

\bibitem{Read2020}
J.~M. Read, J.~R.~E. Bridgen, D.~A.~T. Cummings, A.~Ho, and C.~P. Jewell.
\newblock {Novel coronavirus 2019-nCoV: early estimation of epidemiological
  parameters and epidemic predictions}.
\newblock {\em MedRxiv}, 2020.

\bibitem{Remuzzi2020}
A.~Remuzzi and G.~Remuzzi.
\newblock {COVID-19} and italy: what next?
\newblock {\em The Lancet}, 2020.

\bibitem{Rimmer2020}
.~Rimmer.
\newblock {COVID-19: Disproportionate impact on ethnic minority healthcare
  workers will be explored by government}.
\newblock {\em BMJ}, 369, 2020.

\bibitem{Ross:2009}
J.~V. Ross, T.~House, and M.~J. Keeling.
\newblock Calculation of disease dynamics in a population of households.
\newblock {\em PLOS ONE}, 5(3):e9666, 2010.

\bibitem{Ross1910}
R.~Ross.
\newblock {\em The Prevention of Malaria}.
\newblock Dutton, 1910.

\bibitem{Rubin2009}
G.~J. Rubin, R.~Aml{\^o}t, L.~Page, and S.~Wessely.
\newblock Public perceptions, anxiety, and behaviour change in relation to the
  swine flu outbreak: cross sectional telephone survey.
\newblock {\em BMJ}, 339:b2651, 2009.

\bibitem{Tomba2010}
G.~Scalia~Tomba, {\AA}.~Svensson, T.~Asikainen, and J.~Giesecke.
\newblock Some model based considerations on observing generation times for
  communicable diseases.
\newblock {\em Mathematical Biosciences}, 223(1):24--31, 2010.

\bibitem{sellke_1983}
T.~Sellke.
\newblock On the asymptotic distribution of the size of a stochastic epidemic.
\newblock {\em Journal of Applied Probability}, 20(2):390–394, 1983.

\bibitem{Shaw2016}
L.~M. Shaw.
\newblock {\em {SIR} epidemics in a population of households}.
\newblock PhD thesis, University of Nottingham, 2016.

\bibitem{Su2012}
Y.~Su and J.~Wang.
\newblock Modeling left-truncated and right-censored survival data with
  longitudinal covariates.
\newblock {\em Annals of statistics}, 40(3):1465, 2012.

\bibitem{Sun1995}
J.~Sun.
\newblock Empirical estimation of a distribution function with truncated and
  doubly interval-censored data and its application to {AIDS} studies.
\newblock {\em Biometrics}, pages 1096--1104, 1995.

\bibitem{Sun2020}
K.~Sun, J.~Chen, and C.~Viboud.
\newblock Early epidemiological analysis of the coronavirus disease 2019
  outbreak based on crowdsourced data: a population-level observational study.
\newblock {\em The Lancet Digital Health}, 2020.

\bibitem{Svensson2007}
{\AA}.~Svensson.
\newblock A note on generation times in epidemic models.
\newblock {\em Mathematical Biosciences}, 208(1):300--311, 2007.

\bibitem{Taylor2003}
D.~J. Taylor, M.~A. Weaver, and R.~E. Roddy.
\newblock Evaluating factors associated with {STD} infection in a study with
  interval-censored event times and an unknown proportion of participants not
  at risk for disease.
\newblock {\em Statistics in medicine}, 22(13):2191--2204, 2003.

\bibitem{Valdano2019a}
E.~Valdano, C.~Poletto, P.~Boelle, and V.~Colizza.
\newblock {Reorganization of nurse scheduling reduces the risk of healthcare
  associated infections}.
\newblock {\em medRxiv}, pages 1--18, 2019.

\bibitem{Kerkhove2012}
M.~D. Van~Kerkhove and N.~M. Ferguson.
\newblock Epidemic and intervention modelling: a scientific rationale for
  policy decisions? {L}essons from the 2009 influenza pandemic.
\newblock {\em Bulletin of the World Health Organization}, 90:306--310, 2012.

\bibitem{Varia2003}
M.~Varia, S.~Wilson, S.~Sarwal, A.~McGeer, E.~Gournis, E.~Galanis, and B.~H.
  For.
\newblock {Investigation of a nosocomial outbreak of severe acute respiratory
  syndrome (SARS) in Toronto, Canada}.
\newblock {\em Canadian Medical Association Journal}, 169(4):285--292, 2003.

\bibitem{Virlogeux2016}
V.~Virlogeux, V.~J. Fang, M.~Park, J.~T. Wu, and B.~J. Cowling.
\newblock {Comparison of incubation period distribution of human infections
  with MERS-CoV in South Korea and Saudi Arabia}.
\newblock {\em Scientific Reports}, 6(1):35839, 2016.

\bibitem{Wilder2020}
A.~Wilder-Smith, C.~J. Chiew, and V.~J. Lee.
\newblock Can we contain the {COVID-19} outbreak with the same measures as for
  {SARS}?
\newblock {\em The Lancet Infectious Diseases}, 2020.

\bibitem{WuJT2020}
J.~T. Wu, K.~Leung, M.~Bushman, N.~Kishore, R.~Niehus, P.~M. de~Salazar, B.~J.
  Cowling, M.~Lipsitch, and G.~M. Leung.
\newblock {Estimating clinical severity of COVID-19 from the transmission
  dynamics in Wuhan, China}.
\newblock {\em Nature Medicine}, pages 1--5, 2020.

\bibitem{WuZ2020}
Z.~Wu and J.~M. McGoogan.
\newblock Characteristics of and important lessons from the coronavirus disease
  2019 {(COVID-19)} outbreak in {C}hina: summary of a report of 72 314 cases
  from the {Chinese Center for Disease Control and Prevention}.
\newblock {\em Jama}, 2020.

\bibitem{Yang2020}
J.~Yang, Y.~Zheng, X.~Gou, K.~Pu, Z.~Chen, Q.~Guo, R.~Ji, H.~Wang, Y.~Wang, and
  Y.~Zhou.
\newblock {Prevalence of comorbidities in the novel Wuhan coronavirus
  (COVID-19) infection: a systematic review and meta-analysis}.
\newblock {\em International Journal of Infectious Diseases}, 2020.

\bibitem{Zhou2020}
F.~Zhou, T.~Yu, R.~Du, G.~Fan, Y.~Liu, Z.~Liu, J.~Xiang, Y.~Wang, B.~Song,
  X.~Gu, et~al.
\newblock Clinical course and risk factors for mortality of adult inpatients
  with {COVID-19} in {W}uhan, {C}hina: a retrospective cohort study.
\newblock {\em The Lancet}, 2020.

\end{thebibliography}
\end{document}